\documentclass[reprint,superscriptaddress,amsmath,amssymb,notitlepage,aps,prx,tightenlines]{revtex4-2}

\usepackage[utf8]{inputenc}
\usepackage{graphicx}
\usepackage{bm}
\usepackage{xcolor}
\usepackage{float}
\usepackage{layouts}
\usepackage{booktabs}
\usepackage{braket}
\usepackage{tikz}
\usepackage{siunitx}
\DeclareSIUnit{\atomicunit}{a.u.}
\DeclareSIUnit{\gauss}{G}

\usepackage[colorlinks=True, citecolor=blue, urlcolor=blue, linkcolor=blue]{hyperref}

\newcommand*\circled[1]{\tikz[baseline=(char.base)]{
            \node[shape=circle,draw,inner sep=1.0pt] (char) {#1};}}

\renewcommand{\selectlanguage}[1]{}

\begin{document}
\title{Quantum computer architecture with ions in tweezer arrays}
\author{Benjamin F.~Schiffer}
\email{benjamin.schiffer@mpq.mpg.de}
\affiliation{Max-Planck-Institut für Quantenoptik, Hans-Kopfermann-Str.~1, D-85748~Garching, Germany}%
\affiliation{Munich Center for Quantum Science and Technology (MCQST), Schellingstr.~4, D-80799~Munich, Germany}
\author{Christopher Monroe}
\email{christopher.monroe@duke.edu}
\affiliation{Duke Quantum Center, Department of Electrical and Computer Engineering and Department of Physics, Duke University, Durham, NC 27708, USA}
\affiliation{IonQ, Inc., Boston, MA  02135, USA}
\author{Peter Zoller}
\email{peter.zoller@uibk.ac.at}
\affiliation{Institute for Theoretical Physics, University of Innsbruck, A-6020 Innsbruck, Austria}
\affiliation{Institute for Quantum Optics and Quantum Information (IQOQI), Austrian Academy of Sciences, A-6020 Innsbruck, Austria}
\author{J.~Ignacio Cirac}
\email{ignacio.cirac@mpq.mpg.de}
\affiliation{Max-Planck-Institut für Quantenoptik, Hans-Kopfermann-Str.~1, D-85748~Garching, Germany}%
\affiliation{Munich Center for Quantum Science and Technology (MCQST), Schellingstr.~4, D-80799~Munich, Germany}%

\date{\today}

\begin{abstract}
    We propose a quantum computer architecture based on ions confined in optical tweezer arrays, combining the long coherence times of trapped-ion qubits with the reconfigurability and parallel operation enabled by tweezer platforms. 
    Selected ions are transported to local interaction zones, where excitation to an auxiliary state with a displaced optical potential generates a controllable effective electric dipole. 
    We develop and analyze entangling-gate mechanisms mediated by the Coulomb interaction between such effective dipoles, and show that they enable precise, temperature-robust closure of the center-of-mass and relative motional trajectories, leaving no residual entanglement between the qubits and the motion. 
    We further outline a concrete implementation with barium ions based on state-selective polarizability, and study the suppression of cross-talk during parallel gate execution, with relevance to transversal gates in quantum error correction. 
    Our results thereby establish a realistic route toward scalable ion-tweezer quantum processors.
\end{abstract}

\maketitle 

\section{Introduction}

The development of fault-tolerant quantum computers (FTQCs) has become the defining challenge for all quantum hardware platforms, requiring simultaneous advances in qubit number, gate parallelism, and the architectural connectivity demanded by quantum error-correcting codes~\cite{Nielsen2010Quantum, Preskill1998Faulttolerant, Campbell2017Roads}. 
Among atomic systems, trapped ions have emerged as leaders in gate fidelity and qubit coherence, owing to the near-perfect isolation of their long-lived internal electronic states from the environment, and  Coulomb-mediated motional modes that enable high-quality entangling gates~\cite{Cirac1995Quantum, Sorensen1999Quantum, Cirac2000Scalable, Leibfried2003Quantum, Haffner2008Quantum, Bruzewicz2019TrappedIon}. 
Quantum charge-coupled device (QCCD) architectures have powerfully extended this paradigm by shuttling ions through microtrap networks, enabling flexible connectivity while preserving outstanding gate fidelities~\cite{Kielpinski2002Architecture, Pino2021Demonstration, Ransford202698qubit}. 
In these ion-trap architectures, fixed electrode layouts provide the stable and well-controlled environments underlying their performance, whereas the ion positions in a crystal are only reconfigured indirectly through voltages applied to distant electrodes. 
Scaling hardware platforms toward FTQC naturally places increasing demands on transport speed, gate parallelism, and architectural reconfigurability. 
This motivates the central question addressed here: whether individual ions can be confined and controlled by direct optical methods and thereby open a new and complementary route toward scalable FTQC, following the example set by neutral-atom tweezer platforms.

Neutral-atom tweezer arrays have developed into a powerful and complementary platform, demonstrating highly parallel architectural capabilities~\cite{Endres2016Atombyatom, Barredo2016Atombyatom, Browaeys2020ManyBody, Bluvstein2022Quantum, Graham2022Multiqubit}. 
Freely reconfigurable arrays allow coherent control over thousands of atoms, with parallel transport enabling all-to-all connectivity and naturally matching transversal gate operations for quantum error correction~\cite{Manetsch2025Tweezer, Bluvstein2026Faulttolerant}. 
High-fidelity entangling gates are mediated by laser excitation to Rydberg levels, and the resulting van der Waals interactions enable fast operations on a timescale set by the Rydberg interaction~\cite{Jaksch2000Fast, Levine2019Parallel, Lukin2001Dipole, Evered2023Highfidelity}. 
The performance of these gates reflects a balance between strong Rydberg interactions and incoherent processes associated with the excitation, including Rydberg-state decay and scattering through intermediate states~\cite{Saffman2010Quantum, Zhang2012Fidelity, Saffman2016Quantum, Cohen2021Quantum, Saffman2025Quantum, Norrell2026Entangling}.

In this paper, we propose a qualitatively new architecture for ion-based quantum computing that directly addresses the challenges of previous atomic hardware platforms. 
We propose confining ions directly in optical tweezer arrays, instead of relying only on conventional static or radiofrequency electric fields.
This combines the long-lived internal-state qubits, high-fidelity gates, and long-range Coulomb interactions of trapped ions with the scalability, reconfigurability, and all-to-all connectivity of neutral-atom tweezer platforms. 
Entangling gates are realized through state-dependent tweezer displacements that engineer a large effective electric dipole on demand, in loose analogy to Rydberg dipole-dipole interactions, but involving metastable levels with lifetimes many orders of magnitude longer, thereby strongly suppressing errors from spontaneous emission.  
The long-range Coulomb interaction further enables sympathetic cooling by co-trapped ions, which has no direct neutral-atom counterpart~\cite{Kielpinski2000Sympathetic, Belyansky2019Nondestructive}. 
The remarkable maturation of tweezer technology, which now enables large-scale, individually addressable, reconfigurable atomic arrays, makes the presented architecture particularly timely and practically realizable, opening a concrete engineering roadmap for scaling ion-based quantum computing toward FTQC.

The use of optical potentials in trapped-ion architectures was discussed in early proposals, with ions trapped in nonconsecutive wells and state-dependent pushing gates~\cite{Cirac2000Scalable}. 
A later analysis considered ions loaded into optical-lattice microtraps~\cite{Schmied2008Quantum}, followed by demonstrations of optical dipole trapping of individual ions~\cite{Schneider2010Optical, Huber2014Faroffresonance, Schaetz2017Trapping}. 
State-selective optical potentials have furthermore been experimentally demonstrated for trapping and reshaping ion Coulomb crystals~\cite{Weckesser2021Trapping}. 
In contrast, individually steerable optical tweezers have been explored only to a limited degree in the context of trapped-ion quantum computing. 
For example, tweezers may generate state-dependent forces on ions within a Coulomb crystal~\cite{Mazzanti2023Trapped}, and precisely calibrated momentum kicks can enable fast gates, although their scalability may be challenging~\cite{RobaloPereira2025Fast}. 
Optical tweezers have also been proposed as a tool in combination with conventional traps for manipulating the phonon modes~\cite{Olsacher2020Scalable, Shen2020Scalable, Teoh2021Manipulating}.
Our proposal is therefore naturally connected to earlier work, but follows a distinct route by using steerable optical tweezers as a central resource for trapping, rearranging, and coupling ions.
We emphasize that the conventional ion trap is still useful in the background, in order to preserve very long trapping lifetimes, while the optical tweezers are used for local control of the atomic motion.

This paper is organized as follows.
The ion-tweezer architecture is described in Sec.~\ref{sec:architecture}.
We then develop and analyze a family of two-qubit entangling gates for this architecture, formulated in terms of the center-of-mass and relative motional modes of the two-ion system introduced in Sec.~\ref{ssec:physsetup}. 
We present three \textit{dynamical gates} in Sec.~\ref{ssec:dynamical}. 
A first protocol involves only a single trap displacement and is temperature independent, requiring no ground-state cooling; a second operates for arbitrary Coulomb interaction strengths with similar temperature robustness; a third achieves gate times shorter than the oscillation period of the trap, provided sufficient optical power. 
In addition, we describe a \textit{dipole-blockade gate} in Sec.~\ref{ssec:blockade}, demonstrating that ions can be viewed as artificial Rydberg atoms.
In Sec.~\ref{sec:considerations} we present a concrete implementation based on barium ions and their state-selective polarizability, and show that spatially separated gate zones support parallel operation, enabling efficient transversal gate execution, a key primitive for fault-tolerant quantum error correction.
Finally, we discuss our results in Sec.~\ref{sec:discussion}.

\begin{figure}[t]
    \centering
    \includegraphics[width=\columnwidth]{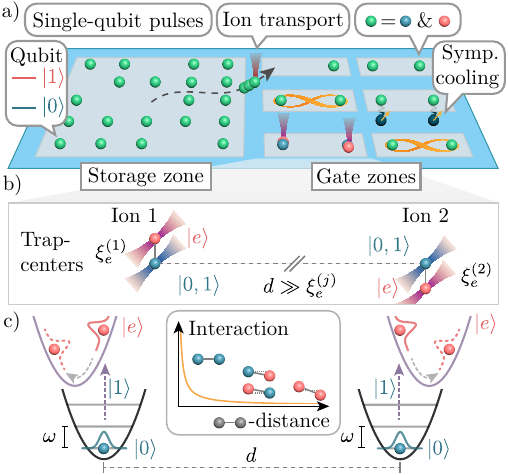}
    \caption{\textbf{Ion-tweezer quantum computing architecture. (a)} Qubits are encoded in atomic levels of Ba$^{+}$, and ions may be transported between a storage zone and gate zones. 
    Entangling gates are realized by state-dependent displaced optical potentials via \textit{push tweezers}. 
    Additional \textit{cooling ions} allow for mid-circuit sympathetic cooling.
    \textbf{(b)} Gate geometry in which the trap centers for the excited states $\ket{e}$ are displaced perpendicular to the main gate axis by $\xi_e^{(j)}$ for ion $j$. 
    The inter-ion distance $d$ is much larger than the trap-center shifts during the gate.
    \textbf{(c)} Illustration of harmonic traps with frequency $\omega$, where displaced wavepackets oscillate in the excited state.
    The different inter-ion separations give rise to state-dependent Coulomb interactions. 
    By jointly engineering the motional dynamics and these interactions, one can realize high-fidelity entangling gates.}
    \label{fig:fig1}
\end{figure}

\section{Ion-tweezer architecture} \label{sec:architecture}

The quantum computer architecture we propose is inspired by neutral-atom optical-tweezer platforms.
The qubits are stored in long-lived internal states $\ket{0}$ and $\ket{1}$ of the ions, which are confined in individual tightly focused optical dipole traps (the \textit{qubit tweezers}).
For example, ions can be loaded by first assembling neutral atoms in optical tweezers and then photoionizing them.
Single-qubit gates can be carried out as in conventional trapped-ion or neutral-atom platforms, either by local addressing or by transport to a dedicated single-qubit zone, where they can be performed in parallel.
A two-ion gate is realized by transporting the selected ions to an interaction region, where they are brought close to one another while remaining sufficiently separated from the rest of the array, which may simultaneously execute other operations, as illustrated in Fig.~\ref{fig:fig1}(a). 
Important advantages of ions compared with neutral atoms include the presence of a deep confinment potential in the background, allowing for very long trapping lifetimes, and the possibility of sympathetic cooling:
the data ions can be coupled to auxiliary \textit{cooling ions} through shared motional modes, allowing entropy to be removed without destroying the stored quantum information~\cite{Kielpinski2000Sympathetic}. 

For two-qubit entangling gates, the ions are excited with a shelving laser to a long-lived state $\ket{e}$, but only when the qubit is in state $\ket{1}$. 
In that state, the ions experience a second tweezer beam (the \textit{push tweezer}), which produces a state-dependent optical dipole potential acting only on $\ket{e}$ in the idealized model and is displaced with respect to the original tweezer beam [Fig.~\ref{fig:fig1}(b,c)].
In this way, we can engineer a large effective electric dipole for a single ion.
The resulting branch-dependent Coulomb interaction generates different phases across the computational basis states.
This can be seen as analogous in spirit to Rydberg interactions, but it also enables new gate mechanisms specific to an ion-tweezer platform that we exploit below.

There are several requirements that must be satisfied in order to obtain a high-fidelity gate.
First, the motional state at the end of the process should not retain any trace of the internal state of the ions.
Ideally, this should hold independently of the initial motional state, so that the gate remains effective at finite temperature. 
Because the Coulomb interaction makes their center-of-mass (CM) and relative (rel) modes oscillate at different frequencies, this requirement is nontrivial.
In addition, the gate should be only weakly sensitive to small imperfections such as control errors.
Since the ions also accumulate single-qubit phase shifts, those shifts should remain moderate so that they can be controlled or compensated. 
Finally, it is desirable for the gate time to be as short as possible.
In the following we will show how these requirements can be met. 

\begin{figure*}[t]
    \centering
    \includegraphics[width=\textwidth]{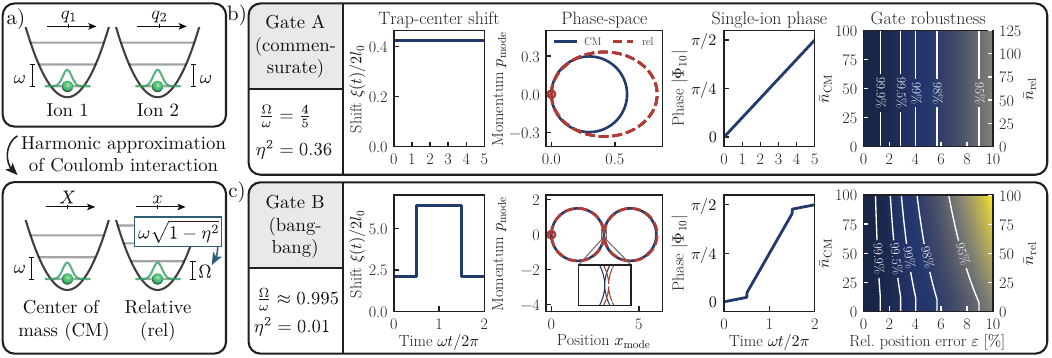}
    \caption{\textbf{Two-ion gate analysis. (a)} For the two-ion geometry, expanding the Coulomb interaction yields a center-of-mass (CM) mode with coordinate $X$ and frequency $\omega$, together with a relative (rel) mode with coordinate $x$ and frequency $\Omega=\omega\sqrt{1-\eta^2}$.
    \textbf{(b)} We analyze the commensurate gate and show the trap-center position shift $\xi(t)/2l_0$, perpendicular to the main axis; the clockwise evolving phase-space trajectories for the $\ket{01}$ branch, starting and ending at the circle marker; the single-ion phase as a function of the gate time; and a fidelity heatmap analyzing the gate robustness for a relative position error $\varepsilon$ of the trap center and an initial thermal state, showing no temperature dependence.
    \textbf{(c)} A three-segment pulse (bang--bang \textit{Gate B}) realizes fast gates for arbitrary values of $\eta^2$. 
    We present the same analysis as above for $\eta^2=0.01$, including an inset highlighting the small difference between the two phase-space trajectories.
    }
    \label{fig:fig2}
\end{figure*}

\section{Two-ion gate} \label{sec:two-qubit}
\subsection{Physical Setup} \label{ssec:physsetup}
Let us consider two ions of mass $m$ confined in respective harmonic traps created by two identical laser tweezers, separated by a distance $d$.
We assume isotropic traps with trap frequency $\omega$ and set $\hbar=1$.
For each ion, the qubit is stored in two ground electronic states; see Sec.~\ref{sec:considerations} for details on the experimental realization.
The gates are carried out by transferring $\ket{1}$ to a long-lived electronic excited state, $\ket{e}$. 
Thus, the laser induces the transition
\begin{equation} \label{Eq:1toe}
    \ket{0}\to\ket{0}, \qquad \ket{1} \to \ket{e}.
\end{equation}
The internal level $\ket{e}$ is subjected to the same harmonic trap as the qubit states, but with its trap center displaced by some amount perpendicular to the line joining the trap centers. 
We can write the motional Hamiltonian as
\begin{equation}
   H_{\rm mot} = \sum_{r,s\in\{0,1,e\}} |r,s\rangle\langle r,s| \otimes  H_{r,s}
\end{equation}
where $H_{r,s}=(p_1^2+p_2^2)/2m + V_{r,s}$ is an effective one-dimensional Hamiltonian that includes the kinetic terms, the Coulomb interaction, as well as the potentials created by the tweezers, which are displaced if the corresponding ion is in the excited state ($r=e$ or $s=e$). 
We refer to the Supplementary Information (SI) for a detailed presentation. 
Further, we assume that the displacements of the ions, $q_1, q_2$, remain small compared with the inter-ion distance, $|q_2-q_1|\ll d$.
The interaction strength may then be quantified by the dimensionless parameter
\begin{equation}
    \eta^2=\frac{e^2}{2\pi\varepsilon_0m d^3\omega^2},
\end{equation}
which is tuned by the distance $d$.

We express the potential part of the Hamiltonian in terms of CM and rel coordinates, with $X=(q_1+q_2)/2$ and $x=q_2-q_1$:
\begin{equation}
    V_{r,s}= \frac{M\omega^2}{2}(X-X^{\mathrm{eq}})^2 +\frac{\mu \Omega^2}{2}(x-x^{\mathrm{eq}})^2 + C_{r,s},
\end{equation}
where $M/2=m=2\mu$, and the rel mode frequency is $\Omega^2 = \omega^2 (1-\eta^2)$. 
The equilibrium positions are denoted by $X^{\mathrm{eq}}$ and $x^{\mathrm{eq}}$, and $C_{r,s}$ is a constant energy offset.
The previous quantities depend on the displacements of the traps, and thus on the internal state of the ions (and we have partially omitted the corresponding subscripts $r,s$ to simplify notation). 
We illustrate the transformation to the new modes in Fig.~\ref{fig:fig2}(a).

The specific form of the Hamiltonian can be more easily expressed in terms of dimensionless parameters. 
We therefore define the ground-state size of the harmonic potentials, $l_0=\sqrt{1/2m\omega}$, and standard annihilation operators $a_{\rm CM}$, $a_{\rm rel}$.
We write
\begin{align}
    H_{r,s} &= D_{r,s}^{} H_0^{} D_{r,s}^\dagger + C_{r,s}^{}, \label{eq:H_displaced}\\
    H_0^{} &= \omega a_{\rm CM}^\dagger a_{\rm CM}^{} + \Omega a_{\rm rel}^\dagger a_{\rm rel}^{},
\end{align}
where we have omitted the zero-point energy, which is common-mode and does not play any role, and where the displacement operator is~\cite{Weedbrook2012Gaussian}:
\begin{equation*}
    D_{r,s}= \exp\!\left[\alpha_{r,s}^{\rm CM} (a_{\rm CM}^\dagger -a_{\rm CM}^{})\right] \exp\!\left[\alpha_{r,s}^{\rm rel} (a_{\rm rel}^\dagger -a_{\rm rel}^{})\right].
\end{equation*}
Here, the coherent-state displacements for the two modes are
\begin{equation}
    \alpha_{r,s}^{\rm CM}=\frac{\xi_r^{(1)}+\xi_s^{(2)}}{2\sqrt{2}\,l_0}, \qquad
    \alpha_{r,s}^{\rm rel}=\frac{\xi_s^{(2)}-\xi_r^{(1)}}{2\sqrt{2}\,l_0(\Omega/\omega)^{3/2}},
\end{equation}
where $\xi_r^{(1)}=0$ for $r=0,1$, $\xi_s^{(2)}=0$ for $s=0,1$, and $\xi_e^{(j)}$ is the excited-state trap displacement of ion $j$.
Importantly, the energy shift 
\begin{equation}  \label{eq.Vshift}
     C_{r,s}=- \frac{\omega \eta^2}{8(1-\eta^2)} \left(\frac{\xi_r^{(1)}-\xi_s^{(2)}}{l_0}\right)^2
\end{equation}
is the branch-dependent scalar offset of the displaced harmonic Hamiltonian in the presence of the Coulomb interaction.

Since the displacements depend on the internal states, this term induces a state-dependent phase shift during the evolution.
The goal is to drive the ions back, $\ket{e}\to\ket{1}$, at the right time so that the total process implements the desired two-qubit gate
\begin{equation} \label{Eq:gate} 
    \ket{i,j} \to e^{i \pi (i-j)^2/2}\ket{i,j}, \quad i,j\in\{0,1\}. 
\end{equation}
Up to local rotations, Eq.~\eqref{Eq:gate} is a controlled-Z gate, so in our analysis we separate correctable single-qubit phases from the non-local entangling phase $\Phi_{ZZ}$.

In practice, the situation is more complicated: once the ions are excited by the laser, they start to oscillate in the displaced potential, so that the internal states become entangled with the motional ones, thus degrading the quality of the gate. 
In the following we explain how this can be avoided.

\subsection{Dynamical gate} \label{ssec:dynamical}

We now describe a family of two-qubit entangling gates for the ion-tweezer architecture in the limit where the excitation in Eq.~\eqref{Eq:1toe} is effectively instantaneous compared with the trap dynamics.
The branch Hamiltonians from Sec.~\ref{ssec:physsetup} describe the CM and rel modes as displaced harmonic oscillators with a branch-dependent scalar shift. 
A high-fidelity gate is obtained when two conditions are satisfied~\cite{Garcia-Ripoll2003Speed}.
First, the CM and rel phase-space trajectories must close exactly at the final time, so that the final positions and momenta coincide with their initial values and the motion disentangles from the qubits.
Second, the phases accumulated in the computational branches $\ket{i,j}$ must generate the non-local phase required by Eq.~\eqref{Eq:gate}.
We introduce the pulse sequences below; additional details are given in the Methods, and full derivations for all protocols are provided in the SI.

\subsubsection*{Gate A (commensurate)}

The simplest protocol is a single-hold pulse, where the excited-state traps are shifted once and then kept fixed for the full gate duration.
Let us consider the case where the perpendicular mode ratio is commensurate,
\begin{equation}
    \frac{\Omega}{\omega}=\sqrt{1-\eta^2}=\frac{p}{q},
\end{equation}
with coprime integers $p<q$. 
The fact that $\eta^2$ can be tuned by choosing the distance between the ions is an advantage over conventional trapped-ion setups, because it allows one to set the factors $p$ and $q$. 
Waiting for a time $T_G=2\pi q/\omega$ then restores the initial motional state exactly.

In order to achieve the desired gate from Eq.~\eqref{Eq:gate}, the two excited-state traps are shifted in opposite directions, $\xi_e^{(1)}=-\xi_{\mathrm{CZ}}$ and $\xi_e^{(2)}=+\xi_{\mathrm{CZ}}$, with
\begin{equation} \label{eq:gateA_xi_z}
    \xi_{\mathrm{CZ}} = 2l_0 z_{\mathrm{CZ}},
    \qquad
    z_{\mathrm{CZ}}^2=\frac{1-\eta^2}{2q\eta^2},
\end{equation}
which yields the entangling phase $\Phi_{ZZ}=\pi$.
We simulate this commensurate protocol (\emph{Gate A}) for $\Omega/\omega=4/5$, corresponding to $\eta^2=0.36$.
For this example one has $z_{\mathrm{CZ}}=\xi_{\mathrm{CZ}}/2l_0\approx0.42$, as shown in Fig.~\ref{fig:fig2}(b).
The leftmost subpanel plots the dimensionless shift $\xi(t)/2l_0 = z g(t)$ during the gate, while the next panel shows, for the $\ket{01}$ branch, the closed phase-space trajectories for a common clock, using the horizontal coordinates $x_{\rm CM}=X/(\sqrt{2}\,l_0)$ and $x_{\rm rel}=(\Omega/\omega)^{1/2}x/(2\sqrt{2}\,l_0)$ and vertical coordinates $p_{\rm mode}=\partial x_{\rm mode}/\partial(\omega t)$.
A third subpanel shows the accumulated single-particle phase, which can be calibrated.
The rightmost subpanel shows the fidelity as a function of the relative position error of the shift, together with its temperature independence for initial thermal states expressed via the expected occupation of each mode at a common temperature. 
Within the harmonic model, a uniform displacement error changes the acquired phase but does not affect closure.
Additional details on the numerical implementation and the fidelity model are given in the SI.

The limitation of \emph{Gate A} is the commensurability requirement, which prevents one from choosing a relatively small $\eta^2$. 
In that regime, $q\gg 1$ and one would have to wait for many trap oscillations, making the gate fidelity more sensitive to imperfections such as trap-frequency drift.
This motivates a more general strategy in which we include additional positional changes of the trap center.

\subsubsection*{Gate B (bang--bang)}

To remove the commensurability constraint, we keep the same perpendicular geometry but allow the excited-state traps to follow a three-segment piecewise-constant waveform.
This allows us to describe a two-ion gate with a fast gate time of $T_G=4\pi/\omega$ for arbitrary values of $\eta^2$, while keeping the single-particle phase small and maintaining robustness against thermal excitation.

We parameterize the physical trap-center shifts as
\begin{equation*}
    \xi_e^{(1)}(t) = -\xi(t), \quad
    \xi_e^{(2)}(t) = +\xi(t), \quad
    \xi(t)=2l_0 z g(t),
\end{equation*}
where $z$ sets the overall dimensionless displacement scale and $g(t)$ is piecewise constant, with values $(1,g_m,1)$ over the three time intervals $(\pi/\omega,\,2\pi/\omega,\,\pi/\omega)$.
Closure of the rel mode then fixes
\begin{equation}
    g_m = 1 - 2\cos\big(\pi\sqrt{1-\eta^2}\big),
\end{equation}
while the CM mode closes automatically.
Compared with \emph{Gate A}, the total phase now contains two ingredients: the integrated branch-dependent scalar shift $\int C_{r,s}[g(t)]\,{\rm d}t$ and additional jump phases generated by the sudden trap-center changes. 
The corresponding phase-space bookkeeping is summarized in the Methods, while the analytical formulas for the single-particle phase $\Phi_{10}$ and the entangling phase $\Phi_{ZZ}$ are given in the SI.

We simulate such a bang--bang sequence in Fig.~\ref{fig:fig2}(c) for $\eta^2=0.01$, corresponding to a larger ion separation than in the commensurate case above; concrete values for barium ions are given in Sec.~\ref{ssec:phys_implement}.
For the plotted example one finds $g_m\approx3.00$ and $z_{\mathrm{CZ}}\approx2.12$, so that the leftmost subpanel, which plots $\xi(t)/2l_0=z g(t)$, has plateau values $(2.12,6.36,2.12)$.
The subpanels are the same as for \emph{Gate A}, now with an inset that highlights the small difference between the CM and rel phase-space trajectories. 
We again consider relative position errors in the trap-center displacements and find that the gate remains highly robust for initial thermal states.

The gates discussed so far have in common that the gate time is limited by the trap frequency.
It turns out that we can achieve gate times that are much shorter with a more flexible piecewise-constant pulse.

\subsubsection*{Gate C (sub-trap-period gates)} 

\begin{figure}[t]
    \centering
    \includegraphics[width=\columnwidth]{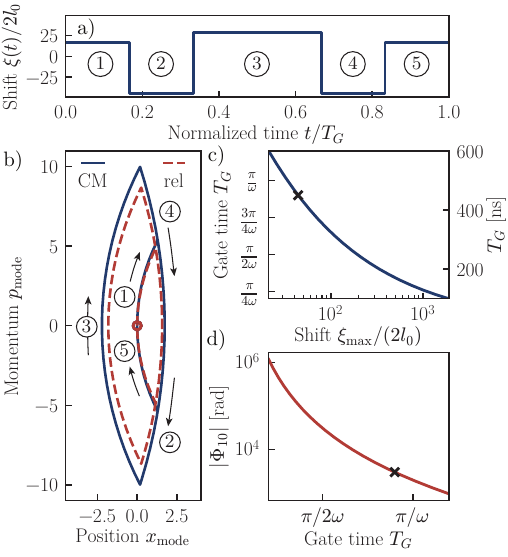}
    \caption{\textbf{Sub-trap-period entangling gates. (a)} A five-segment pulse is compatible with arbitrary values of $\eta^2$ and allows one to reduce the gate time $T_G$ below the trap period.
    Here, $T_G=\qty{450}{\nano\second}$ for $\omega/2\pi=\qty{1}{\mega\hertz}$ and $\eta^2=0.36$.
    \textbf{(b)} Coherent-state evolution in phase space for the $\ket{01}$ branch during the five segments.
    \textbf{(c)} Gate time $T_G$ as a function of the maximum shift $\xi_{\rm max} = \max_t{|\xi(t)|}$. 
    A marker indicates the value of $T_G$ considered in panels (a,b).
    \textbf{(d)} Magnitude of the single-ion phase as a function of the gate time.
    }
    \label{fig:fig3}
\end{figure}

We can engineer a very fast gate by not following full oscillation periods in phase space, but instead combining arcs of oscillations at different trap-center positions.
This allows us to find phase-space trajectories where both modes close jointly, while achieving a sub-trap-period gate time $T_G<2\pi/\omega$.
Ion gates that are faster than the trap period have already been proposed~\cite{Garcia-Ripoll2003Speed, Garcia-Ripoll2005Coherent}, but those proposals focused on ions trapped in Coulomb crystals.

Here, we show that a five-segment pulse is sufficient to achieve exact mode closure for fast gates and arbitrary values of $\eta^2$.
The trap center shifts for such a pulse are shown in Fig.~\ref{fig:fig3}(a) and we identify phase-space arcs with the segment number in Fig.~\ref{fig:fig3}(b).
This symmetric pulse scheme is fully determined by the three trap-center positions during the segments \circled{1}, \circled{2}, and \circled{3}; the analytic expressions for the positions are included in the Methods.
The required maximum shift $\xi_{\rm max} = \max_t{|\xi(t)|}$ is smaller when the two ions are placed closer to each other, so that their interaction is stronger. 
Fig.~\ref{fig:fig3}(c,d) shows the gate time as a function of the maximum shift and the magnitude of the accumulated single-particle phase as a function of the gate time.
A marker highlights the $T_G=\qty{450}{\nano\second}$ gate at $\eta^2=0.36$, to which the other panels correspond.

Although the five-segment pulse allows, in principle, for arbitrarily short gate times, in practice the speed is limited by the anharmonicity of the tweezer potential and by the precision with which the resulting single-particle phase can be calibrated. 
The first limitation is controlled by how far the trap center can be shifted while remaining in the approximately harmonic regime. 
The extent of this regime is set by the shape of the tweezer potential and must be compared with the displacements required by a given pulse. 
We discuss this constraint concretely for the case of barium ions in Sec.~\ref{sec:considerations} and in more detail in the SI, and show that there are realistic parameter regimes for sub-trap-period gates. 
Importantly, as long as the ion wavepacket is small compared with the beam waist throughout the pulse, the motion can still be described locally by a harmonic problem, with effective mode frequencies that become time dependent when anharmonicity is present and reduce to constants in the harmonic limit. 
The dynamics then remain Gaussian and can be incorporated exactly into the gate evolution, using standard optimal-control methods~\cite{Garcia-Ripoll2005Coherent, Glaser2015Training}. 
Alternatively, the accessible regime can be extended by engineering traps with a larger harmonic region, for example by superimposing an additional potential, adding a Laguerre--Gaussian beam, or using a spatial light modulator. 
The second limitation arises from the large single-particle phase accumulated for large trap displacements. 
Such phases can be mitigated, for example, by using an echo sequence in which the gate is repeated with permuted computational-basis states, or by employing additional optical tweezers.

\subsection{Dipole-blockade gate} \label{ssec:blockade}

Beyond the gates discussed above, the same platform also supports a conceptually distinct blockade-type gate.
In this case, the ion acts as an \emph{artificial Rydberg atom}, directly inspired by the Rydberg-blockade mechanism~\cite{Jaksch2000Fast}:
a $2\pi$ pulse is resonant in one branch while another branch experiences an interaction-induced detuning to suppress the excitation.

In the same spirit, we describe a dipole-blockade gate for two ions that proceeds in a sequence of three steps: 
\begin{itemize}
    \item[(1)] a laser drives the first ion $\ket{1}_1\to \ket{e}_1$; 
    \item[(2)] a second pulse drives the second (target) ion\\
                $\ket{1}_2\to \ket{e}_2 \to \ket{1}_2$; 
    \item[(3)] the first step is reversed so that $\ket{e}_1\to\ket{1}_1$.
\end{itemize}
As in the dynamical gates above, the state $\ket{e}$ experiences a displaced optical potential.
If ion~1 is effectively held in the shifted potential, it acts as an effective dipole and changes the spectrum for exciting the target ion.
A target pulse can then be chosen to drive a resonant $2\pi$-rotation in the unblocked branch so that this branch accumulates a $\pi$-phase.
After accounting for local phases, this realizes a dipole-blockade phase gate.

The analogy with Rydberg atoms becomes more explicit when the motional states of the ions are viewed as the analog of the Rydberg levels. 
We assume that the ions are initially in the motional ground state.
Then, they will have to be driven resonantly to a joint Fock state of the CM and rel modes, which requires spectral resolution to realize the gate.
The corresponding Rabi frequency must therefore remain smaller than the level spacing between the relevant Fock states, which keeps the pulse duration on the scale of several trap periods.

\begin{figure}[t]
    \centering
    \includegraphics[width=\columnwidth]{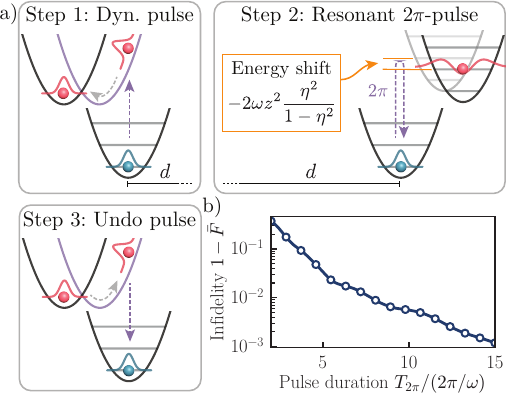}
    \caption{
    \textbf{Dipole-blockade gate. (a)} Schematic of the three-step gate sequence. 
    The excitation of the first ion can be implemented, for example, by a dynamical pulse, as visualized here. 
    The middle step is a resonant $2\pi$ pulse, where the interaction energy shifts the Fock states off resonance, resulting in a branch-dependent phase.
    A third step reverses the initial pulse.
    \textbf{(b)} Gate infidelity as a function of the $2\pi$ pulse duration for an initial vacuum state. 
    We choose $\Omega/\omega=4/5$ and $z=2/3$ and assume a perfect execution of Steps 1 \& 3.
    A resonant pulse with a duration of several trap periods allows one to realize a high-fidelity gate.
    }
    \label{fig:fig4}
\end{figure}

We may implement the gate with a combination of dynamical operations and resonant pulses:
the same concepts for piecewise displaced oscillators used in \textit{Gate A} and \textit{Gate B} can realize steps (1) and (3) in a dynamical way.
Ion~1 is then first displaced and subsequently parked at a shifted position, such that the CM and rel phase-space trajectories close.
Fig.~\ref{fig:fig4}(a) illustrates the three-step sequence schematically.
We include a numerical example in Fig.~\ref{fig:fig4}(b), using the commensurate perpendicular geometry with $\Omega/\omega=4/5$, corresponding to $\eta^2=0.36$, because this provides a larger spectral separation between the relevant Fock states than for smaller $\eta^2$.
For the same perpendicular geometry as in Fig.~\ref{fig:fig1}(b), the blocked target transition in the dynamical protocol is detuned by
\begin{equation}
    \nu_z^{\mathrm{(dyn)}} = -2\omega z^2\frac{\eta^2}{1-\eta^2}.
    \label{eq:nuz-def}
\end{equation}
In the numerical benchmark, we assume that the dynamical preparation and its reversal in the first and third steps are perfect. 
We therefore simulate only the second step as a smooth resonant target $2\pi$ pulse for $z=2/3$, starting from the motional vacuum state. 
Additional details are included in the Methods and the SI.

Compared with the previous dynamical gates, this blockade protocol is temperature sensitive, since the coupling to higher-energy motional states depends on the initial Fock state. 
In practice this favors a gate operation after sympathetic cooling, or alternatively the use of more robust adiabatic or composite target pulses.

\section{Experimental and Scaling Considerations} \label{sec:considerations}

\subsection{Physical implementation and imperfections} \label{ssec:phys_implement}

A central ingredient of the entangling gate is the state-dependent addressing of the ions, and we consider barium ions to be well suited to the ion-tweezer platform described here.
Although Ba$^+$ is relatively heavy compared with other alkaline-earth ions, this is offset by its higher polarizability and longer transition wavelengths, which are more compatible with practical laser systems.
Specifically, we consider encoding the qubit in the $6S_{1/2}$ ground-state hyperfine levels of $^{137}$Ba$^+$, or in the Zeeman levels of $^{138}$Ba$^+$, while using the $5D_{5/2}$ manifold for the push state.
A detailed description of the proposed atomic levels is given in the Methods.

Specializing to $^{138}$Ba$^+$, we can convert the dimensionless parameters in the protocols into concrete distances and displacements.
For $\omega/2\pi=\qty{1}{\mega\hertz}$, the parameter $\eta^2=0.36$ corresponds to an ion distance $d\approx\qty{5.2}{\micro\metre}$, while for the weaker interaction at $\eta^2=0.01$ the ions are further separated at $d\approx\qty{17.2}{\micro\metre}$.
Taking the beam waist as $w_0=\qty{1}{\micro\metre}$, we then compare the gate displacements in the protocols with the available trap-center displacements.
We impose that the ion may not be displaced beyond $\xi_{1/2}\approx \qty{600}{\nano\metre}$, corresponding to half of the trap depth.
This provides a simple benchmark for remaining in the approximately harmonic regime while preventing the ion from being lost from the trap.
The required maximum displacement $\xi_{\rm max}$ is around $\qty{5}{\nano\metre}$ for \emph{Gate A} in Fig.~\ref{fig:fig2}(b) and $\qty{80}{\nano\metre}$ for \emph{Gate B} in Fig.~\ref{fig:fig2}(c), placing the protocols safely within the margins.
For the sub-trap-period gate in Fig.~\ref{fig:fig3}, the marker at $T_G=\qty{450}{\nano\second}$ and $\eta^2=0.36$ requires $\xi_{\max}\approx\qty{530}{\nano\metre}$, and was chosen to remain below $\xi_{1/2}$.
Substantially shorter gate times would entail correspondingly larger displacements and would push the dynamics beyond the simple harmonic regime, requiring either an explicit treatment of anharmonicity in the gate analysis or stronger confinement.
Nevertheless, the \emph{Gate~C} protocol points to a clear route toward substantially faster gates as the tweezer power and beam focusing are improved.

Next, we consider the optical power requirements. 
For fixed beam waist and trap frequency, the polarizability $\alpha$ effectively sets the optical power, which scales as $P\propto1/\alpha$~\cite{Grimm2000Optical}.
Based on the levels available with Ba$^+$, we consider a representative scenario with effective polarizability $\qty{500}{\atomicunit}$~\footnote{To convert from atomic units (\si{\atomicunit}) of polarizability to physical units, multiply by $4\pi\varepsilon_0a_B^3$, where $a_B=\qty{5.29e-11}{\metre}$ is the Bohr radius.},
requiring a power per tweezer of $\qty{2.3}{\watt}$.
More details on the required displacements and optical power for the different gates are given in the SI.

Beyond steering errors of the trap center and thermal excitations included in the analysis of the gate in Fig.~\ref{fig:fig2}, additional error sources exist.
One error source is the anharmonicity of the trapping potential, also discussed in the SI.
For weak anharmonicity, the leading-order effect is a shift of the effective center-of-mass and relative mode frequencies that depends on how far the ions are displaced from the trap minimum, which can be absorbed into effective values for $\omega$ and $\Omega$. 
Although our emphasis was on analytical understanding, the gates can be further optimized with standard optimal-control techniques to improve the robustness against such imperfections.
Another class of hardware noise arises from fluctuations of the optical tweezers themselves. 
In particular, the dominant trap-noise channels in optical tweezers are fundamentally different in character from those in Paul traps: here the noise is optical, arising from laser-intensity jitter and photon recoil heating.
We do not discuss these effects in detail here, as they are largely the same as in neutral-atom tweezer platforms.

\subsection{Parallel gate operations} \label{ssec:parallel_gates}

\begin{figure}[t]
    \centering
    \includegraphics[width=\columnwidth]{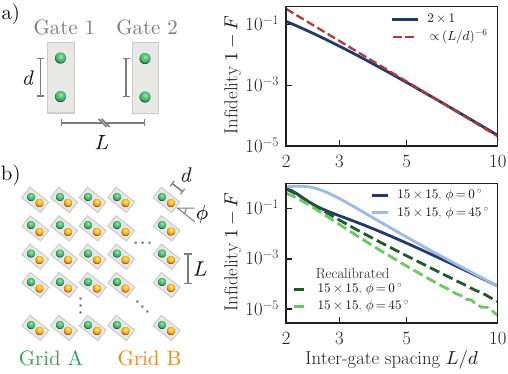}
    \caption{\textbf{Parallelizing two-ion gates. (a)} Two perpendicular bang--bang gates executed in parallel. 
    The plotted quantity is the average infidelity of one ion pair after tracing out the neighboring simultaneously driven pair and the motion, shown as a function of the inter-gate distance $L/d$, with $d$ the intra-gate distance.
    The cross-talk is a dipolar interaction, so the infidelity decays asymptotically as a power law $\propto (L/d)^{-6}$.
    Simulations are performed with the same gate parameters as in Fig.~\ref{fig:fig2}(c) for an initial vacuum state.
    \textbf{(b)} Gate infidelity of the central gate when executing a large square array ($15\times15$) of parallel CZ gates.
    Two gate orientations with an angle $\phi=0^\circ$ or $45^\circ$ are probed.
    Blue curves show the infidelity without recalibration, while green curves include an additional optimization over local $\sigma^z$ rotations.
    }
    \label{fig:fig5}
\end{figure}

An important property of the proposed architecture is that the tweezers allow for natural parallelization of quantum gates by imposing a sufficient distance between interaction zones.
We analyze the effect of cross-talk between quantum gates that are separated by an inter-gate distance $L$.
First, we consider a pair of gates in Fig.~\ref{fig:fig5}(a) that simultaneously execute the previously described \emph{Gate B} (bang--bang) at $\eta^2=0.01$.
We observe that a ratio of $L/d\gtrsim 4$ is sufficient to reduce the gate infidelity to below $1\%$.
For larger separations the asymptotic behavior follows $1-F\propto(L/d)^{-6}$, as expected for dipolar cross-talk. 

Next, we consider the parallel execution of a square grid of these gates as a primitive for transversal logical operations in FTQC.
A full grid of ions (Grid A) may be jointly and coherently transported such that each ion is brought into pairwise proximity with an ion of another grid (Grid B), as visualized in Fig.~\ref{fig:fig5}(b).
We simulate the infidelity of the same \emph{Gate B} for a $15\times15$ array of simultaneously executed two-ion gates.
Within the harmonic model, the average over the computational-basis states of the surrounding gates entering the central-gate fidelity factorizes into pairwise contributions, avoiding an explicit sum over exponentially many basis states.
A derivation of this factorized expression is given in the SI.
We choose the common gate-orientation angle as $\phi=0^\circ$ or $\phi=45^\circ$, illustrating that cross-talk can be reduced by tuning this angle.
Additionally, we allow for recalibration of the central gate by local $\sigma^z$~rotations, which can significantly improve the fidelity.
To reduce the infidelity to a level of $1\%$ after calibration, we require an inter-gate spacing of $L/d\gtrsim 4$ even for very large square arrays.
In fact, the asymptotic scaling remains $A(\phi)\times(L/d)^{-6}$ for arbitrarily large arrays.
In two dimensions, the $j$th shell around the central gate contains $\sim \lambda_j$ neighboring gates, each contributing quadratically in the dipolar coupling, i.e., scaling as $\lambda_j^{-6}$, such that the prefactor $A(\phi)\sim\sum_j \lambda_j^{-5}$ is finite.
Finally, we note that scheduling can further reduce the phase error for fixed $L/d$ at the expense of a longer gate sequence.

\section{Discussion and Outlook} \label{sec:discussion}

We have proposed a quantum-computing architecture based on ions trapped in optical tweezers.
This platform offers a scalable route to FTQC with trapped ions while leveraging the rapid progress of modern tweezer technology.
Optical tweezers are used not only to trap and transport ions, but also to implement entangling gates.
We view this proposal as a realistic blueprint based on current technological capabilities.
To this end, we have identified barium as a promising ion species and explicitly included its polarizability, as well as constraints associated with optical power and beam shaping, into our analysis.
The most challenging regime for two-ion gates corresponds to gate times significantly below the trap period, which require high laser power to produce the large trap displacements involved.
Nevertheless, our results indicate that even nanosecond-scale gate times are, in principle, achievable in an ion-tweezer quantum-computing platform.
We generally expect that continued progress in laser technology will further strengthen the prospects of tweezer-based platforms for quantum computation.

Compared with neutral atoms, ions offer several additional properties for gate design.
Their relevant states can be long-lived, with decay times of the order of minutes and therefore much longer than those of Rydberg states. 
The displacement can also be tuned to scale the size of the dipole, providing extra freedom in gate design. In addition, the directional effective dipoles of ions allow the cross-talk to be tuned through the gate orientation, a kind of control that neutral atoms do not naturally provide. 
Ion tweezers may also be combined with a deep global trap, as in standard trapped-ion experiments. 
The main drawback, however, is that ion motion is more sensitive to the gate procedure, which can reduce the gate fidelity. 
While a related effect also exists for Rydberg atoms, the relevant time scales there are typically much longer than the gate time and can usually be ignored.

At the architectural level, ions may also offer new opportunities for transport. 
In neutral-atom tweezer arrays, reconfiguring the geometry is limited by the transport times, since finite trap depths and available laser power constrain the speed even when shuttling is parallelized.
In ion-tweezer arrays, the charge degree of freedom provides an additional control handle: electric fields could move larger blocks collectively and faster, while the optical tweezers retain microscopic site-resolved control. 
It will also be interesting to explore hybrid ion-atom architectures in which an optically trapped ion is combined with a neutral atom or polar molecule in tweezers, potentially enabling hybrid gate schemes.
More broadly, the ion charge may enable possibilities that have not been available in neutral-atom architectures.

\begin{acknowledgments}
We thank Sirui Lu, Simon Schaffrath, Tao Shi, and Yichao Yu for insightful discussions.
B.F.S.~and J.I.C.~acknowledge funding from the German Federal Ministry of Research, Technology and Space (BMFTR) through the funded project ALMANAQC, grant number 13N17236 within the research program ``Quantum Systems''. 
C.M.~acknowledges funding from the US Department of Energy Quantum Systems Accelerator (DE-FOA-0002253).
P.Z.~gratefully acknowledges the hospitality of MPQ during extended stays where this work was performed. 
Work at MPQ is part of the Munich Quantum Valley, which is supported by the Bavarian state government with funds from the Hightech Agenda Bayern Plus.
\end{acknowledgments}

\appendix

\section*{Methods}

\subsection{Details on the experimental considerations} \label{methods:experiment}

We model each optical tweezer as a monochromatic running-wave Gaussian beam with electric field
\begin{equation}
    \mathbf{E}=\frac{1}{2}\mathcal{E}(\rho)
    \big[\boldsymbol{\epsilon}\,e^{i(\mathbf{k}\cdot\mathbf{r}-\omega_L t)} +\boldsymbol{\epsilon}^{*}e^{-i(\mathbf{k}\cdot\mathbf{r}-\omega_L t)} \big],
\end{equation}
where $\rho$ is the transverse distance from the beam axis, $\mathbf{k}$ is the propagation wavevector, $\omega_L=c|\mathbf{k}|$ is the optical frequency, and $\boldsymbol{\epsilon}$ is the polarization unit vector. 
The Gaussian field envelope is $\mathcal{E}(\rho)=\mathcal{E}_0 e^{-\rho^2/w_0^2}$, where $w_0$ is the beam waist, so that the corresponding intensity profile is $I(\rho)=I_0 e^{-2\rho^2/w_0^2}$ with peak intensity $I_0=c\varepsilon_0\mathcal{E}_0^2/2$.

The optical potential on an alkali-like ion is ${U(\rho) = -\frac{1}{4}\mathcal{E}(\rho)^2\langle{n\ell jm}|\hat{\alpha}_{n\ell j}|n\ell jm\rangle}$, where $n$, $\ell$, $j$, and $m\equiv m_J$ denote the principal, orbital, total angular momentum, and magnetic quantum numbers of the relevant state. 
The polarizability operator is~\cite{LeKien2013Dynamical}
\begin{align*}
    \hat{\alpha}_{n\ell j} 
    &= \alpha^{S}_{n\ell j}
    + \alpha^{V}_{n\ell j}
    \frac{(-i\boldsymbol{\epsilon}^{*}\times\boldsymbol{\epsilon})\cdot\boldsymbol{\hat{J}}}{2j}
    \notag\\
    &\quad+
    \alpha^{T}_{n\ell j}
    \frac{3[(\boldsymbol{\epsilon}\cdot\boldsymbol{\hat{J}})(\boldsymbol{\epsilon}^{*}\cdot\boldsymbol{\hat{J}})
    +(\boldsymbol{\epsilon}^{*}\cdot\boldsymbol{\hat{J}})(\boldsymbol{\epsilon}\cdot\boldsymbol{\hat{J}})]
    -2\boldsymbol{\hat{J}}^2}{2j(2j-1)}.
\end{align*}
Here $\alpha^{S}_{n\ell j}$, $\alpha^{V}_{n\ell j}$, and $\alpha^{T}_{n\ell j}$ are the scalar, vector, and tensor dynamic polarizabilities, and $\boldsymbol{\hat{J}}=\boldsymbol{\hat{L}}+\boldsymbol{\hat{S}}$ is the electronic total angular momentum. 
For $j=1/2$, the tensor term is omitted.
The quantization axis is chosen to be aligned with an external magnetic field $\boldsymbol{B}$, and the effective internal Hamiltonian is ${H_{\rm eff}=-\tfrac14 \mathcal{E}(\rho)^2\hat{\alpha}_{n\ell j}+g_j\mu_B \boldsymbol{B}\cdot\boldsymbol{\hat{J}}}$, where $g_j$ is the Land\'e $g$-factor and $\mu_B$ the Bohr magneton. 
In the estimates below we focus on $^{138}$Ba$^+$ and do not consider hyperfine structure, although the same approach can be extended to isotopes with nuclear spin.

\begin{figure}[t]
    \centering
    \includegraphics[width=\columnwidth]{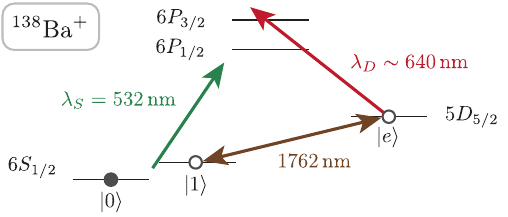}
    \caption{Relevant energy levels in $^{138}$Ba$^+$, including the qubit states $\ket{0}=\ket{6S_{1/2},m_J=-1/2}$ and $\ket{1}=\ket{6S_{1/2},m_J=+1/2}$, and the shelved \textit{push} state $\ket{e}=\ket{5D_{5/2},m_J=+5/2}$. 
    The shelving step addresses the $6S_{1/2}\leftrightarrow 5D_{5/2}$ quadrupole transition at $\lambda_e=\qty{1762}{\nano\metre}$.
    We consider an off-resonant beam at $\lambda_S=\qty{532}{\nano\metre}$ for trapping the qubit states, and a beam near $\lambda_D=\qty{640}{\nano\metre}$ for the push tweezer.
    }
    \label{fig:BaLevels}
\end{figure}

\subsubsection{Barium atomic ion tweezer potentials}

We consider barium as a suitable ion species for the gates described in this paper.
For $^{138}$Ba$^+$, we encode the qubit in the Zeeman sublevels $\ket{0}=\ket{6S_{1/2},m_J=-1/2}$ and $\ket{1}=\ket{6S_{1/2},m_J=+1/2}$, and use a metastable sublevel of the $5D_{5/2}$ manifold as the auxiliary \textit{push} state $\ket{e}=\ket{5D_{5/2},m_J=+5/2}$.
The $5D_{5/2}$ lifetime of approximately $\qty{30}{\second}$ allows shelving during the two-ion gate with negligible spontaneous decay~\cite{Zhang2020Branching}. 
The shelving step addresses the $\ket{1}\rightarrow\ket{e}$ branch on the electric-quadrupole transition $6S_{1/2}\leftrightarrow 5D_{5/2}$ at $\lambda_e=\qty{1762}{\nano\metre}$. 
A bias magnetic field defines the quantization axis, and the shelving-beam geometry and polarization are chosen to drive the desired $\Delta m_J=+2$ component.
The relevant level structure and wavelengths are summarized in Fig.~\ref{fig:BaLevels}.

\begin{figure}[t]
    \centering
    \includegraphics[width=\columnwidth]{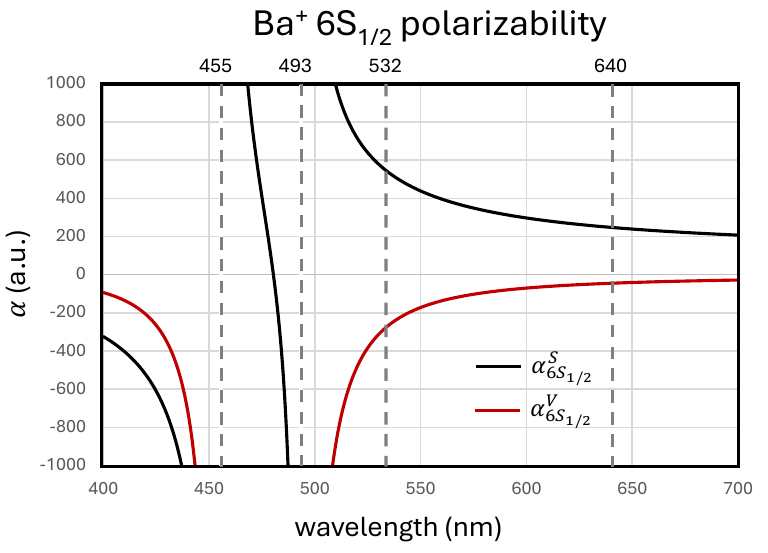}
    \caption{Scalar and vector polarizabilities of the Ba$^+$ $6S_{1/2}$ state as a function of the laser wavelength. 
    For $6S_{1/2}$ the tensor contribution vanishes, and for linearly polarized light the vector contribution is also zero, so the response is purely scalar. 
    Vertical dashed lines indicate the trapping wavelength $\lambda_S=\qty{532}{\nano\metre}$, push wavelength $\lambda_D=\qty{640}{\nano\metre}$ and the two resonant poles at $\qty{455}{\nano\metre}$ and $\qty{493}{\nano\metre}$. 
    Atomic parameters from Ref.~\cite{DelawareDatabase}.}
    \label{fig:Spol}
\end{figure}

\begin{figure}[t]
    \centering
    \includegraphics[width=\columnwidth]{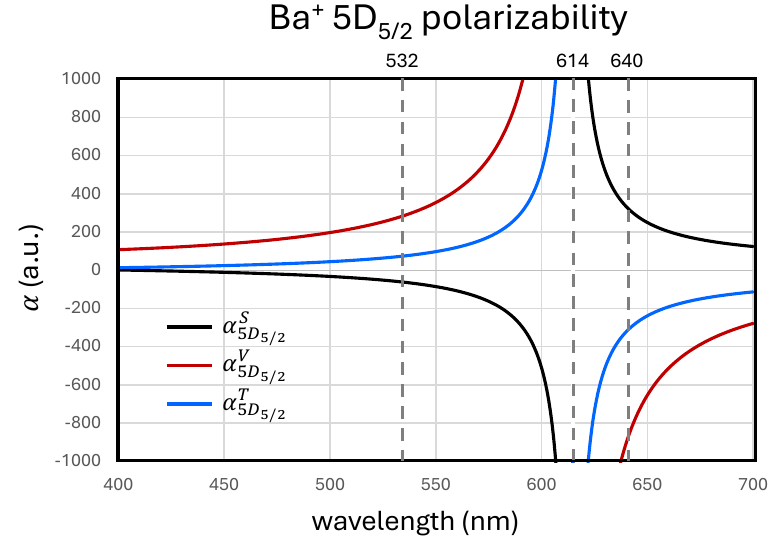}
    \caption{Scalar, vector, and tensor polarizabilities of the Ba$^+$ $5D_{5/2}$ manifold versus the laser wavelength. 
    Vertical dashed lines indicate the trapping wavelength $\lambda_S=\qty{532}{\nano\metre}$, push wavelength $\lambda_D=\qty{640}{\nano\metre}$ and the resonant pole at $\qty{614}{\nano\metre}$. Atomic parameters from Ref.~\cite{DelawareDatabase}.}
    \label{fig:Dpol}
\end{figure}

We consider two optical potentials whose centers can be controlled independently, one for the $6S_{1/2}$ qubit manifold and one for the shelved $5D_{5/2}$ push state.
Figures~\ref{fig:Spol} and~\ref{fig:Dpol} show the wavelength dependence of the polarizabilities relevant to the states.
We choose the main trapping wavelength to be $\lambda_S=\qty{532}{\nano\metre}$, where high optical power is available from Nd-doped solid-state lasers.
The $6S_{1/2}$ ground state has no tensor shift and the vector shift vanishes for linearly polarized light. 
The remaining scalar polarizability at this wavelength is $\alpha_{6S_{1/2}}(\lambda_S)\approx +\qty{560}{\atomicunit}$, so the qubit manifold is strongly and equally trapped. 
For a linearly polarized beam along the magnetic field quantization axis, the polarizability for the $\ket{5D_{5/2},m_J}$ state is
\begin{equation}
\alpha^{\text{lin}}_{5D_{5/2}}(\lambda) =\alpha^S_{5D_{5/2}}(\lambda) + \alpha^T_{5D_{5/2}}(\lambda)\left(\frac{3m_J^2}{10}-\frac{7}{8}\right).
\end{equation}
We find a near cancellation of scalar and tensor contributions at $\lambda_S=\qty{532}{\nano\metre}$, with the net polarizability of the $\ket{5D_{5/2},m_J=\pm5/2}$ states $\alpha_{5D_{5/2}}(\lambda_S)=+\qty{13}{\atomicunit}$, which is a factor of $43$ less than $\alpha_{6S_{1/2}}(\lambda_S)$.

For the push trap one possibility is to work at a wavelength of \qty{480.4}{\nano\metre}, where the polarizability of the $6S_{1/2}$ manifold cancels due to interference between the $P_{1/2}$ and $P_{3/2}$ excited-state contributions, allowing the push beam to act exclusively on the $D$-state. 
However, the vector polarizability of the ground states is large, so that the linear polarization needs to be very pure. 
Moreover, the polarizabilities of the $5D_{5/2}$ states themselves are only around \qty{20}{\atomicunit} for linearly polarized light, thus requiring substantial optical power.

Instead, we consider a push beam that is red-detuned from the $5D_{5/2}\leftrightarrow6P_{3/2}$ transition at \qty{614}{\nano\metre}.
Near this resonance, the largest polarizability stems from circularly polarized light propagating along the quantization axis, with a net polarizability of the $\ket{5D_{5/2},m_J}$ state of
\begin{align}
    \alpha^{\text{circ}\pm}_{5D_{5/2}}(\lambda) &=\alpha^S_{5D_{5/2}}(\lambda) \pm \alpha^V_{5D_{5/2}}(\lambda)\left(\frac{m_J}{5}\right) \notag\\
    &+ \alpha^T_{5D_{5/2}}(\lambda)\left(\frac{7}{16}-\frac{3m_J^2}{20}\right).
\end{align}
For concreteness, we take a push trap wavelength of $\lambda_D=\qty{640}{\nano\metre}$, which yields a net polarizability of the $\ket{5D_{5/2},m_J=\pm5/2}$ states of $\alpha_e(\lambda_D)\approx +\qty{950}{\atomicunit}$, with the sign matching the sign of circular polarization.
The push beam also impresses a residual polarizability on the $6S_{1/2}$ qubit states of $\qty{270}{\atomicunit}$, which is less than half that of the qubit tweezer. 
This is not a fundamental obstacle for the state-dependent two-ion gates as long as there is a sufficient difference between the effective polarizabilities of the states. 
Cross-talk between the trapping potentials may then be absorbed into recalibrated trap-center shifts, as discussed in the SI. 

\subsubsection{Barium atomic ion scattering decoherence} \label{methods:ssec_scatter}

We now consider decoherence from light scattering.
The scattering rate is set by the imaginary part of the polarizability and therefore scales with the light intensity in the same way as the optical potential, but with a different dependence on the detuning.
A simple estimate for the far-detuned case relates the total photon scattering rate from a given intermediate state $i$ to the tweezer depth $U(0)$ following the relation ${\Gamma_{\mathrm{sc}}\simeq (\gamma_i/|\Delta_i|)|U(0)|/\hbar}$~\cite{Grimm2000Optical}.
This expression is valid when $\gamma_i \ll \Delta_i \ll \omega_L$, or when the detuning $\Delta_i$ from the intermediate scattering state is much larger than its linewidth $\gamma_i$, but much smaller than the optical frequency itself. 
The probability of scattering during one trap period $2\pi/\omega$ in the tweezer is then
\begin{align}
    p_{\text{sc}}=\sum_i\frac{2\pi\gamma_i}{\hbar \omega}\left|\frac{U(0)}{\Delta_i}\right|.    
    \label{eq:tot_scatter}
\end{align}
Additionally, one may resolve the decay channels and include interference between intermediate states in the scattering estimate to obtain more precise estimates~\cite{Moore2023}.

The total scattering rate as estimated here overstates the decoherence rate, which is dominated by Raman scattering. 
In some atomic species, Raman scattering is a small fraction of the total scattering process~\cite{Ozeri2007, Moore2023}. 
In $^{138}$Ba$^+$, the high branching ratios of the $6P$ states to $S$ and $D$ states lead to smaller suppression of Raman scattering for some states~\cite{Zhang2020Branching,Moore2023}, so we use the total scattering as a conservative bound for decoherence for all combinations of states and wavelengths. 

In Table~\ref{ScatteringTable}, we report the total scattering probability per trap period in several scenarios. 
We include both the qubit and push states for the qubit tweezer ($\lambda=\qty{532}{\nano\metre}$) and the push tweezers ($\lambda=\qty{640}{\nano\metre}$ or $\qty{700}{\nano\metre}$) in the $^{138}$Ba$^+$ system.
The beam waist is set to $w_0=\qty{1}{\micro\metre}$, as in the main text. 
In addition to the trap frequency $\omega/2\pi=\qty{1}{\mega\hertz}$, we include $\omega/2\pi=\qty{100}{\kilo\hertz}$, which is more experimentally accessible and also has the advantage of lower scattering rates.
The dominant scattering occurs for the $5D_{5/2}$ state in the push tweezer at \qty{640}{\nano\metre}, which one may suppress by moving the wavelength of the push tweezer further into the red, at the cost of requiring more optical power per tweezer.

\begin{table}[htb]
    \begin{tabular}{c c c c c}
    \toprule
    $\omega/2\pi$ & State &\qty{532}{\nano\metre} ($\pi$)   & \qty{640}{\nano\metre} ($\sigma^{-}$)   & \qty{700}{\nano\metre} ($\sigma^{-}$)  \\ \midrule
    \qty{100}{\kilo\hertz} & $6S_{1/2}$&$5\times 10^{-4}$ &$3\times 10^{-5}$&$4\times 10^{-5}$ \\
    \qty{100}{\kilo\hertz} & $5D_{5/2}$&---$^{\rm (a)}$ &$3\times 10^{-3}$&$9\times 10^{-4}$ \\ 
    \midrule
    \qty{1}{\mega\hertz} & $6S_{1/2}$&$5\times 10^{-3}$ &$3\times 10^{-4}$&$4\times 10^{-4}$ \\
    \qty{1}{\mega\hertz} & $5D_{5/2}$&---$^{\rm (a)}$ &$3\times 10^{-2}$&$9\times 10^{-3}$ \\ 
    \bottomrule
    \end{tabular}
    \caption{Total scattering probability per oscillation period $p_{\text{sc}}$, given a tweezer beam waist $w_0=\qty{1}{\micro\metre}$ and trap frequencies $\omega/2\pi=\qty{100}{\kilo\hertz}$ or $\omega/2\pi=\qty{1}{\mega\hertz}$.
    We consider both qubit ($6S_{1/2}$) and push ($5D_{5/2}$) states in $^{138}$Ba$^+$, 
    subject to potentials for the qubit (\qty{532}{\nano\metre}, linear polarization) and push (\qty{640}{\nano\metre} or \qty{700}{\nano\metre}, circular polarization) tweezers. 
    The $5D_{5/2}$ values for the \qty{640}{\nano\metre} and \qty{700}{\nano\metre} push tweezers are single-pole estimates using Eq.~\eqref{eq:tot_scatter}, where for the $\ket{e}$ state and $\sigma^-$ light, the dominant pole is $5D_{5/2}\to6P_{3/2}$. 
    The $6S_{1/2}$ values are channel-resolved total scattering probabilities, including coherent contributions from the $6P_{1/2}$ and $6P_{3/2}$ intermediate states following Ref.~\cite{Moore2023}.
    (a) For $5D_{5/2},m_J=+5/2$ under \qty{532}{\nano\metre} $\pi$ light, the nearby $5D_{5/2}\to6P_{3/2}$ transition is forbidden and remaining higher-level contributions are negligible.}
    \label{ScatteringTable}
\end{table}

\subsection{Gate descriptions} \label{methods:gates}

We summarize the most important formulas used for the gate panels of Figs.~\ref{fig:fig2}, \ref{fig:fig3} and \ref{fig:fig4}, with additional details on the numerical simulations included in the SI.
We keep the conventions of the main text and set $\hbar=1$. 
For the perpendicular geometry we write $\Omega/\omega=\sqrt{1-\eta^2}$.
We further use the notation:
\[
 l_0 = \sqrt{\frac{1}{2m\omega}},\qquad
 \xi(t) = 2 l_0 z g(t),\qquad
 \eta^2 = \frac{e^2}{2\pi\varepsilon_0 m\omega^2 d^3}.
\]
Here $\xi(t)$ is the physical trap-center shift, while $z g(t)=\xi(t)/2l_0$ is the corresponding normalized shift.
The dimensionless control waveform is therefore $g(t)$, while $z$ fixes the overall displacement scale.

\subsubsection{Closure conditions} \label{methods:displacement}

First, we introduce notation for bookkeeping of the quantum state during the gates.
We consider one normal mode of frequency $\nu$ whose equilibrium position is shifted by real values $\delta_j$ during time intervals $t_j$, with $j=1,\ldots,N$. 
We assume that the trap center is zero before the first jump and after the last one, so $\delta_{N+1}=0$. 
If $\alpha_j^+$ denotes the coherent amplitude immediately after the jump into segment $j$, and $\alpha_j^-$ the amplitude immediately before the next jump, then
\begin{align}
    \alpha_1^+
    &= -\delta_1,
    \notag\\
    \alpha_j^-
    &= \alpha_j^+\,\mathrm e^{-i\nu t_j},
    \notag\\
    \alpha_{j+1}^+
    &= \alpha_j^- + \delta_j-\delta_{j+1},
    \qquad
    j=1,\ldots,N.
    \label{eq:app-recursion}
\end{align}

The motional phase acquired at the jumps follows from the identity~\cite{Gerry2004Introductory}:
\begin{equation}
    D(\Delta\alpha)\ket{\alpha} = \exp\!\Big[\frac{\Delta\alpha\,\alpha^* - \Delta\alpha^*\alpha}{2}\Big]\ket{\alpha+\Delta\alpha}.
    \label{eq:app-weyl}
\end{equation}
For a jump $\delta_j\to\delta_{j+1}$ one therefore gets
\begin{equation}
    \delta\phi_j(\nu) = \frac{(\delta_j-\delta_{j+1})\alpha_j^{-*}-(\delta_j-\delta_{j+1})\alpha_j^-}{2i}.
 \label{eq:app-jump-phase}
\end{equation}
The final state of an initial vacuum mode after the gate unitary $U_\nu$ is therefore
\begin{equation}
    U_\nu\ket{0}
    = \exp\!\Big[i\sum_{j=1}^{N}\delta\phi_j(\nu)\Big]
      \ket{\alpha_{N+1}^+},
\end{equation}
and its overlap with the vacuum is
\begin{equation}
    \bra{0}U_\nu\ket{0}
    = \exp\!\Big[-\frac{1}{2}\left|\alpha_{N+1}^+\right|^2
    + i\sum_{j=1}^{N}\delta\phi_j(\nu)\Big].
    \label{eq:app-return-amp}
\end{equation}
Exact mode closure corresponds to $\alpha_{N+1}^+=0$. 
For an initial vacuum mode this means that the mode returns to the vacuum up to a phase. More generally, if the corresponding closure condition holds for every mode and computational branch, the branch-dependent displacement vanishes and the gate becomes insensitive to the initial motional state, including thermal states~\cite{Garcia-Ripoll2005Coherent}.

\subsubsection{Commensurate (Gate A)}

The upper row of Fig.~\ref{fig:fig2} uses the simplest pulse,
\begin{equation}
     g(t) = 1, \qquad 0<t<T_G.
    \label{eq:app-single-hold}
\end{equation}
For one mode with coherent shift $\beta$, the sequence is $0\to\beta\to0$. 
Eq.~\eqref{eq:app-recursion} gives
\[
    \alpha_1^+ = -\beta, \quad
    \alpha_1^- = -\beta\,\mathrm e^{-i\nu T_G}, \quad 
    \alpha_f   = \beta\left(1-\mathrm e^{-i\nu T_G}\right).
\]
Thus the mode is exactly restored whenever $\nu T_G=2\pi n$, requiring commensurability of the mode frequencies:
\begin{equation}
     \omega T_G = 2\pi q, \qquad
     \Omega T_G = 2\pi p, \qquad
     \frac{\Omega}{\omega} = \frac{p}{q}.
     \label{eq:app-single-commensurate}
\end{equation}
The jump phase when the trap center is returned to the origin then also vanishes, as Eq.~\eqref{eq:app-jump-phase} yields
\begin{align}
     \delta \phi(\nu)
     &= -\beta^2\sin(\nu T_G)=0.
     \label{eq:app-single-jump}
\end{align}
The particular commensurate example shown in the upper row of Fig.~\ref{fig:fig2} uses $\Omega/\omega=4/5$, so $p=4$, $q=5$, $T_G=10\pi/\omega$, and $\eta^2=1-(4/5)^2=9/25$.
Imposing the CZ condition yields $z_{\mathrm{CZ}}^2=(1-\eta^2)/(2q\eta^2)=8/45$, hence
\begin{align}
    \beta_{\mathrm{CM}}
    &= \frac{z_{\mathrm{CZ}}}{\sqrt{2}}
    = \sqrt{\frac{4}{45}}
    \approx 0.298, \quad \text{and}
    \notag\\
    \beta_{\mathrm{rel}}
    &= \frac{z_{\mathrm{CZ}}}{\sqrt{2}\,(\Omega/\omega)^{3/2}}
    = \frac{5}{12}
    \approx 0.417,
\end{align}
are the amplitudes used in Fig.~\ref{fig:fig2}(b).

\subsubsection{Bang--bang (Gate B)}

The gate in Fig.~\ref{fig:fig2}(c) uses the exact perpendicular closure pulse at fixed total time $T_G = \frac{4\pi}{\omega}$. 
Its waveform is
\begin{align}
    g(t)
    &=
    \begin{cases}
    1, & 0<t<\tau_3 \text{ or } 3\tau_3<t<4\tau_3, \\
    g_m, & \tau_3<t<3\tau_3.
    \end{cases}
    \label{eq:app-three-step-pulse}
\end{align}
with $\tau_3 = \pi/\omega$.
For one mode with frequency $\nu$ and coherent shift $\beta$, the sequence is $0\to\beta\to g_m\beta\to\beta\to0$. 
Defining $u= \mathrm e^{-i\nu\tau_3}$, the recursion in Eq.~\eqref{eq:app-recursion} gives
\begin{align}
    \alpha_1^-
    &= -\beta u,
    \notag\\
    \alpha_2^+
    &= -\beta u + (1-g_m)\beta,
    \notag\\
    \alpha_2^-
    &= -\beta u^2\left(u+g_m-1\right),
    \notag\\
    \alpha_3^-
    &= -\beta u^4 + (1-g_m)\beta\left(u^3-u\right).
    \label{eq:app-three-step-recursion}
\end{align}
The final coherent amplitude is then $\alpha_f=\alpha_3^-+\beta$. 
For the CM mode one has $u=\mathrm e^{-i\pi}=-1$, so $\alpha_f=0$ for any $g_m$. 
For the rel mode, $\nu=\Omega$, and the exact closure condition $\alpha_f=0$ gives
\begin{equation}
    g_m = 1-2\cos\big(\pi \sqrt{1-\eta^2}\big).
    \label{eq:app-three-step-gm}
\end{equation}
Thus the three-segment pulse closes both perpendicular modes exactly for arbitrary $\eta^2<1$ within the harmonic approximation.

\subsubsection{Sub-trap-period gate (Gate C)}

The protocol used in Fig.~\ref{fig:fig3} is a symmetric five-segment pulse, described by the waveform
\begin{align}
    g(t)
    &=
    \begin{cases}
        1, & 0<t<\tau_5, \\
        b, & \tau_5<t<2\tau_5, \\
        c, & 2\tau_5<t<4\tau_5, \\
        b, & 4\tau_5<t<5\tau_5, \\
        1, & 5\tau_5<t<6\tau_5,
    \end{cases}
    \qquad
    \tau_5
    = \frac{T_G}{6}.
    \label{eq:app-five-pulse}
\end{align}
For a mode of frequency $\nu$, exact closure is equivalent to the vanishing of the Fourier component at frequency $\nu$~\cite{Garcia-Ripoll2005Coherent}, $I_{\nu}[g] = \int_0^{T_G} g(t)\,\mathrm e^{-i\nu t}\,\mathrm dt = 0$.
Evaluating the integral for the symmetric pulse in Eq.~\eqref{eq:app-five-pulse} yields the linear equations for the CM and rel mode:
\begin{align}
    \cos\!\left(\frac{5\omega\tau_5}{2}\right)
    + b\cos\!\left(\frac{3\omega\tau_5}{2}\right)
    + c\cos\!\left(\frac{\omega\tau_5}{2}\right)
    &= 0,
    \nonumber\\
    \cos\!\left(\frac{5\Omega\tau_5}{2}\right)
    + b\cos\!\left(\frac{3\Omega\tau_5}{2}\right)
    + c\cos\!\left(\frac{\Omega\tau_5}{2}\right)
    &= 0. 
\end{align}
The solution to these equations gives the plateau values
\begin{equation}
    b = \frac{u_1v_5-u_5v_1}{u_3v_1-u_1v_3},
    \qquad
    c = \frac{u_5v_3-u_3v_5}{u_3v_1-u_1v_3},
    \label{eq:app-five-bc}
\end{equation}
where we introduced the shorthands $u_n = \cos(n\omega T_G/12)$ and $v_n = \cos(n\Omega T_G/12)$ with $n\in\{1,3,5\}$.
For the example shown in Fig.~\ref{fig:fig3}, we take $\eta^2=0.36$ and $T_G=\qty{450}{\nano\second}$ at $\omega/2\pi=\qty{1}{\mega\hertz}$.
Eq.~\eqref{eq:app-five-bc} yields $b\approx -2.64$ and $c\approx 1.67$.
For a CZ gate (up to local rotations), we require $z_{\mathrm{CZ}}\approx 16.7$, so the normalized shift values in Fig.~\ref{fig:fig3}(a) are $\xi(t)/(2l_0)=z_{\mathrm{CZ}}(1,b,c,b,1)\approx (16.7,-44.2,28.0,-44.2,16.7)$.

\subsubsection{Dipole-blockade gate}

The dipole-blockade protocol in Fig.~\ref{fig:fig4} combines an exact dynamical preparation pulse of the control ion, a resonant target $2\pi$ pulse, and the reversed preparation pulse on the control ion.
For the middle pulse, the coherent-state displacements induced by exciting ion~2 are
\begin{equation}
    \alpha_{\mathrm{CM}}=\frac{z}{\sqrt{2}}, \qquad
    \alpha_{\mathrm{rel}}=\frac{z}{\sqrt{2}(\Omega/\omega)^{3/2}}.
    \label{eq:app-blockade-alphas}
\end{equation}
The interaction shift of the displaced geometry is
\begin{equation}
    \nu_z=-\omega z^2\frac{\eta^2}{1-\eta^2}. \label{eq:app-blockade-nuz}
\end{equation}
In the convention of a dynamical pulse on the control ion that we use here, this ion is placed at rest at a displacement $2\xi$ with $\xi=2l_0z$ after the pulse.
Then, the unblocked transition changes the displacement from $0$ to $\xi$, whereas the blocked one changes it from $2\xi$ to $3\xi$.
The detuning relevant for the target pulse is therefore twice the standard interaction shift
\begin{equation}
    \nu_z^{(\mathrm{dyn})}=2\nu_z=-2\omega z^2\frac{\eta^2}{1-\eta^2}.
    \label{eq:app-blockade-nuz-catch}
\end{equation}
This is the detuning that allows one to realize the two-ion gate using the blockade mechanism.

\bibliography{refs}

\begin{thebibliography}{54}%
\makeatletter
\providecommand \@ifxundefined [1]{%
 \@ifx{#1\undefined}
}%
\providecommand \@ifnum [1]{%
 \ifnum #1\expandafter \@firstoftwo
 \else \expandafter \@secondoftwo
 \fi
}%
\providecommand \@ifx [1]{%
 \ifx #1\expandafter \@firstoftwo
 \else \expandafter \@secondoftwo
 \fi
}%
\providecommand \natexlab [1]{#1}%
\providecommand \enquote  [1]{``#1''}%
\providecommand \bibnamefont  [1]{#1}%
\providecommand \bibfnamefont [1]{#1}%
\providecommand \citenamefont [1]{#1}%
\providecommand \href@noop [0]{\@secondoftwo}%
\providecommand \href [0]{\begingroup \@sanitize@url \@href}%
\providecommand \@href[1]{\@@startlink{#1}\@@href}%
\providecommand \@@href[1]{\endgroup#1\@@endlink}%
\providecommand \@sanitize@url [0]{\catcode `\\12\catcode `\$12\catcode `\&12\catcode `\#12\catcode `\^12\catcode `\_12\catcode `\%12\relax}%
\providecommand \@@startlink[1]{}%
\providecommand \@@endlink[0]{}%
\providecommand \url  [0]{\begingroup\@sanitize@url \@url }%
\providecommand \@url [1]{\endgroup\@href {#1}{\urlprefix }}%
\providecommand \urlprefix  [0]{URL }%
\providecommand \Eprint [0]{\href }%
\providecommand \doibase [0]{https://doi.org/}%
\providecommand \selectlanguage [0]{\@gobble}%
\providecommand \bibinfo  [0]{\@secondoftwo}%
\providecommand \bibfield  [0]{\@secondoftwo}%
\providecommand \translation [1]{[#1]}%
\providecommand \BibitemOpen [0]{}%
\providecommand \bibitemStop [0]{}%
\providecommand \bibitemNoStop [0]{.\EOS\space}%
\providecommand \EOS [0]{\spacefactor3000\relax}%
\providecommand \BibitemShut  [1]{\csname bibitem#1\endcsname}%
\let\auto@bib@innerbib\@empty
\bibitem [{\citenamefont {Nielsen}\ and\ \citenamefont {Chuang}(2010)}]{Nielsen2010Quantum}%
  \BibitemOpen
  \bibfield  {author} {\bibinfo {author} {\bibfnamefont {M.~A.}\ \bibnamefont {Nielsen}}\ and\ \bibinfo {author} {\bibfnamefont {I.~L.}\ \bibnamefont {Chuang}},\ }\href {https://doi.org/10.1017/cbo9780511976667} {\emph {\bibinfo {title} {Quantum {{Computation}} and {{Quantum Information}} (10th {{Anniversary}} Edition)}}}\ (\bibinfo  {publisher} {Cambridge University Press},\ \bibinfo {year} {2010})\BibitemShut {NoStop}%
\bibitem [{\citenamefont {Preskill}(1998)}]{Preskill1998Faulttolerant}%
  \BibitemOpen
  \bibfield  {author} {\bibinfo {author} {\bibfnamefont {J.}~\bibnamefont {Preskill}},\ }\bibfield  {title} {\bibinfo {title} {Fault-tolerant quantum computation},\ }in\ \href {https://doi.org/10.1142/9789812385253_0008} {\emph {\bibinfo {booktitle} {Introduction to Quantum Computation and Information}}}\ (\bibinfo  {publisher} {WORLD SCIENTIFIC},\ \bibinfo {year} {1998})\ pp.\ \bibinfo {pages} {213--269}\BibitemShut {NoStop}%
\bibitem [{\citenamefont {Campbell}\ \emph {et~al.}(2017)\citenamefont {Campbell}, \citenamefont {Terhal},\ and\ \citenamefont {Vuillot}}]{Campbell2017Roads}%
  \BibitemOpen
  \bibfield  {author} {\bibinfo {author} {\bibfnamefont {E.~T.}\ \bibnamefont {Campbell}}, \bibinfo {author} {\bibfnamefont {B.~M.}\ \bibnamefont {Terhal}},\ and\ \bibinfo {author} {\bibfnamefont {C.}~\bibnamefont {Vuillot}},\ }\bibfield  {title} {\bibinfo {title} {Roads towards fault-tolerant universal quantum computation},\ }\href {https://doi.org/10.1038/nature23460} {\bibfield  {journal} {\bibinfo  {journal} {Nature}\ }\textbf {\bibinfo {volume} {549}},\ \bibinfo {pages} {172} (\bibinfo {year} {2017})}\BibitemShut {NoStop}%
\bibitem [{\citenamefont {Cirac}\ and\ \citenamefont {Zoller}(1995)}]{Cirac1995Quantum}%
  \BibitemOpen
  \bibfield  {author} {\bibinfo {author} {\bibfnamefont {J.~I.}\ \bibnamefont {Cirac}}\ and\ \bibinfo {author} {\bibfnamefont {P.}~\bibnamefont {Zoller}},\ }\bibfield  {title} {\bibinfo {title} {Quantum {{Computations}} with {{Cold Trapped Ions}}},\ }\href {https://doi.org/10.1103/PhysRevLett.74.4091} {\bibfield  {journal} {\bibinfo  {journal} {Physical Review Letters}\ }\textbf {\bibinfo {volume} {74}},\ \bibinfo {pages} {4091} (\bibinfo {year} {1995})}\BibitemShut {NoStop}%
\bibitem [{\citenamefont {S{\o}rensen}\ and\ \citenamefont {M{\o}lmer}(1999)}]{Sorensen1999Quantum}%
  \BibitemOpen
  \bibfield  {author} {\bibinfo {author} {\bibfnamefont {A.}~\bibnamefont {S{\o}rensen}}\ and\ \bibinfo {author} {\bibfnamefont {K.}~\bibnamefont {M{\o}lmer}},\ }\bibfield  {title} {\bibinfo {title} {Quantum {{Computation}} with {{Ions}} in {{Thermal Motion}}},\ }\href {https://doi.org/10.1103/PhysRevLett.82.1971} {\bibfield  {journal} {\bibinfo  {journal} {Physical Review Letters}\ }\textbf {\bibinfo {volume} {82}},\ \bibinfo {pages} {1971} (\bibinfo {year} {1999})}\BibitemShut {NoStop}%
\bibitem [{\citenamefont {Cirac}\ and\ \citenamefont {Zoller}(2000)}]{Cirac2000Scalable}%
  \BibitemOpen
  \bibfield  {author} {\bibinfo {author} {\bibfnamefont {J.~I.}\ \bibnamefont {Cirac}}\ and\ \bibinfo {author} {\bibfnamefont {P.}~\bibnamefont {Zoller}},\ }\bibfield  {title} {\bibinfo {title} {A scalable quantum computer with ions in an array of microtraps},\ }\href {https://doi.org/10.1038/35007021} {\bibfield  {journal} {\bibinfo  {journal} {Nature}\ }\textbf {\bibinfo {volume} {404}},\ \bibinfo {pages} {579} (\bibinfo {year} {2000})}\BibitemShut {NoStop}%
\bibitem [{\citenamefont {Leibfried}\ \emph {et~al.}(2003)\citenamefont {Leibfried}, \citenamefont {Blatt}, \citenamefont {Monroe},\ and\ \citenamefont {Wineland}}]{Leibfried2003Quantum}%
  \BibitemOpen
  \bibfield  {author} {\bibinfo {author} {\bibfnamefont {D.}~\bibnamefont {Leibfried}}, \bibinfo {author} {\bibfnamefont {R.}~\bibnamefont {Blatt}}, \bibinfo {author} {\bibfnamefont {C.}~\bibnamefont {Monroe}},\ and\ \bibinfo {author} {\bibfnamefont {D.}~\bibnamefont {Wineland}},\ }\bibfield  {title} {\bibinfo {title} {Quantum dynamics of single trapped ions},\ }\href {https://doi.org/10.1103/RevModPhys.75.281} {\bibfield  {journal} {\bibinfo  {journal} {Reviews of Modern Physics}\ }\textbf {\bibinfo {volume} {75}},\ \bibinfo {pages} {281} (\bibinfo {year} {2003})}\BibitemShut {NoStop}%
\bibitem [{\citenamefont {H{\"a}ffner}\ \emph {et~al.}(2008)\citenamefont {H{\"a}ffner}, \citenamefont {Roos},\ and\ \citenamefont {Blatt}}]{Haffner2008Quantum}%
  \BibitemOpen
  \bibfield  {author} {\bibinfo {author} {\bibfnamefont {H.}~\bibnamefont {H{\"a}ffner}}, \bibinfo {author} {\bibfnamefont {C.~F.}\ \bibnamefont {Roos}},\ and\ \bibinfo {author} {\bibfnamefont {R.}~\bibnamefont {Blatt}},\ }\bibfield  {title} {\bibinfo {title} {Quantum computing with trapped ions},\ }\href {https://doi.org/10.1016/j.physrep.2008.09.003} {\bibfield  {journal} {\bibinfo  {journal} {Physics Reports}\ }\textbf {\bibinfo {volume} {469}},\ \bibinfo {pages} {155} (\bibinfo {year} {2008})}\BibitemShut {NoStop}%
\bibitem [{\citenamefont {Bruzewicz}\ \emph {et~al.}(2019)\citenamefont {Bruzewicz}, \citenamefont {Chiaverini}, \citenamefont {McConnell},\ and\ \citenamefont {Sage}}]{Bruzewicz2019TrappedIon}%
  \BibitemOpen
  \bibfield  {author} {\bibinfo {author} {\bibfnamefont {C.~D.}\ \bibnamefont {Bruzewicz}}, \bibinfo {author} {\bibfnamefont {J.}~\bibnamefont {Chiaverini}}, \bibinfo {author} {\bibfnamefont {R.}~\bibnamefont {McConnell}},\ and\ \bibinfo {author} {\bibfnamefont {J.~M.}\ \bibnamefont {Sage}},\ }\bibfield  {title} {\bibinfo {title} {Trapped-{{Ion Quantum Computing}}: {{Progress}} and {{Challenges}}},\ }\href {https://doi.org/10.1063/1.5088164} {\bibfield  {journal} {\bibinfo  {journal} {Applied Physics Reviews}\ }\textbf {\bibinfo {volume} {6}},\ \bibinfo {pages} {021314} (\bibinfo {year} {2019})}\BibitemShut {NoStop}%
\bibitem [{\citenamefont {Kielpinski}\ \emph {et~al.}(2002)\citenamefont {Kielpinski}, \citenamefont {Monroe},\ and\ \citenamefont {Wineland}}]{Kielpinski2002Architecture}%
  \BibitemOpen
  \bibfield  {author} {\bibinfo {author} {\bibfnamefont {D.}~\bibnamefont {Kielpinski}}, \bibinfo {author} {\bibfnamefont {C.}~\bibnamefont {Monroe}},\ and\ \bibinfo {author} {\bibfnamefont {D.~J.}\ \bibnamefont {Wineland}},\ }\bibfield  {title} {\bibinfo {title} {Architecture for a large-scale ion-trap quantum computer},\ }\href {https://doi.org/10.1038/nature00784} {\bibfield  {journal} {\bibinfo  {journal} {Nature}\ }\textbf {\bibinfo {volume} {417}},\ \bibinfo {pages} {709} (\bibinfo {year} {2002})}\BibitemShut {NoStop}%
\bibitem [{\citenamefont {Pino}\ \emph {et~al.}(2021)\citenamefont {Pino}, \citenamefont {Dreiling}, \citenamefont {Figgatt}, \citenamefont {Gaebler}, \citenamefont {Moses}, \citenamefont {Allman}, \citenamefont {Baldwin}, \citenamefont {{Foss-Feig}}, \citenamefont {Hayes}, \citenamefont {Mayer}, \citenamefont {{Ryan-Anderson}},\ and\ \citenamefont {Neyenhuis}}]{Pino2021Demonstration}%
  \BibitemOpen
  \bibfield  {author} {\bibinfo {author} {\bibfnamefont {J.~M.}\ \bibnamefont {Pino}}, \bibinfo {author} {\bibfnamefont {J.~M.}\ \bibnamefont {Dreiling}}, \bibinfo {author} {\bibfnamefont {C.}~\bibnamefont {Figgatt}}, \bibinfo {author} {\bibfnamefont {J.~P.}\ \bibnamefont {Gaebler}}, \bibinfo {author} {\bibfnamefont {S.~A.}\ \bibnamefont {Moses}}, \bibinfo {author} {\bibfnamefont {M.~S.}\ \bibnamefont {Allman}}, \bibinfo {author} {\bibfnamefont {C.~H.}\ \bibnamefont {Baldwin}}, \bibinfo {author} {\bibfnamefont {M.}~\bibnamefont {{Foss-Feig}}}, \bibinfo {author} {\bibfnamefont {D.}~\bibnamefont {Hayes}}, \bibinfo {author} {\bibfnamefont {K.}~\bibnamefont {Mayer}}, \bibinfo {author} {\bibfnamefont {C.}~\bibnamefont {{Ryan-Anderson}}},\ and\ \bibinfo {author} {\bibfnamefont {B.}~\bibnamefont {Neyenhuis}},\ }\bibfield  {title} {\bibinfo {title} {Demonstration of the trapped-ion quantum {{CCD}} computer architecture},\ }\href {https://doi.org/10.1038/s41586-021-03318-4} {\bibfield  {journal} {\bibinfo  {journal}
  {Nature}\ }\textbf {\bibinfo {volume} {592}},\ \bibinfo {pages} {209} (\bibinfo {year} {2021})}\BibitemShut {NoStop}%
\bibitem [{\citenamefont {Ransford}\ \emph {et~al.}(2026)\citenamefont {Ransford}, \citenamefont {Allman}, \citenamefont {Arkinstall}, \citenamefont {Campora}, \citenamefont {Cooper}, \citenamefont {Delaney}, \citenamefont {Dreiling}, \citenamefont {Estey}, \citenamefont {Figgatt}, \citenamefont {Hall}, \citenamefont {Husain}, \citenamefont {Isanaka}, \citenamefont {Kennedy}, \citenamefont {Kotibhaskar} \emph {et~al.}}]{Ransford202698qubit}%
  \BibitemOpen
  \bibfield  {author} {\bibinfo {author} {\bibfnamefont {A.}~\bibnamefont {Ransford}}, \bibinfo {author} {\bibfnamefont {M.~S.}\ \bibnamefont {Allman}}, \bibinfo {author} {\bibfnamefont {J.}~\bibnamefont {Arkinstall}}, \bibinfo {author} {\bibfnamefont {J.~P.}\ \bibnamefont {Campora}}, \bibinfo {author} {\bibfnamefont {S.~F.}\ \bibnamefont {Cooper}}, \bibinfo {author} {\bibfnamefont {R.~D.}\ \bibnamefont {Delaney}}, \bibinfo {author} {\bibfnamefont {J.~M.}\ \bibnamefont {Dreiling}}, \bibinfo {author} {\bibfnamefont {B.}~\bibnamefont {Estey}}, \bibinfo {author} {\bibfnamefont {C.}~\bibnamefont {Figgatt}}, \bibinfo {author} {\bibfnamefont {A.}~\bibnamefont {Hall}}, \bibinfo {author} {\bibfnamefont {A.~A.}\ \bibnamefont {Husain}}, \bibinfo {author} {\bibfnamefont {A.}~\bibnamefont {Isanaka}}, \bibinfo {author} {\bibfnamefont {C.~J.}\ \bibnamefont {Kennedy}}, \bibinfo {author} {\bibfnamefont {N.}~\bibnamefont {Kotibhaskar}}, \emph {et~al.},\ }\bibfield  {title} {\bibinfo {title} {A 98-qubit trapped-ion quantum
  computer with all-to-all connectivity},\ }\bibfield  {journal} {\bibinfo  {journal} {Nature}\ }\href {https://doi.org/10.1038/s41586-026-10676-4} {10.1038/s41586-026-10676-4} (\bibinfo {year} {2026})\BibitemShut {NoStop}%
\bibitem [{\citenamefont {Endres}\ \emph {et~al.}(2016)\citenamefont {Endres}, \citenamefont {Bernien}, \citenamefont {Keesling}, \citenamefont {Levine}, \citenamefont {Anschuetz}, \citenamefont {Krajenbrink}, \citenamefont {Senko}, \citenamefont {Vuletic}, \citenamefont {Greiner},\ and\ \citenamefont {Lukin}}]{Endres2016Atombyatom}%
  \BibitemOpen
  \bibfield  {author} {\bibinfo {author} {\bibfnamefont {M.}~\bibnamefont {Endres}}, \bibinfo {author} {\bibfnamefont {H.}~\bibnamefont {Bernien}}, \bibinfo {author} {\bibfnamefont {A.}~\bibnamefont {Keesling}}, \bibinfo {author} {\bibfnamefont {H.}~\bibnamefont {Levine}}, \bibinfo {author} {\bibfnamefont {E.~R.}\ \bibnamefont {Anschuetz}}, \bibinfo {author} {\bibfnamefont {A.}~\bibnamefont {Krajenbrink}}, \bibinfo {author} {\bibfnamefont {C.}~\bibnamefont {Senko}}, \bibinfo {author} {\bibfnamefont {V.}~\bibnamefont {Vuletic}}, \bibinfo {author} {\bibfnamefont {M.}~\bibnamefont {Greiner}},\ and\ \bibinfo {author} {\bibfnamefont {M.~D.}\ \bibnamefont {Lukin}},\ }\bibfield  {title} {\bibinfo {title} {Atom-by-atom assembly of defect-free one-dimensional cold atom arrays},\ }\href {https://doi.org/10.1126/science.aah3752} {\bibfield  {journal} {\bibinfo  {journal} {Science}\ }\textbf {\bibinfo {volume} {354}},\ \bibinfo {pages} {1024} (\bibinfo {year} {2016})}\BibitemShut {NoStop}%
\bibitem [{\citenamefont {Barredo}\ \emph {et~al.}(2016)\citenamefont {Barredo}, \citenamefont {{de L{\'e}s{\'e}leuc}}, \citenamefont {Lienhard}, \citenamefont {Lahaye},\ and\ \citenamefont {Browaeys}}]{Barredo2016Atombyatom}%
  \BibitemOpen
  \bibfield  {author} {\bibinfo {author} {\bibfnamefont {D.}~\bibnamefont {Barredo}}, \bibinfo {author} {\bibfnamefont {S.}~\bibnamefont {{de L{\'e}s{\'e}leuc}}}, \bibinfo {author} {\bibfnamefont {V.}~\bibnamefont {Lienhard}}, \bibinfo {author} {\bibfnamefont {T.}~\bibnamefont {Lahaye}},\ and\ \bibinfo {author} {\bibfnamefont {A.}~\bibnamefont {Browaeys}},\ }\bibfield  {title} {\bibinfo {title} {An atom-by-atom assembler of defect-free arbitrary two-dimensional atomic arrays},\ }\href {https://doi.org/10.1126/science.aah3778} {\bibfield  {journal} {\bibinfo  {journal} {Science}\ }\textbf {\bibinfo {volume} {354}},\ \bibinfo {pages} {1021} (\bibinfo {year} {2016})}\BibitemShut {NoStop}%
\bibitem [{\citenamefont {Browaeys}\ and\ \citenamefont {Lahaye}(2020)}]{Browaeys2020ManyBody}%
  \BibitemOpen
  \bibfield  {author} {\bibinfo {author} {\bibfnamefont {A.}~\bibnamefont {Browaeys}}\ and\ \bibinfo {author} {\bibfnamefont {T.}~\bibnamefont {Lahaye}},\ }\bibfield  {title} {\bibinfo {title} {Many-body physics with individually-controlled {{Rydberg}} atoms},\ }\href {https://doi.org/10.1038/s41567-019-0733-z} {\bibfield  {journal} {\bibinfo  {journal} {Nature Physics}\ }\textbf {\bibinfo {volume} {16}},\ \bibinfo {pages} {132} (\bibinfo {year} {2020})}\BibitemShut {NoStop}%
\bibitem [{\citenamefont {Bluvstein}\ \emph {et~al.}(2022)\citenamefont {Bluvstein}, \citenamefont {Levine}, \citenamefont {Semeghini}, \citenamefont {Wang}, \citenamefont {Ebadi}, \citenamefont {Kalinowski}, \citenamefont {Keesling}, \citenamefont {Maskara}, \citenamefont {Pichler}, \citenamefont {Greiner}, \citenamefont {Vuleti{\'c}},\ and\ \citenamefont {Lukin}}]{Bluvstein2022Quantum}%
  \BibitemOpen
  \bibfield  {author} {\bibinfo {author} {\bibfnamefont {D.}~\bibnamefont {Bluvstein}}, \bibinfo {author} {\bibfnamefont {H.}~\bibnamefont {Levine}}, \bibinfo {author} {\bibfnamefont {G.}~\bibnamefont {Semeghini}}, \bibinfo {author} {\bibfnamefont {T.~T.}\ \bibnamefont {Wang}}, \bibinfo {author} {\bibfnamefont {S.}~\bibnamefont {Ebadi}}, \bibinfo {author} {\bibfnamefont {M.}~\bibnamefont {Kalinowski}}, \bibinfo {author} {\bibfnamefont {A.}~\bibnamefont {Keesling}}, \bibinfo {author} {\bibfnamefont {N.}~\bibnamefont {Maskara}}, \bibinfo {author} {\bibfnamefont {H.}~\bibnamefont {Pichler}}, \bibinfo {author} {\bibfnamefont {M.}~\bibnamefont {Greiner}}, \bibinfo {author} {\bibfnamefont {V.}~\bibnamefont {Vuleti{\'c}}},\ and\ \bibinfo {author} {\bibfnamefont {M.~D.}\ \bibnamefont {Lukin}},\ }\bibfield  {title} {\bibinfo {title} {A quantum processor based on coherent transport of entangled atom arrays},\ }\href {https://doi.org/10.1038/s41586-022-04592-6} {\bibfield  {journal} {\bibinfo  {journal} {Nature}\
  }\textbf {\bibinfo {volume} {604}},\ \bibinfo {pages} {451} (\bibinfo {year} {2022})}\BibitemShut {NoStop}%
\bibitem [{\citenamefont {Graham}\ \emph {et~al.}(2022)\citenamefont {Graham}, \citenamefont {Song}, \citenamefont {Scott}, \citenamefont {Poole}, \citenamefont {Phuttitarn}, \citenamefont {Jooya}, \citenamefont {Eichler}, \citenamefont {Jiang}, \citenamefont {Marra}, \citenamefont {Grinkemeyer}, \citenamefont {Kwon}, \citenamefont {Ebert}, \citenamefont {Cherek}, \citenamefont {Lichtman}, \citenamefont {Gillette} \emph {et~al.}}]{Graham2022Multiqubit}%
  \BibitemOpen
  \bibfield  {author} {\bibinfo {author} {\bibfnamefont {T.~M.}\ \bibnamefont {Graham}}, \bibinfo {author} {\bibfnamefont {Y.}~\bibnamefont {Song}}, \bibinfo {author} {\bibfnamefont {J.}~\bibnamefont {Scott}}, \bibinfo {author} {\bibfnamefont {C.}~\bibnamefont {Poole}}, \bibinfo {author} {\bibfnamefont {L.}~\bibnamefont {Phuttitarn}}, \bibinfo {author} {\bibfnamefont {K.}~\bibnamefont {Jooya}}, \bibinfo {author} {\bibfnamefont {P.}~\bibnamefont {Eichler}}, \bibinfo {author} {\bibfnamefont {X.}~\bibnamefont {Jiang}}, \bibinfo {author} {\bibfnamefont {A.}~\bibnamefont {Marra}}, \bibinfo {author} {\bibfnamefont {B.}~\bibnamefont {Grinkemeyer}}, \bibinfo {author} {\bibfnamefont {M.}~\bibnamefont {Kwon}}, \bibinfo {author} {\bibfnamefont {M.}~\bibnamefont {Ebert}}, \bibinfo {author} {\bibfnamefont {J.}~\bibnamefont {Cherek}}, \bibinfo {author} {\bibfnamefont {M.~T.}\ \bibnamefont {Lichtman}}, \bibinfo {author} {\bibfnamefont {M.}~\bibnamefont {Gillette}}, \emph {et~al.},\ }\bibfield  {title} {\bibinfo {title}
  {Multi-qubit entanglement and algorithms on a neutral-atom quantum computer},\ }\href {https://doi.org/10.1038/s41586-022-04603-6} {\bibfield  {journal} {\bibinfo  {journal} {Nature}\ }\textbf {\bibinfo {volume} {604}},\ \bibinfo {pages} {457} (\bibinfo {year} {2022})}\BibitemShut {NoStop}%
\bibitem [{\citenamefont {Manetsch}\ \emph {et~al.}(2025)\citenamefont {Manetsch}, \citenamefont {Nomura}, \citenamefont {Bataille}, \citenamefont {Lv}, \citenamefont {Leung},\ and\ \citenamefont {Endres}}]{Manetsch2025Tweezer}%
  \BibitemOpen
  \bibfield  {author} {\bibinfo {author} {\bibfnamefont {H.~J.}\ \bibnamefont {Manetsch}}, \bibinfo {author} {\bibfnamefont {G.}~\bibnamefont {Nomura}}, \bibinfo {author} {\bibfnamefont {E.}~\bibnamefont {Bataille}}, \bibinfo {author} {\bibfnamefont {X.}~\bibnamefont {Lv}}, \bibinfo {author} {\bibfnamefont {K.~H.}\ \bibnamefont {Leung}},\ and\ \bibinfo {author} {\bibfnamefont {M.}~\bibnamefont {Endres}},\ }\bibfield  {title} {\bibinfo {title} {A tweezer array with 6,100 highly coherent atomic qubits},\ }\href {https://doi.org/10.1038/s41586-025-09641-4} {\bibfield  {journal} {\bibinfo  {journal} {Nature}\ }\textbf {\bibinfo {volume} {647}},\ \bibinfo {pages} {60} (\bibinfo {year} {2025})}\BibitemShut {NoStop}%
\bibitem [{\citenamefont {Bluvstein}\ \emph {et~al.}(2026)\citenamefont {Bluvstein}, \citenamefont {Geim}, \citenamefont {Li}, \citenamefont {Evered}, \citenamefont {Bonilla~Ataides}, \citenamefont {Baranes}, \citenamefont {Gu}, \citenamefont {Manovitz}, \citenamefont {Xu}, \citenamefont {Kalinowski}, \citenamefont {Majidy}, \citenamefont {Kokail}, \citenamefont {Maskara}, \citenamefont {Trapp} \emph {et~al.}}]{Bluvstein2026Faulttolerant}%
  \BibitemOpen
  \bibfield  {author} {\bibinfo {author} {\bibfnamefont {D.}~\bibnamefont {Bluvstein}}, \bibinfo {author} {\bibfnamefont {A.~A.}\ \bibnamefont {Geim}}, \bibinfo {author} {\bibfnamefont {S.~H.}\ \bibnamefont {Li}}, \bibinfo {author} {\bibfnamefont {S.~J.}\ \bibnamefont {Evered}}, \bibinfo {author} {\bibfnamefont {J.~P.}\ \bibnamefont {Bonilla~Ataides}}, \bibinfo {author} {\bibfnamefont {G.}~\bibnamefont {Baranes}}, \bibinfo {author} {\bibfnamefont {A.}~\bibnamefont {Gu}}, \bibinfo {author} {\bibfnamefont {T.}~\bibnamefont {Manovitz}}, \bibinfo {author} {\bibfnamefont {M.}~\bibnamefont {Xu}}, \bibinfo {author} {\bibfnamefont {M.}~\bibnamefont {Kalinowski}}, \bibinfo {author} {\bibfnamefont {S.}~\bibnamefont {Majidy}}, \bibinfo {author} {\bibfnamefont {C.}~\bibnamefont {Kokail}}, \bibinfo {author} {\bibfnamefont {N.}~\bibnamefont {Maskara}}, \bibinfo {author} {\bibfnamefont {E.~C.}\ \bibnamefont {Trapp}}, \emph {et~al.},\ }\bibfield  {title} {\bibinfo {title} {A fault-tolerant neutral-atom architecture for
  universal quantum computation},\ }\href {https://doi.org/10.1038/s41586-025-09848-5} {\bibfield  {journal} {\bibinfo  {journal} {Nature}\ }\textbf {\bibinfo {volume} {649}},\ \bibinfo {pages} {39} (\bibinfo {year} {2026})}\BibitemShut {NoStop}%
\bibitem [{\citenamefont {Jaksch}\ \emph {et~al.}(2000)\citenamefont {Jaksch}, \citenamefont {Cirac}, \citenamefont {Zoller}, \citenamefont {Rolston}, \citenamefont {C{\^o}t{\'e}},\ and\ \citenamefont {Lukin}}]{Jaksch2000Fast}%
  \BibitemOpen
  \bibfield  {author} {\bibinfo {author} {\bibfnamefont {D.}~\bibnamefont {Jaksch}}, \bibinfo {author} {\bibfnamefont {J.~I.}\ \bibnamefont {Cirac}}, \bibinfo {author} {\bibfnamefont {P.}~\bibnamefont {Zoller}}, \bibinfo {author} {\bibfnamefont {S.~L.}\ \bibnamefont {Rolston}}, \bibinfo {author} {\bibfnamefont {R.}~\bibnamefont {C{\^o}t{\'e}}},\ and\ \bibinfo {author} {\bibfnamefont {M.~D.}\ \bibnamefont {Lukin}},\ }\bibfield  {title} {\bibinfo {title} {Fast {{Quantum Gates}} for {{Neutral Atoms}}},\ }\href {https://doi.org/10.1103/PhysRevLett.85.2208} {\bibfield  {journal} {\bibinfo  {journal} {Physical Review Letters}\ }\textbf {\bibinfo {volume} {85}},\ \bibinfo {pages} {2208} (\bibinfo {year} {2000})}\BibitemShut {NoStop}%
\bibitem [{\citenamefont {Levine}\ \emph {et~al.}(2019)\citenamefont {Levine}, \citenamefont {Keesling}, \citenamefont {Semeghini}, \citenamefont {Omran}, \citenamefont {Wang}, \citenamefont {Ebadi}, \citenamefont {Bernien}, \citenamefont {Greiner}, \citenamefont {Vuleti{\'c}}, \citenamefont {Pichler},\ and\ \citenamefont {Lukin}}]{Levine2019Parallel}%
  \BibitemOpen
  \bibfield  {author} {\bibinfo {author} {\bibfnamefont {H.}~\bibnamefont {Levine}}, \bibinfo {author} {\bibfnamefont {A.}~\bibnamefont {Keesling}}, \bibinfo {author} {\bibfnamefont {G.}~\bibnamefont {Semeghini}}, \bibinfo {author} {\bibfnamefont {A.}~\bibnamefont {Omran}}, \bibinfo {author} {\bibfnamefont {T.~T.}\ \bibnamefont {Wang}}, \bibinfo {author} {\bibfnamefont {S.}~\bibnamefont {Ebadi}}, \bibinfo {author} {\bibfnamefont {H.}~\bibnamefont {Bernien}}, \bibinfo {author} {\bibfnamefont {M.}~\bibnamefont {Greiner}}, \bibinfo {author} {\bibfnamefont {V.}~\bibnamefont {Vuleti{\'c}}}, \bibinfo {author} {\bibfnamefont {H.}~\bibnamefont {Pichler}},\ and\ \bibinfo {author} {\bibfnamefont {M.~D.}\ \bibnamefont {Lukin}},\ }\bibfield  {title} {\bibinfo {title} {Parallel implementation of high-fidelity multiqubit gates with neutral atoms},\ }\href {https://doi.org/10.1103/PhysRevLett.123.170503} {\bibfield  {journal} {\bibinfo  {journal} {Physical Review Letters}\ }\textbf {\bibinfo {volume} {123}},\ \bibinfo
  {pages} {170503} (\bibinfo {year} {2019})}\BibitemShut {NoStop}%
\bibitem [{\citenamefont {Lukin}\ \emph {et~al.}(2001)\citenamefont {Lukin}, \citenamefont {Fleischhauer}, \citenamefont {Cote}, \citenamefont {Duan}, \citenamefont {Jaksch}, \citenamefont {Cirac},\ and\ \citenamefont {Zoller}}]{Lukin2001Dipole}%
  \BibitemOpen
  \bibfield  {author} {\bibinfo {author} {\bibfnamefont {M.~D.}\ \bibnamefont {Lukin}}, \bibinfo {author} {\bibfnamefont {M.}~\bibnamefont {Fleischhauer}}, \bibinfo {author} {\bibfnamefont {R.}~\bibnamefont {Cote}}, \bibinfo {author} {\bibfnamefont {L.~M.}\ \bibnamefont {Duan}}, \bibinfo {author} {\bibfnamefont {D.}~\bibnamefont {Jaksch}}, \bibinfo {author} {\bibfnamefont {J.~I.}\ \bibnamefont {Cirac}},\ and\ \bibinfo {author} {\bibfnamefont {P.}~\bibnamefont {Zoller}},\ }\bibfield  {title} {\bibinfo {title} {Dipole {{Blockade}} and {{Quantum Information Processing}} in {{Mesoscopic Atomic Ensembles}}},\ }\href {https://doi.org/10.1103/PhysRevLett.87.037901} {\bibfield  {journal} {\bibinfo  {journal} {Physical Review Letters}\ }\textbf {\bibinfo {volume} {87}},\ \bibinfo {pages} {037901} (\bibinfo {year} {2001})}\BibitemShut {NoStop}%
\bibitem [{\citenamefont {Evered}\ \emph {et~al.}(2023)\citenamefont {Evered}, \citenamefont {Bluvstein}, \citenamefont {Kalinowski}, \citenamefont {Ebadi}, \citenamefont {Manovitz}, \citenamefont {Zhou}, \citenamefont {Li}, \citenamefont {Geim}, \citenamefont {Wang}, \citenamefont {Maskara}, \citenamefont {Levine}, \citenamefont {Semeghini}, \citenamefont {Greiner}, \citenamefont {Vuleti{\'c}},\ and\ \citenamefont {Lukin}}]{Evered2023Highfidelity}%
  \BibitemOpen
  \bibfield  {author} {\bibinfo {author} {\bibfnamefont {S.~J.}\ \bibnamefont {Evered}}, \bibinfo {author} {\bibfnamefont {D.}~\bibnamefont {Bluvstein}}, \bibinfo {author} {\bibfnamefont {M.}~\bibnamefont {Kalinowski}}, \bibinfo {author} {\bibfnamefont {S.}~\bibnamefont {Ebadi}}, \bibinfo {author} {\bibfnamefont {T.}~\bibnamefont {Manovitz}}, \bibinfo {author} {\bibfnamefont {H.}~\bibnamefont {Zhou}}, \bibinfo {author} {\bibfnamefont {S.~H.}\ \bibnamefont {Li}}, \bibinfo {author} {\bibfnamefont {A.~A.}\ \bibnamefont {Geim}}, \bibinfo {author} {\bibfnamefont {T.~T.}\ \bibnamefont {Wang}}, \bibinfo {author} {\bibfnamefont {N.}~\bibnamefont {Maskara}}, \bibinfo {author} {\bibfnamefont {H.}~\bibnamefont {Levine}}, \bibinfo {author} {\bibfnamefont {G.}~\bibnamefont {Semeghini}}, \bibinfo {author} {\bibfnamefont {M.}~\bibnamefont {Greiner}}, \bibinfo {author} {\bibfnamefont {V.}~\bibnamefont {Vuleti{\'c}}},\ and\ \bibinfo {author} {\bibfnamefont {M.~D.}\ \bibnamefont {Lukin}},\ }\bibfield  {title} {\bibinfo {title}
  {High-fidelity parallel entangling gates on a neutral-atom quantum computer},\ }\href {https://doi.org/10.1038/s41586-023-06481-y} {\bibfield  {journal} {\bibinfo  {journal} {Nature}\ }\textbf {\bibinfo {volume} {622}},\ \bibinfo {pages} {268} (\bibinfo {year} {2023})}\BibitemShut {NoStop}%
\bibitem [{\citenamefont {Saffman}\ \emph {et~al.}(2010)\citenamefont {Saffman}, \citenamefont {Walker},\ and\ \citenamefont {M{\o}lmer}}]{Saffman2010Quantum}%
  \BibitemOpen
  \bibfield  {author} {\bibinfo {author} {\bibfnamefont {M.}~\bibnamefont {Saffman}}, \bibinfo {author} {\bibfnamefont {T.~G.}\ \bibnamefont {Walker}},\ and\ \bibinfo {author} {\bibfnamefont {K.}~\bibnamefont {M{\o}lmer}},\ }\bibfield  {title} {\bibinfo {title} {Quantum information with {{Rydberg}} atoms},\ }\href {https://doi.org/10.1103/RevModPhys.82.2313} {\bibfield  {journal} {\bibinfo  {journal} {Reviews of Modern Physics}\ }\textbf {\bibinfo {volume} {82}},\ \bibinfo {pages} {2313} (\bibinfo {year} {2010})}\BibitemShut {NoStop}%
\bibitem [{\citenamefont {Zhang}\ \emph {et~al.}(2012)\citenamefont {Zhang}, \citenamefont {Gill}, \citenamefont {Isenhower}, \citenamefont {Walker},\ and\ \citenamefont {Saffman}}]{Zhang2012Fidelity}%
  \BibitemOpen
  \bibfield  {author} {\bibinfo {author} {\bibfnamefont {X.~L.}\ \bibnamefont {Zhang}}, \bibinfo {author} {\bibfnamefont {A.~T.}\ \bibnamefont {Gill}}, \bibinfo {author} {\bibfnamefont {L.}~\bibnamefont {Isenhower}}, \bibinfo {author} {\bibfnamefont {T.~G.}\ \bibnamefont {Walker}},\ and\ \bibinfo {author} {\bibfnamefont {M.}~\bibnamefont {Saffman}},\ }\bibfield  {title} {\bibinfo {title} {Fidelity of a {{Rydberg}}-blockade quantum gate from simulated quantum process tomography},\ }\href {https://doi.org/10.1103/PhysRevA.85.042310} {\bibfield  {journal} {\bibinfo  {journal} {Physical Review A}\ }\textbf {\bibinfo {volume} {85}},\ \bibinfo {pages} {042310} (\bibinfo {year} {2012})}\BibitemShut {NoStop}%
\bibitem [{\citenamefont {Saffman}(2016)}]{Saffman2016Quantum}%
  \BibitemOpen
  \bibfield  {author} {\bibinfo {author} {\bibfnamefont {M.}~\bibnamefont {Saffman}},\ }\bibfield  {title} {\bibinfo {title} {Quantum computing with atomic qubits and {{Rydberg}} interactions: Progress and challenges},\ }\href {https://doi.org/10.1088/0953-4075/49/20/202001} {\bibfield  {journal} {\bibinfo  {journal} {Journal of Physics B: Atomic, Molecular and Optical Physics}\ }\textbf {\bibinfo {volume} {49}},\ \bibinfo {pages} {202001} (\bibinfo {year} {2016})}\BibitemShut {NoStop}%
\bibitem [{\citenamefont {Cohen}\ and\ \citenamefont {Thompson}(2021)}]{Cohen2021Quantum}%
  \BibitemOpen
  \bibfield  {author} {\bibinfo {author} {\bibfnamefont {S.~R.}\ \bibnamefont {Cohen}}\ and\ \bibinfo {author} {\bibfnamefont {J.~D.}\ \bibnamefont {Thompson}},\ }\bibfield  {title} {\bibinfo {title} {Quantum {{Computing}} with {{Circular Rydberg Atoms}}},\ }\href {https://doi.org/10.1103/PRXQuantum.2.030322} {\bibfield  {journal} {\bibinfo  {journal} {PRX Quantum}\ }\textbf {\bibinfo {volume} {2}},\ \bibinfo {pages} {030322} (\bibinfo {year} {2021})}\BibitemShut {NoStop}%
\bibitem [{\citenamefont {Saffman}(2025)}]{Saffman2025Quantum}%
  \BibitemOpen
  \bibfield  {author} {\bibinfo {author} {\bibfnamefont {M.}~\bibnamefont {Saffman}},\ }\href {https://doi.org/10.48550/arXiv.2505.11218} {\bibinfo {title} {Quantum computing with atomic qubit arrays: Confronting the cost of connectivity}} (\bibinfo {year} {2025}),\ \Eprint {https://arxiv.org/abs/2505.11218} {arXiv:2505.11218} \BibitemShut {NoStop}%
\bibitem [{\citenamefont {Norrell}\ \emph {et~al.}(2026)\citenamefont {Norrell}, \citenamefont {Shen}, \citenamefont {Saffman},\ and\ \citenamefont {Otten}}]{Norrell2026Entangling}%
  \BibitemOpen
  \bibfield  {author} {\bibinfo {author} {\bibfnamefont {S.~A.}\ \bibnamefont {Norrell}}, \bibinfo {author} {\bibfnamefont {Y.}~\bibnamefont {Shen}}, \bibinfo {author} {\bibfnamefont {M.}~\bibnamefont {Saffman}},\ and\ \bibinfo {author} {\bibfnamefont {M.}~\bibnamefont {Otten}},\ }\href {https://doi.org/10.48550/arXiv.2605.19245} {\bibinfo {title} {Entangling gate performance and fidelity limits with neutral atom {{F\"orster}} resonances}} (\bibinfo {year} {2026}),\ \Eprint {https://arxiv.org/abs/2605.19245} {arXiv:2605.19245} \BibitemShut {NoStop}%
\bibitem [{\citenamefont {Kielpinski}\ \emph {et~al.}(2000)\citenamefont {Kielpinski}, \citenamefont {King}, \citenamefont {Myatt}, \citenamefont {Sackett}, \citenamefont {Turchette}, \citenamefont {Itano}, \citenamefont {Monroe}, \citenamefont {Wineland},\ and\ \citenamefont {Zurek}}]{Kielpinski2000Sympathetic}%
  \BibitemOpen
  \bibfield  {author} {\bibinfo {author} {\bibfnamefont {D.}~\bibnamefont {Kielpinski}}, \bibinfo {author} {\bibfnamefont {B.~E.}\ \bibnamefont {King}}, \bibinfo {author} {\bibfnamefont {C.~J.}\ \bibnamefont {Myatt}}, \bibinfo {author} {\bibfnamefont {C.~A.}\ \bibnamefont {Sackett}}, \bibinfo {author} {\bibfnamefont {Q.~A.}\ \bibnamefont {Turchette}}, \bibinfo {author} {\bibfnamefont {W.~M.}\ \bibnamefont {Itano}}, \bibinfo {author} {\bibfnamefont {C.}~\bibnamefont {Monroe}}, \bibinfo {author} {\bibfnamefont {D.~J.}\ \bibnamefont {Wineland}},\ and\ \bibinfo {author} {\bibfnamefont {W.~H.}\ \bibnamefont {Zurek}},\ }\bibfield  {title} {\bibinfo {title} {Sympathetic cooling of trapped ions for quantum logic},\ }\href {https://doi.org/10.1103/PhysRevA.61.032310} {\bibfield  {journal} {\bibinfo  {journal} {Physical Review A}\ }\textbf {\bibinfo {volume} {61}},\ \bibinfo {pages} {032310} (\bibinfo {year} {2000})}\BibitemShut {NoStop}%
\bibitem [{\citenamefont {Belyansky}\ \emph {et~al.}(2019)\citenamefont {Belyansky}, \citenamefont {Young}, \citenamefont {Bienias}, \citenamefont {Eldredge}, \citenamefont {Kaufman}, \citenamefont {Zoller},\ and\ \citenamefont {Gorshkov}}]{Belyansky2019Nondestructive}%
  \BibitemOpen
  \bibfield  {author} {\bibinfo {author} {\bibfnamefont {R.}~\bibnamefont {Belyansky}}, \bibinfo {author} {\bibfnamefont {J.~T.}\ \bibnamefont {Young}}, \bibinfo {author} {\bibfnamefont {P.}~\bibnamefont {Bienias}}, \bibinfo {author} {\bibfnamefont {Z.}~\bibnamefont {Eldredge}}, \bibinfo {author} {\bibfnamefont {A.~M.}\ \bibnamefont {Kaufman}}, \bibinfo {author} {\bibfnamefont {P.}~\bibnamefont {Zoller}},\ and\ \bibinfo {author} {\bibfnamefont {A.~V.}\ \bibnamefont {Gorshkov}},\ }\bibfield  {title} {\bibinfo {title} {Nondestructive cooling of an atomic quantum register via state-insensitive {{Rydberg}} interactions},\ }\href {https://doi.org/10.1103/PhysRevLett.123.213603} {\bibfield  {journal} {\bibinfo  {journal} {Physical Review Letters}\ }\textbf {\bibinfo {volume} {123}},\ \bibinfo {pages} {213603} (\bibinfo {year} {2019})}\BibitemShut {NoStop}%
\bibitem [{\citenamefont {Schmied}\ \emph {et~al.}(2008)\citenamefont {Schmied}, \citenamefont {Roscilde}, \citenamefont {Murg}, \citenamefont {Porras},\ and\ \citenamefont {Cirac}}]{Schmied2008Quantum}%
  \BibitemOpen
  \bibfield  {author} {\bibinfo {author} {\bibfnamefont {R.}~\bibnamefont {Schmied}}, \bibinfo {author} {\bibfnamefont {T.}~\bibnamefont {Roscilde}}, \bibinfo {author} {\bibfnamefont {V.}~\bibnamefont {Murg}}, \bibinfo {author} {\bibfnamefont {D.}~\bibnamefont {Porras}},\ and\ \bibinfo {author} {\bibfnamefont {J.~I.}\ \bibnamefont {Cirac}},\ }\bibfield  {title} {\bibinfo {title} {Quantum phases of trapped ions in an optical lattice},\ }\href {https://doi.org/10.1088/1367-2630/10/4/045017} {\bibfield  {journal} {\bibinfo  {journal} {New Journal of Physics}\ }\textbf {\bibinfo {volume} {10}},\ \bibinfo {pages} {045017} (\bibinfo {year} {2008})}\BibitemShut {NoStop}%
\bibitem [{\citenamefont {Schneider}\ \emph {et~al.}(2010)\citenamefont {Schneider}, \citenamefont {Enderlein}, \citenamefont {Huber},\ and\ \citenamefont {Schaetz}}]{Schneider2010Optical}%
  \BibitemOpen
  \bibfield  {author} {\bibinfo {author} {\bibfnamefont {{\relax Ch}.}~\bibnamefont {Schneider}}, \bibinfo {author} {\bibfnamefont {M.}~\bibnamefont {Enderlein}}, \bibinfo {author} {\bibfnamefont {T.}~\bibnamefont {Huber}},\ and\ \bibinfo {author} {\bibfnamefont {T.}~\bibnamefont {Schaetz}},\ }\bibfield  {title} {\bibinfo {title} {Optical trapping of an ion},\ }\href {https://doi.org/10.1038/nphoton.2010.236} {\bibfield  {journal} {\bibinfo  {journal} {Nature Photonics}\ }\textbf {\bibinfo {volume} {4}},\ \bibinfo {pages} {772} (\bibinfo {year} {2010})}\BibitemShut {NoStop}%
\bibitem [{\citenamefont {Huber}\ \emph {et~al.}(2014)\citenamefont {Huber}, \citenamefont {Lambrecht}, \citenamefont {Schmidt}, \citenamefont {Karpa},\ and\ \citenamefont {Schaetz}}]{Huber2014Faroffresonance}%
  \BibitemOpen
  \bibfield  {author} {\bibinfo {author} {\bibfnamefont {T.}~\bibnamefont {Huber}}, \bibinfo {author} {\bibfnamefont {A.}~\bibnamefont {Lambrecht}}, \bibinfo {author} {\bibfnamefont {J.}~\bibnamefont {Schmidt}}, \bibinfo {author} {\bibfnamefont {L.}~\bibnamefont {Karpa}},\ and\ \bibinfo {author} {\bibfnamefont {T.}~\bibnamefont {Schaetz}},\ }\bibfield  {title} {\bibinfo {title} {A far-off-resonance optical trap for a {{Ba}}+ ion},\ }\href {https://doi.org/10.1038/ncomms6587} {\bibfield  {journal} {\bibinfo  {journal} {Nature Communications}\ }\textbf {\bibinfo {volume} {5}},\ \bibinfo {pages} {5587} (\bibinfo {year} {2014})}\BibitemShut {NoStop}%
\bibitem [{\citenamefont {Schaetz}(2017)}]{Schaetz2017Trapping}%
  \BibitemOpen
  \bibfield  {author} {\bibinfo {author} {\bibfnamefont {T.}~\bibnamefont {Schaetz}},\ }\bibfield  {title} {\bibinfo {title} {Trapping ions and atoms optically},\ }\href {https://doi.org/10.1088/1361-6455/aa69b2} {\bibfield  {journal} {\bibinfo  {journal} {Journal of Physics B: Atomic, Molecular and Optical Physics}\ }\textbf {\bibinfo {volume} {50}},\ \bibinfo {pages} {102001} (\bibinfo {year} {2017})}\BibitemShut {NoStop}%
\bibitem [{\citenamefont {Weckesser}\ \emph {et~al.}(2021)\citenamefont {Weckesser}, \citenamefont {Thielemann}, \citenamefont {Hoenig}, \citenamefont {Lambrecht}, \citenamefont {Karpa},\ and\ \citenamefont {Schaetz}}]{Weckesser2021Trapping}%
  \BibitemOpen
  \bibfield  {author} {\bibinfo {author} {\bibfnamefont {P.}~\bibnamefont {Weckesser}}, \bibinfo {author} {\bibfnamefont {F.}~\bibnamefont {Thielemann}}, \bibinfo {author} {\bibfnamefont {D.}~\bibnamefont {Hoenig}}, \bibinfo {author} {\bibfnamefont {A.}~\bibnamefont {Lambrecht}}, \bibinfo {author} {\bibfnamefont {L.}~\bibnamefont {Karpa}},\ and\ \bibinfo {author} {\bibfnamefont {T.}~\bibnamefont {Schaetz}},\ }\bibfield  {title} {\bibinfo {title} {Trapping, shaping and isolating of ion {{Coulomb}} crystals via state-selective optical potentials},\ }\href {https://doi.org/10.1103/PhysRevA.103.013112} {\bibfield  {journal} {\bibinfo  {journal} {Physical Review A}\ }\textbf {\bibinfo {volume} {103}},\ \bibinfo {pages} {1} (\bibinfo {year} {2021})}\BibitemShut {NoStop}%
\bibitem [{\citenamefont {Mazzanti}\ \emph {et~al.}(2023)\citenamefont {Mazzanti}, \citenamefont {Gerritsma}, \citenamefont {Spreeuw},\ and\ \citenamefont {{Safavi-Naini}}}]{Mazzanti2023Trapped}%
  \BibitemOpen
  \bibfield  {author} {\bibinfo {author} {\bibfnamefont {M.}~\bibnamefont {Mazzanti}}, \bibinfo {author} {\bibfnamefont {R.}~\bibnamefont {Gerritsma}}, \bibinfo {author} {\bibfnamefont {R.~J.~C.}\ \bibnamefont {Spreeuw}},\ and\ \bibinfo {author} {\bibfnamefont {A.}~\bibnamefont {{Safavi-Naini}}},\ }\bibfield  {title} {\bibinfo {title} {Trapped ions quantum logic gate with optical tweezers and the {{Magnus}} effect},\ }\href {https://doi.org/10.1103/PhysRevResearch.5.033036} {\bibfield  {journal} {\bibinfo  {journal} {Physical Review Research}\ }\textbf {\bibinfo {volume} {5}},\ \bibinfo {pages} {033036} (\bibinfo {year} {2023})}\BibitemShut {NoStop}%
\bibitem [{\citenamefont {Robalo~Pereira}\ \emph {et~al.}(2025)\citenamefont {Robalo~Pereira}, \citenamefont {Bond}, \citenamefont {Mazzanti}, \citenamefont {Gerritsma},\ and\ \citenamefont {{Safavi-Naini}}}]{RobaloPereira2025Fast}%
  \BibitemOpen
  \bibfield  {author} {\bibinfo {author} {\bibfnamefont {C.}~\bibnamefont {Robalo~Pereira}}, \bibinfo {author} {\bibfnamefont {L.~J.}\ \bibnamefont {Bond}}, \bibinfo {author} {\bibfnamefont {M.}~\bibnamefont {Mazzanti}}, \bibinfo {author} {\bibfnamefont {R.}~\bibnamefont {Gerritsma}},\ and\ \bibinfo {author} {\bibfnamefont {A.}~\bibnamefont {{Safavi-Naini}}},\ }\bibfield  {title} {\bibinfo {title} {Fast {{Quantum Gates}} with {{Electric Field Pulses}} and {{Optical Tweezers}} in {{Trapped Ions}}},\ }\href {https://doi.org/10.3390/e27060595} {\bibfield  {journal} {\bibinfo  {journal} {Entropy}\ }\textbf {\bibinfo {volume} {27}},\ \bibinfo {pages} {595} (\bibinfo {year} {2025})}\BibitemShut {NoStop}%
\bibitem [{\citenamefont {Olsacher}\ \emph {et~al.}(2020)\citenamefont {Olsacher}, \citenamefont {Postler}, \citenamefont {Schindler}, \citenamefont {Monz}, \citenamefont {Zoller},\ and\ \citenamefont {Sieberer}}]{Olsacher2020Scalable}%
  \BibitemOpen
  \bibfield  {author} {\bibinfo {author} {\bibfnamefont {T.}~\bibnamefont {Olsacher}}, \bibinfo {author} {\bibfnamefont {L.}~\bibnamefont {Postler}}, \bibinfo {author} {\bibfnamefont {P.}~\bibnamefont {Schindler}}, \bibinfo {author} {\bibfnamefont {T.}~\bibnamefont {Monz}}, \bibinfo {author} {\bibfnamefont {P.}~\bibnamefont {Zoller}},\ and\ \bibinfo {author} {\bibfnamefont {L.~M.}\ \bibnamefont {Sieberer}},\ }\bibfield  {title} {\bibinfo {title} {Scalable and {{Parallel Tweezer Gates}} for {{Quantum Computing}} with {{Long Ion Strings}}},\ }\href {https://doi.org/10.1103/PRXQuantum.1.020316} {\bibfield  {journal} {\bibinfo  {journal} {PRX Quantum}\ }\textbf {\bibinfo {volume} {1}},\ \bibinfo {pages} {020316} (\bibinfo {year} {2020})}\BibitemShut {NoStop}%
\bibitem [{\citenamefont {Shen}\ and\ \citenamefont {Lin}(2020)}]{Shen2020Scalable}%
  \BibitemOpen
  \bibfield  {author} {\bibinfo {author} {\bibfnamefont {Y.-C.}\ \bibnamefont {Shen}}\ and\ \bibinfo {author} {\bibfnamefont {G.-D.}\ \bibnamefont {Lin}},\ }\bibfield  {title} {\bibinfo {title} {Scalable quantum computing stabilised by optical tweezers on an ion crystal},\ }\href {https://doi.org/10.1088/1367-2630/ab84b6} {\bibfield  {journal} {\bibinfo  {journal} {New Journal of Physics}\ }\textbf {\bibinfo {volume} {22}},\ \bibinfo {pages} {053032} (\bibinfo {year} {2020})}\BibitemShut {NoStop}%
\bibitem [{\citenamefont {Teoh}\ \emph {et~al.}(2021)\citenamefont {Teoh}, \citenamefont {Sajjan}, \citenamefont {Sun}, \citenamefont {Rajabi},\ and\ \citenamefont {Islam}}]{Teoh2021Manipulating}%
  \BibitemOpen
  \bibfield  {author} {\bibinfo {author} {\bibfnamefont {Y.~H.}\ \bibnamefont {Teoh}}, \bibinfo {author} {\bibfnamefont {M.}~\bibnamefont {Sajjan}}, \bibinfo {author} {\bibfnamefont {Z.}~\bibnamefont {Sun}}, \bibinfo {author} {\bibfnamefont {F.}~\bibnamefont {Rajabi}},\ and\ \bibinfo {author} {\bibfnamefont {R.}~\bibnamefont {Islam}},\ }\bibfield  {title} {\bibinfo {title} {Manipulating phonons of a trapped-ion system using optical tweezers},\ }\href {https://doi.org/10.1103/PhysRevA.104.022420} {\bibfield  {journal} {\bibinfo  {journal} {Physical Review A}\ }\textbf {\bibinfo {volume} {104}},\ \bibinfo {pages} {022420} (\bibinfo {year} {2021})}\BibitemShut {NoStop}%
\bibitem [{\citenamefont {Weedbrook}\ \emph {et~al.}(2012)\citenamefont {Weedbrook}, \citenamefont {Pirandola}, \citenamefont {{Garcia-Patron}}, \citenamefont {Cerf}, \citenamefont {Ralph}, \citenamefont {Shapiro},\ and\ \citenamefont {Lloyd}}]{Weedbrook2012Gaussian}%
  \BibitemOpen
  \bibfield  {author} {\bibinfo {author} {\bibfnamefont {C.}~\bibnamefont {Weedbrook}}, \bibinfo {author} {\bibfnamefont {S.}~\bibnamefont {Pirandola}}, \bibinfo {author} {\bibfnamefont {R.}~\bibnamefont {{Garcia-Patron}}}, \bibinfo {author} {\bibfnamefont {N.~J.}\ \bibnamefont {Cerf}}, \bibinfo {author} {\bibfnamefont {T.~C.}\ \bibnamefont {Ralph}}, \bibinfo {author} {\bibfnamefont {J.~H.}\ \bibnamefont {Shapiro}},\ and\ \bibinfo {author} {\bibfnamefont {S.}~\bibnamefont {Lloyd}},\ }\bibfield  {title} {\bibinfo {title} {Gaussian quantum information},\ }\href {https://doi.org/10.1103/RevModPhys.84.621} {\bibfield  {journal} {\bibinfo  {journal} {Reviews of Modern Physics}\ }\textbf {\bibinfo {volume} {84}},\ \bibinfo {pages} {621} (\bibinfo {year} {2012})}\BibitemShut {NoStop}%
\bibitem [{\citenamefont {{Garc{\'i}a-Ripoll}}\ \emph {et~al.}(2003)\citenamefont {{Garc{\'i}a-Ripoll}}, \citenamefont {Zoller},\ and\ \citenamefont {Cirac}}]{Garcia-Ripoll2003Speed}%
  \BibitemOpen
  \bibfield  {author} {\bibinfo {author} {\bibfnamefont {J.~J.}\ \bibnamefont {{Garc{\'i}a-Ripoll}}}, \bibinfo {author} {\bibfnamefont {P.}~\bibnamefont {Zoller}},\ and\ \bibinfo {author} {\bibfnamefont {J.~I.}\ \bibnamefont {Cirac}},\ }\bibfield  {title} {\bibinfo {title} {Speed optimized two-qubit gates with laser coherent control techniques for ion trap quantum computing},\ }\href {https://doi.org/10.1103/PhysRevLett.91.157901} {\bibfield  {journal} {\bibinfo  {journal} {Physical Review Letters}\ }\textbf {\bibinfo {volume} {91}},\ \bibinfo {pages} {157901} (\bibinfo {year} {2003})}\BibitemShut {NoStop}%
\bibitem [{\citenamefont {{Garc{\'i}a-Ripoll}}\ \emph {et~al.}(2005)\citenamefont {{Garc{\'i}a-Ripoll}}, \citenamefont {Zoller},\ and\ \citenamefont {Cirac}}]{Garcia-Ripoll2005Coherent}%
  \BibitemOpen
  \bibfield  {author} {\bibinfo {author} {\bibfnamefont {J.~J.}\ \bibnamefont {{Garc{\'i}a-Ripoll}}}, \bibinfo {author} {\bibfnamefont {P.}~\bibnamefont {Zoller}},\ and\ \bibinfo {author} {\bibfnamefont {J.~I.}\ \bibnamefont {Cirac}},\ }\bibfield  {title} {\bibinfo {title} {Coherent control of trapped ions using off-resonant lasers},\ }\href {https://doi.org/10.1103/PhysRevA.71.062309} {\bibfield  {journal} {\bibinfo  {journal} {Physical Review A}\ }\textbf {\bibinfo {volume} {71}},\ \bibinfo {pages} {062309} (\bibinfo {year} {2005})}\BibitemShut {NoStop}%
\bibitem [{\citenamefont {Glaser}\ \emph {et~al.}(2015)\citenamefont {Glaser}, \citenamefont {Boscain}, \citenamefont {Calarco}, \citenamefont {Koch}, \citenamefont {K{\"o}ckenberger}, \citenamefont {Kosloff}, \citenamefont {Kuprov}, \citenamefont {Luy}, \citenamefont {Schirmer}, \citenamefont {{Schulte-Herbr{\"u}ggen}}, \citenamefont {Sugny},\ and\ \citenamefont {Wilhelm}}]{Glaser2015Training}%
  \BibitemOpen
  \bibfield  {author} {\bibinfo {author} {\bibfnamefont {S.~J.}\ \bibnamefont {Glaser}}, \bibinfo {author} {\bibfnamefont {U.}~\bibnamefont {Boscain}}, \bibinfo {author} {\bibfnamefont {T.}~\bibnamefont {Calarco}}, \bibinfo {author} {\bibfnamefont {C.~P.}\ \bibnamefont {Koch}}, \bibinfo {author} {\bibfnamefont {W.}~\bibnamefont {K{\"o}ckenberger}}, \bibinfo {author} {\bibfnamefont {R.}~\bibnamefont {Kosloff}}, \bibinfo {author} {\bibfnamefont {I.}~\bibnamefont {Kuprov}}, \bibinfo {author} {\bibfnamefont {B.}~\bibnamefont {Luy}}, \bibinfo {author} {\bibfnamefont {S.}~\bibnamefont {Schirmer}}, \bibinfo {author} {\bibfnamefont {T.}~\bibnamefont {{Schulte-Herbr{\"u}ggen}}}, \bibinfo {author} {\bibfnamefont {D.}~\bibnamefont {Sugny}},\ and\ \bibinfo {author} {\bibfnamefont {F.~K.}\ \bibnamefont {Wilhelm}},\ }\bibfield  {title} {\bibinfo {title} {Training {{Schr\"odinger}}'s cat: Quantum optimal control},\ }\href {https://doi.org/10.1140/epjd/e2015-60464-1} {\bibfield  {journal} {\bibinfo  {journal} {The European
  Physical Journal D}\ }\textbf {\bibinfo {volume} {69}},\ \bibinfo {pages} {279} (\bibinfo {year} {2015})}\BibitemShut {NoStop}%
\bibitem [{\citenamefont {Grimm}\ \emph {et~al.}(2000)\citenamefont {Grimm}, \citenamefont {Weidem{\"u}ller},\ and\ \citenamefont {Ovchinnikov}}]{Grimm2000Optical}%
  \BibitemOpen
  \bibfield  {author} {\bibinfo {author} {\bibfnamefont {R.}~\bibnamefont {Grimm}}, \bibinfo {author} {\bibfnamefont {M.}~\bibnamefont {Weidem{\"u}ller}},\ and\ \bibinfo {author} {\bibfnamefont {Y.~B.}\ \bibnamefont {Ovchinnikov}},\ }\bibfield  {title} {\bibinfo {title} {Optical dipole traps for neutral atoms},\ }in\ \href {https://doi.org/10.1016/S1049-250X(08)60186-X} {\emph {\bibinfo {booktitle} {Advances In Atomic, Molecular, and Optical Physics}}},\ Vol.~\bibinfo {volume} {42}\ (\bibinfo  {publisher} {Elsevier},\ \bibinfo {year} {2000})\ pp.\ \bibinfo {pages} {95--170}\BibitemShut {NoStop}%
\bibitem [{Note1()}]{Note1}%
  \BibitemOpen
  \bibinfo {note} {To convert from atomic units (\si {\atomicunit }) of polarizability to physical units, multiply by $4\pi \varepsilon _0a_B^3$, where $a_B=\qty {5.29e-11}{\metre }$ is the Bohr radius.}\BibitemShut {Stop}%
\bibitem [{\citenamefont {Le~Kien}\ \emph {et~al.}(2013)\citenamefont {Le~Kien}, \citenamefont {Schneeweiss},\ and\ \citenamefont {Rauschenbeutel}}]{LeKien2013Dynamical}%
  \BibitemOpen
  \bibfield  {author} {\bibinfo {author} {\bibfnamefont {F.}~\bibnamefont {Le~Kien}}, \bibinfo {author} {\bibfnamefont {P.}~\bibnamefont {Schneeweiss}},\ and\ \bibinfo {author} {\bibfnamefont {A.}~\bibnamefont {Rauschenbeutel}},\ }\bibfield  {title} {\bibinfo {title} {Dynamical polarizability of atoms in arbitrary light fields: General theory and application to cesium},\ }\href {https://doi.org/10.1140/epjd/e2013-30729-x} {\bibfield  {journal} {\bibinfo  {journal} {The European Physical Journal D}\ }\textbf {\bibinfo {volume} {67}},\ \bibinfo {pages} {92} (\bibinfo {year} {2013})}\BibitemShut {NoStop}%
\bibitem [{\citenamefont {Zhang}\ \emph {et~al.}(2020)\citenamefont {Zhang}, \citenamefont {Arnold}, \citenamefont {Chanu}, \citenamefont {Kaewuam}, \citenamefont {Safronova},\ and\ \citenamefont {Barrett}}]{Zhang2020Branching}%
  \BibitemOpen
  \bibfield  {author} {\bibinfo {author} {\bibfnamefont {Z.}~\bibnamefont {Zhang}}, \bibinfo {author} {\bibfnamefont {K.~J.}\ \bibnamefont {Arnold}}, \bibinfo {author} {\bibfnamefont {S.~R.}\ \bibnamefont {Chanu}}, \bibinfo {author} {\bibfnamefont {R.}~\bibnamefont {Kaewuam}}, \bibinfo {author} {\bibfnamefont {M.~S.}\ \bibnamefont {Safronova}},\ and\ \bibinfo {author} {\bibfnamefont {M.~D.}\ \bibnamefont {Barrett}},\ }\bibfield  {title} {\bibinfo {title} {Branching fractions for {$P_{3/2}$} decays in {$\mathrm{Ba}^+$}},\ }\href {https://doi.org/10.1103/PhysRevA.101.062515} {\bibfield  {journal} {\bibinfo  {journal} {Physical Review A}\ }\textbf {\bibinfo {volume} {101}},\ \bibinfo {pages} {062515} (\bibinfo {year} {2020})}\BibitemShut {NoStop}%
\bibitem [{\citenamefont {Kiruga}\ \emph {et~al.}(2026)\citenamefont {Kiruga}, \citenamefont {Cheung}, \citenamefont {Filin}, \citenamefont {Barakhshan}, \citenamefont {Bhosale}, \citenamefont {Badhan}, \citenamefont {Arora}, \citenamefont {Eigenmann},\ and\ \citenamefont {Safronova}}]{DelawareDatabase}%
  \BibitemOpen
  \bibfield  {author} {\bibinfo {author} {\bibfnamefont {A.}~\bibnamefont {Kiruga}}, \bibinfo {author} {\bibfnamefont {C.}~\bibnamefont {Cheung}}, \bibinfo {author} {\bibfnamefont {D.}~\bibnamefont {Filin}}, \bibinfo {author} {\bibfnamefont {P.}~\bibnamefont {Barakhshan}}, \bibinfo {author} {\bibfnamefont {A.}~\bibnamefont {Bhosale}}, \bibinfo {author} {\bibfnamefont {V.}~\bibnamefont {Badhan}}, \bibinfo {author} {\bibfnamefont {B.}~\bibnamefont {Arora}}, \bibinfo {author} {\bibfnamefont {R.}~\bibnamefont {Eigenmann}},\ and\ \bibinfo {author} {\bibfnamefont {M.~S.}\ \bibnamefont {Safronova}},\ }\bibfield  {title} {\bibinfo {title} {Portal for high-precision atomic data and computation},\ }\href {https://doi.org/https://doi.org/10.1016/j.cpc.2025.109951} {\bibfield  {journal} {\bibinfo  {journal} {Computer Physics Communications}\ }\textbf {\bibinfo {volume} {319}},\ \bibinfo {pages} {109951} (\bibinfo {year} {2026})}\BibitemShut {NoStop}%
\bibitem [{\citenamefont {Moore}\ \emph {et~al.}(2023)\citenamefont {Moore}, \citenamefont {Campbell}, \citenamefont {Hudson}, \citenamefont {Boguslawski}, \citenamefont {Wineland},\ and\ \citenamefont {Allcock}}]{Moore2023}%
  \BibitemOpen
  \bibfield  {author} {\bibinfo {author} {\bibfnamefont {I.~D.}\ \bibnamefont {Moore}}, \bibinfo {author} {\bibfnamefont {W.~C.}\ \bibnamefont {Campbell}}, \bibinfo {author} {\bibfnamefont {E.~R.}\ \bibnamefont {Hudson}}, \bibinfo {author} {\bibfnamefont {M.~J.}\ \bibnamefont {Boguslawski}}, \bibinfo {author} {\bibfnamefont {D.~J.}\ \bibnamefont {Wineland}},\ and\ \bibinfo {author} {\bibfnamefont {D.~T.~C.}\ \bibnamefont {Allcock}},\ }\bibfield  {title} {\bibinfo {title} {Photon scattering errors during stimulated raman transitions in trapped-ion qubits},\ }\href {https://doi.org/10.1103/PhysRevA.107.032413} {\bibfield  {journal} {\bibinfo  {journal} {Phys. Rev. A}\ }\textbf {\bibinfo {volume} {107}},\ \bibinfo {pages} {032413} (\bibinfo {year} {2023})}\BibitemShut {NoStop}%
\bibitem [{\citenamefont {Ozeri}\ \emph {et~al.}(2007)\citenamefont {Ozeri}, \citenamefont {Itano}, \citenamefont {Blakestad}, \citenamefont {Britton}, \citenamefont {Chiaverini}, \citenamefont {Jost}, \citenamefont {Langer}, \citenamefont {Leibfried}, \citenamefont {Reichle}, \citenamefont {Seidelin}, \citenamefont {Wesenberg},\ and\ \citenamefont {Wineland}}]{Ozeri2007}%
  \BibitemOpen
  \bibfield  {author} {\bibinfo {author} {\bibfnamefont {R.}~\bibnamefont {Ozeri}}, \bibinfo {author} {\bibfnamefont {W.~M.}\ \bibnamefont {Itano}}, \bibinfo {author} {\bibfnamefont {R.~B.}\ \bibnamefont {Blakestad}}, \bibinfo {author} {\bibfnamefont {J.}~\bibnamefont {Britton}}, \bibinfo {author} {\bibfnamefont {J.}~\bibnamefont {Chiaverini}}, \bibinfo {author} {\bibfnamefont {J.~D.}\ \bibnamefont {Jost}}, \bibinfo {author} {\bibfnamefont {C.}~\bibnamefont {Langer}}, \bibinfo {author} {\bibfnamefont {D.}~\bibnamefont {Leibfried}}, \bibinfo {author} {\bibfnamefont {R.}~\bibnamefont {Reichle}}, \bibinfo {author} {\bibfnamefont {S.}~\bibnamefont {Seidelin}}, \bibinfo {author} {\bibfnamefont {J.~H.}\ \bibnamefont {Wesenberg}},\ and\ \bibinfo {author} {\bibfnamefont {D.~J.}\ \bibnamefont {Wineland}},\ }\bibfield  {title} {\bibinfo {title} {Errors in trapped-ion quantum gates due to spontaneous photon scattering},\ }\href {https://doi.org/10.1103/PhysRevA.75.042329} {\bibfield  {journal} {\bibinfo  {journal}
  {Phys. Rev. A}\ }\textbf {\bibinfo {volume} {75}},\ \bibinfo {pages} {042329} (\bibinfo {year} {2007})}\BibitemShut {NoStop}%
\bibitem [{\citenamefont {Gerry}\ and\ \citenamefont {Knight}(2004)}]{Gerry2004Introductory}%
  \BibitemOpen
  \bibfield  {author} {\bibinfo {author} {\bibfnamefont {C.}~\bibnamefont {Gerry}}\ and\ \bibinfo {author} {\bibfnamefont {P.}~\bibnamefont {Knight}},\ }\href@noop {} {\emph {\bibinfo {title} {Introductory Quantum Optics}}}\ (\bibinfo  {publisher} {Cambridge University Press},\ \bibinfo {address} {Cambridge},\ \bibinfo {year} {2004})\BibitemShut {NoStop}%
\bibitem [{\citenamefont {Nielsen}(2002)}]{Nielsen2002Simple}%
  \BibitemOpen
  \bibfield  {author} {\bibinfo {author} {\bibfnamefont {M.~A.}\ \bibnamefont {Nielsen}},\ }\bibfield  {title} {\bibinfo {title} {A simple formula for the average gate fidelity of a quantum dynamical operation},\ }\href {https://doi.org/10.1016/S0375-9601(02)01272-0} {\bibfield  {journal} {\bibinfo  {journal} {Physics Letters A}\ }\textbf {\bibinfo {volume} {303}},\ \bibinfo {pages} {249} (\bibinfo {year} {2002})}\BibitemShut {NoStop}%
\end{thebibliography}%

\clearpage
\onecolumngrid
\section*{Supplementary Information}
\setcounter{subsection}{0}
\setcounter{subsubsection}{0}

This Supplement complements the main text and the Methods. 
It introduces the physical model for the two-ion gate in Sec.~\ref{sec:two-qubit}, as well as the description and analysis of the gates in Secs.~\ref{ssec:dynamical} and~\ref{ssec:blockade} of the main text.
We include the intermediate steps and highlight the underlying assumptions of the model. 
We also provide details on the numerical simulation of the robustness analysis for initial thermal states and for studying cross-talk in Sec.~\ref{sec:considerations}.

\subsection{Two-ion Hamiltonian} \label{si:hamiltonian}

In this section we derive the formulas quoted in Sec.~\ref{sec:two-qubit} for the perpendicular two-qubit gate. 
Throughout we set $\hbar=1$ and use a \textit{one-dimensional} motional model along the push direction, which is perpendicular to the gate axis. 
The push direction is thus the one directly affected by the state-dependent tweezer displacement and therefore the coordinate that controls the gate mechanism in the main text.
For Gates~A and~B this geometry is advantageous because it yields a smaller single-particle phase at fixed entangling phase. 
A parallel geometry, in which the push is along the ion--ion axis, is briefly discussed in Sec.~\ref{si:parallel-geometry} of this Supplement.

Initially, the qubit is encoded in the long-lived states $\ket{0}$ and $\ket{1}$, with $\ket{e}$ serving as an additional shelving state.
Pulses between the states are described by the laser Hamiltonian
\begin{equation} \label{eq:si-H_las}
    H_{\mathrm{las}}(t)
    =-\sum_{j=1}^{2}\Delta_j(t)\ket{e}_j\bra{e}_j +\sum_{j=1}^{2}\frac{\Omega_{R,j}(t)}{2}
    \left(\ket{e}_j\bra{1}_j+\ket{1}_j\bra{e}_j\right),
\end{equation}
where $\Delta_j(t)$ is the detuning from the $\ket{1}\leftrightarrow\ket{e}$ transition on ion $j$ and $\Omega_{R,j}(t)$ is the corresponding Rabi frequency.
The laser Hamiltonian acts together with a branch-dependent motional Hamiltonian that we introduce in the next paragraphs.
For the analytical description we assume that the pulses act in the Lamb--Dicke regime, so that recoil is ignored in Eq.~\eqref{eq:si-H_las}.
We also assume for Gates~A--C that the Rabi frequencies $\Omega_{R,j}(t)$ are much larger than the trap frequency, so that we may ignore the duration of these pulses from the analysis.
For the dipole-blockade gate in Sec.~\ref{ssec:blockade} we explicitly include the pulse duration of the second step.

Next, we describe the motional Hamiltonian.
We approximate each branch-dependent tweezer potential by a harmonic potential around its local minimum, which is accurate when the wavepacket remains close to that minimum. 
We discuss corrections to the harmonic potentials later in Sec.~\ref{si:gaussian-nonlinearity}.
For simplicity we take the trap frequencies seen by $\ket{0}$, $\ket{1}$, and $\ket{e}$ to be the same within a gate zone, so that the internal-state dependence enters only through the trap-center shift.
Residual cross-talk of the tweezers between $\ket{0}$, $\ket{1}$, and $\ket{e}$ would amount to state-dependent variations of the mode frequencies that can be included in a more detailed treatment or numerically optimized gates.
Here, we focus on the idealized case where all states are subject to identical harmonic potentials.

Let $q_1$ and $q_2$ denote the coordinates of the two ions along the push direction.
In the internal branch $\ket{r,s}$ the trap minima of the ions are shifted by the physical distances $\xi_r^{(1)}$ and $\xi_s^{(2)}$.
We denote by $d$ the equilibrium ion separation along the gate axis before the branch-dependent perpendicular push is applied, so that any static Coulomb-induced axial shift common to all branches is absorbed into the definition of $d$. 
When the gate is off, the equilibrium position is therefore simply $q_1=q_2=0$.
The corresponding one-dimensional motional Hamiltonian is then
\begin{equation}
   H_{\rm mot} = \sum_{r,s\in\{0,1,e\}} |r,s\rangle\langle r,s| \otimes  H_{r,s}
\end{equation}
with
\begin{equation}
    H_{r,s} = \frac{p_1^2+p_2^2}{2m} + \frac{m\omega^2}{2}\left(q_1-\xi_r^{(1)}\right)^2 + \frac{m\omega^2}{2}\left(q_2-\xi_s^{(2)}\right)^2 + \frac{e^2}{4\pi\varepsilon_0\sqrt{d^2+(q_2-q_1)^2}}.
    \label{eq:si-Hrs-start}
\end{equation}
The last term in $H_{r,s}$ describes the Coulomb interaction between the ions.
Only the shelved state $\ket{e}$ experiences the push displacement, i.e.~ $\xi_r^{(1)}=0$ for $r=0,1$ and $\xi_s^{(2)}=0$ for $s=0,1$.
We further choose opposite transverse shifts for the two ions, i.e., $\xi_e^{(1)}<0$ and $\xi_e^{(2)}>0$.

As a next step we consider the harmonic approximation of the Coulomb potential.
The approximation is reasonable as long as one can assume that we operate in a regime where the displacement is small compared with the inter-ion separation,
\begin{equation}
    |q_2-q_1|\ll d.    
\end{equation}
In that regime the Coulomb term can be expanded to second order in $|q_2-q_1|/d$ as
\begin{equation}
    \frac{1}{\sqrt{d^2+(q_2-q_1)^2}}
    \approx \frac{1}{d}
    - \frac{(q_2-q_1)^2}{2d^3}.
    \label{eq:si-coulomb-expand}
\end{equation}
In addition, we note that the excursion of the ion wavefunction from the local trap minimum should remain small compared to the tweezer waist $w_0$.
The corresponding leading correction to the harmonic potential is summarized later in Sec.~\ref{si:gaussian-nonlinearity}.

For convenience we may write the Hamiltonian within the harmonic approximation in new coordinates, and introduce center-of-mass (CM) and relative (rel) coordinates,
\begin{align}
    X &= \frac{q_1+q_2}{2}, & x&= q_2-q_1, \\
    P &= p_1+p_2,           & p&= \frac{p_2-p_1}{2}, \\
    M &= 2m,                & \mu&= \frac{m}{2}.
\end{align}
Using Eq.~\eqref{eq:si-coulomb-expand}, the Hamiltonian then becomes
\begin{equation}
    H_{r,s}
    = \frac{P^2}{2M} + \frac{p^2}{2\mu}
    + \frac{M\omega^2}{2}\Big(X-\frac{\xi_r^{(1)}+\xi_s^{(2)}}{2}\Big)^2
    + \frac{\mu\omega^2}{2}
    \left[\big(x-(\xi_s^{(2)}-\xi_r^{(1)})\big)^2 - \eta^2 x^2\right]
    + \text{const},
    \label{eq:si-Hcmrel-pre}
\end{equation}
where we defined the dimensionless interaction parameter
\begin{align}
    \eta^2
    &= \frac{e^2}{2\pi\varepsilon_0 m\omega^2 d^3},
    \label{eq:si-eta-def}
\end{align}
which becomes a measure for the interaction strength between the ions and can be tuned by changing the inter-ion distance $d$.
We can already see from Eq.~\eqref{eq:si-Hcmrel-pre} that the center-of-mass mode is unaffected by the Coulomb interaction, whereas the relative mode becomes softer.
Completing the square in the relative coordinate yields the following form for the Hamiltonian
\begin{equation}
    H_{r,s}
    = \frac{P^2}{2M} + \frac{M\omega^2}{2}\left(X-X_{r,s}^{\mathrm{eq}}\right)^2 + \frac{p^2}{2\mu} + \frac{\mu\Omega^2}{2}\left(x-x_{r,s}^{\mathrm{eq}}\right)^2 + C_{r,s}, 
    \label{eq:si-Hcmrel-square}
\end{equation}
with the equilibrium positions for the CM and rel modes
\begin{align}
    X_{r,s}^{\mathrm{eq}} = \frac{\xi_r^{(1)}+\xi_s^{(2)}}{2}, \qquad
    x_{r,s}^{\mathrm{eq}} = \frac{\xi_s^{(2)}-\xi_r^{(1)}}{1-\eta^2}
\end{align}
and the rel mode frequency
\begin{align}
    \Omega = \omega\sqrt{1-\eta^2},
\end{align}
which depends on the interaction parameter $\eta^2$, defined in Eq.~\eqref{eq:si-eta-def}.
We define the single-ion oscillator length
\begin{equation}
    l_0=\sqrt{\frac{1}{2m\omega}}
\end{equation}
as a reference length scale which we use to state the branch-dependent scalar energy shift that generates the entangling phase,
\begin{align}
    C_{r,s}
    & = -\omega\frac{\eta^2}{8(1-\eta^2)}
    \left(\frac{\xi_s^{(2)}-\xi_r^{(1)}}{l_0}\right)^2,
    \label{eq:si-Crs}
\end{align}
as stated in the main text Eq.~\eqref{eq.Vshift} in Sec.~\ref{ssec:physsetup}.

We now bring the Hamiltonian into the form of a displaced oscillator, Eq.~\eqref{eq:H_displaced} in the main text.
We therefore quantize the two normal modes.
Their oscillator lengths are rescaled as
\begin{equation}
    l_0^{\mathrm{CM}}
    = \sqrt{\frac{1}{2M\omega}}
    = \frac{l_0}{\sqrt{2}},
    \qquad
    l_0^{\mathrm{rel}}
    = \sqrt{\frac{1}{2\mu\Omega}}
    = \frac{\sqrt{2}\,l_0}{(\Omega/\omega)^{1/2}}.
\end{equation}
We then define the corresponding ladder operators for both modes as
\begin{align}
    X &= l_0^{\mathrm{CM}}\left(a_{\rm CM}^{}+a_{\rm CM}^\dagger\right), &
    P&= \frac{i}{2l_0^{\mathrm{CM}}}\left(a_{\rm CM}^\dagger-a_{\rm CM}^{}\right),
    \notag\\
    x &= l_0^{\mathrm{rel}}\left(a_{\rm rel}^{}+a_{\rm rel}^\dagger\right), &
    p&= \frac{i}{2l_0^{\mathrm{rel}}}\left(a_{\rm rel}^\dagger-a_{\rm rel}^{}\right).
    \label{eq:si-ladder-defs}
\end{align}
Up to the branch-independent zero-point energy, Eq.~\eqref{eq:si-Hcmrel-square} can then be written as
\begin{equation}
    H_{r,s}
    = D_{r,s}^{}H_0^{}D_{r,s}^\dagger + C_{r,s}^{},
    \qquad
    H_0
    = \omega a_{\rm CM}^\dagger a_{\rm CM}^{}+\Omega a_{\rm rel}^\dagger a_{\rm rel}^{},
    \label{eq:si-H-displaced}
\end{equation}
with displacement operator
\begin{align}
    D_{r,s}
    &= \exp\!\left[\alpha_{r,s}^{\rm CM}\left(a_{\rm CM}^\dagger-a_{\rm CM}\right)\right]
    \exp\!\left[\alpha_{r,s}^{\rm rel}\left(a_{\rm rel}^\dagger-a_{\rm rel}\right)\right].
    \label{eq:si-Drs}
\end{align}
The coherent-state displacements of both modes are therefore given as
\begin{equation}
    \alpha_{r,s}^{\rm CM}
    = \frac{X_{r,s}^{\mathrm{eq}}}{2l_0^{\mathrm{CM}}}
    = \frac{\xi_r^{(1)}+\xi_s^{(2)}}{2\sqrt{2}\,l_0},
    \qquad
    \alpha_{r,s}^{\rm rel}
    = \frac{x_{r,s}^{\mathrm{eq}}}{2l_0^{\mathrm{rel}}}
    = \frac{\xi_s^{(2)}-\xi_r^{(1)}}{2\sqrt{2}\,l_0(\Omega/\omega)^{3/2}}.
    \label{eq:si-Zz}
\end{equation}
This is the Hamiltonian form given in Sec.~\ref{ssec:physsetup}, which shows that the two-ion system can be described by two harmonic oscillators for the CM and the rel mode, and the scalar potential term $C_{r,s}$.

\subsection{Piecewise-displaced oscillators}

The dynamical gates in Sec.~\ref{ssec:dynamical} reduce, mode by mode, to harmonic oscillators whose equilibrium position is changed in time. 
The displaced-oscillator form in Eq.~\eqref{eq:si-H-displaced} therefore allows for a particularly simple analysis of the gate.

Below, we introduce the bookkeeping to describe the evolution of an initial coherent state during the gate evolution, which allows us to design the concrete protocols for the dynamical gates and analyze their gate fidelity.
A residual branch-dependent displacement at the end of the gate entangles the qubits with motion, which is why we impose exact closure of the phase-space trajectories for both modes. 
If every computational branch closes in every mode, the gate acts on the qubits only through phases and is therefore independent of the initial motional density matrix, including the thermal states analyzed later in Sec.~\ref{si:thermal-fidelity}.

For the gate description it is convenient to parameterize the physical trap-center shift by
\begin{align}
    \xi(t)
    &= 2l_0 z g(t),
    \label{eq:si-agt}
\end{align}
where $g(t)$ is a dimensionless waveform, typically of order unity, and where the dimensionless, positive parameter $z$ fixes the overall displacement scale.
In the sign convention adopted here we take the trap-center shifts to be
\begin{align}
    \xi_e^{(1)}(t)
    &= -\xi(t),
    &
    \xi_e^{(2)}(t)
    &= +\xi(t).
\end{align}
For the $\ket{01}$ and $\ket{10}$ branches, the resulting coherent displacements are obtained from Eq.~\eqref{eq:si-Zz}:
\begin{equation}
    \delta_{\mathrm{CM}}(t)
    = \frac{\xi(t)}{2\sqrt{2}\,l_0}
    = \frac{zg(t)}{\sqrt{2}},
    \qquad
    \delta_{\mathrm{rel}}(t)
    = \frac{\xi(t)}{2\sqrt{2}\,l_0\,(\Omega/\omega)^{3/2}}
    = \frac{zg(t)}{\sqrt{2}\,(\Omega/\omega)^{3/2}},
    \label{eq:si-mixed-deltas}
\end{equation}
so that the $\ket{01}$ branch uses $+\delta_{\mathrm{CM}}(t)$, the $\ket{10}$ branch uses $-\delta_{\mathrm{CM}}(t)$, and both use the same $\delta_{\mathrm{rel}}(t)$. The $\ket{11}$ branch has no center-of-mass shift, and its relative displacement is 
\begin{equation}
    \widetilde\delta_{\mathrm{rel}}(t) = 2\delta_{\mathrm{rel}}(t)
    = \frac{\sqrt{2}\,zg(t)}{(\Omega/\omega)^{3/2}}.
    \label{eq:si-double-delta}
\end{equation}

For a piecewise-constant pulse the evolution of each normal mode can be followed exactly in the displaced frame. 
Consider one mode of frequency $\nu$ whose equilibrium position is shifted by real values $\delta_j$ during time interval $t_j$, with $j=1,\ldots,N$.
We assume that the trap center is at the origin before the first change and after the last one, so $\delta_0=\delta_{N+1}=0$. 
If $\alpha_j^+$ denotes the coherent amplitude immediately after entering segment $j$, and $\alpha_j^-$ the amplitude immediately before the next change of trap center, then we can write a recursion for the evolution of the coherent amplitude, as
\begin{align}
    \alpha_1^+
    &= -\delta_1,
    \notag\\
    \alpha_j^-
    &= \alpha_j^+ e^{-i\nu t_j},
    \notag\\
    \alpha_{j+1}^+
    &= \alpha_j^- + \delta_j-\delta_{j+1},
    \qquad
    j=1,\ldots,N.
    \label{eq:si-recursion}
\end{align}
These relations are simply free harmonic rotation within each time segment followed by a translation when the trap center is changed. 

The phase associated with a sudden change of trap center follows from the identity~\cite{Weedbrook2012Gaussian}
\begin{align}
    D(\Delta\alpha)\ket{\alpha}
    &= \exp\!\left[\frac{\Delta\alpha\,\alpha^* - \Delta\alpha^*\alpha}{2}\right]\ket{\alpha+\Delta\alpha},
    \label{eq:si-weyl}
\end{align}
which yields, for a change $\delta_j\to\delta_{j+1}$,
\begin{align}
    \delta\phi_j(\nu)
    &= \operatorname{Im}\!\left[(\delta_j-\delta_{j+1})\alpha_j^{-*}\right],
    \label{eq:si-jump-phase}
\end{align}
because the physical trap-center displacements $\delta_j$ are taken to be real.

For an initial vacuum state and a gate that is described by $U_\nu$, the final state is therefore
\begin{equation}
    U_\nu\ket{0}
    = \exp\!\Big[i\sum_{j=1}^{N}\delta\phi_j(\nu)\Big]
      \ket{\alpha_{N+1}^+}.
\end{equation}
The vacuum overlap is then given as
\begin{equation}
    \langle 0|U_\nu |0\rangle
    = \exp\!\Big[-\frac{1}{2}\left|\alpha_{N+1}^+\right|^2
    + i\sum_{j=1}^{N}\delta\phi_j(\nu)\Big].
\end{equation}
Exact mode closure is equivalent to $\alpha_{N+1}^+=0$, in which case the branch-dependent displacement vanishes.
The mode may still undergo the same branch-independent harmonic evolution as any undriven oscillator, but it no longer remains entangled with the internal states. 
An exactly closed harmonic gate is therefore insensitive to the initial motional state.
We also refer to Refs.~\cite{Garcia-Ripoll2003Speed, Garcia-Ripoll2005Coherent} for a related discussion.

The phase entering the two-ion gate for branch $b\in\{00,01,10,11\}$ is determined by the scalar shift $C_b(t)$ together with the phases accumulated due to the trap-center changes of both modes, as stated in Eq.~\eqref{eq:si-jump-phase}. 
We note that the scalar term enters the time evolution through $\exp[-i\int_0^{T_G} C_b(t)\,\mathrm{d}t]$ and therefore contributes with a minus sign.
Therefore, the total phase accumulated for branch $b$ is
\begin{equation}
    \Phi_b = -\int_0^{T_G} C_b(t)\,\mathrm{d}t
    + \sum_j \delta\phi_{j,b}^{\rm CM}
    + \sum_j \delta\phi_{j,b}^{\rm rel}.
    \label{eq:si-total-phase}
\end{equation}
For the symmetric perpendicular protocols used in the main text the scalar shifts are:
\begin{align}
    C_{00}(t)
    &= 0,
    &
    C_{01}(t)=C_{10}(t)
    &= -\omega\frac{\eta^2}{2(1-\eta^2)}z^2 g(t)^2,
    &
    C_{11}(t)
    &= 4C_{10}(t).
    \label{eq:si-branch-Ct}
\end{align}
We thus have $\Phi_{01}=\Phi_{10}$ and the entangling phase is
\begin{align}
    \Phi_{ZZ}
    &= \Phi_{11}-\Phi_{10}-\Phi_{01}+\Phi_{00}
    = \Phi_{11}-2\Phi_{10},
    \label{eq:si-phizz-def}
\end{align}
where the global phase and local phases are subtracted, as they can be calibrated or compensated.

\subsection{Description of the two-ion gates}

We now provide a full description of the four gate protocols presented in the main text, starting with the three dynamical gates and then presenting the dipole-blockade gate.

\subsubsection{Commensurate single-hold gate (Gate A).}
The conceptually simplest pulse is the commensurate gate, which requires only a single hold during the gate duration.
For this pulse we therefore set the waveform to
\begin{align}
    g(t)
    &= 1,
    \qquad
    0<t<T_G,
    \label{eq:si-gateA-pulse}
\end{align}
i.e.~for the full duration of the pulse.
For one mode with coherent shift $\beta$, the sequence is $0\to\beta\to0$.
The recursion in Eq.~\eqref{eq:si-recursion} then gives a final coherent amplitude
\begin{align}
    \alpha_f
    &= \beta\left(1-e^{-i\nu T_G}\right).
    \label{eq:si-gateA-alpha}
\end{align}
Therefore the mode closes whenever $\nu T_G=2\pi n$, i.e., the time is chosen such that the mode completes a full circle in phase space.
To close both modes simultaneously one then chooses the gate time in a commensurate way, such that both modes complete the circle in phase space at the same instant. 
We therefore select
\begin{align}
    \omega T_G
    &= 2\pi q,
    &
    \Omega T_G
    &= 2\pi p,
    \label{eq:si-gateA-comm}
\end{align}
with coprime integers $p$ and $q$. 
We observe that under Eq.~\eqref{eq:si-gateA-comm} all phases from sudden trap-center changes vanish for this choice of $T_G$, so only the static shift $C_{r,s}$ contributes to the entangling phase.
For the $\ket{10}$ branch phase $\Phi_{10}^{(A)}$ and for the entangling phase $\Phi_{ZZ}^{(A)}$ one then obtains with Eq.~\eqref{eq:si-branch-Ct} the expressions
\begin{equation}
    \Phi_{10}^{(A)}
    = \pi q\frac{\eta^2}{1-\eta^2}z^2,
    \qquad
    \Phi_{ZZ}^{(A)}
    = \Phi_{11}^{(A)}-2\Phi_{10}^{(A)}
    = \pi q\frac{2\eta^2}{1-\eta^2}z^2.
    \label{eq:si-gateA-phases}
\end{equation}
Imposing an entangling phase $\Phi_{ZZ}^{(A)}=\pi$, corresponding to a CZ gate, fixes the required displacement amplitude to
\begin{equation}
    z_{\mathrm{CZ}}^2
    = \frac{1-\eta^2}{2q\eta^2},
       \label{eq:si-gateA-zcz}
\end{equation}
and the trap-center shift to $\xi_{\mathrm{CZ}} = 2l_0 z_{\mathrm{CZ}}$.
These are the amplitude and shift quoted in Eq.~\eqref{eq:gateA_xi_z} in the Gate~A discussion of Sec.~\ref{ssec:dynamical}.
For the example used in Fig.~\ref{fig:fig2}(b), we consider $\Omega/\omega=4/5$, so $q=5$ and $\eta^2=1-(4/5)^2=9/25$. 
The plotted constant shift is $\xi_{\mathrm{CZ}}/2l_0=z_{\mathrm{CZ}}=\sqrt{8/45}\approx0.422$. 
With these parameters, the single-particle phase is correspondingly $\Phi_{10}^{(A)}=\pi/2$.

We also consider the stability of the gate in the presence of a uniform relative displacement error of the trap center,
\begin{equation}
    \xi(t) \to (1+\varepsilon)\xi(t),
\end{equation}
which could arise due to imperfect steering of the tweezer.
While such a displacement rescales the gate phase, it does not affect the closure of the modes, because Eq.~\eqref{eq:si-gateA-comm} depends only on the frequencies. 
Since the entangling phase is quadratic in the displacement amplitude, we obtain
\begin{equation}
    \Phi_{ZZ}^{(A)}(\varepsilon)
    = \pi(1+\varepsilon)^2,
    \label{eq:app-single-error}
\end{equation}
so the phase error is $\delta\Phi(\varepsilon) = \pi\big[(1+\varepsilon)^2-1\big]$ independent of the initial coherent state, and thus independent of the initial temperature.

As discussed in the main text in Sec.~\ref{ssec:dynamical}, a disadvantage of the commensurate pulse is that it requires long gate times when $\eta^2$ is small, motivating more flexible dynamical gates that do not require the commensurability condition.

\subsubsection{Three-segment bang--bang gate (Gate B).}
We describe the fast pulse used in Fig.~\ref{fig:fig2}(c) that has fixed total duration $T_G=4\pi/\omega$, which is independent of the value for $\eta^2$.
This implies that the distance $d$ between the ions does not have to be chosen to be compatible with an $\eta^2$ value suitable for the commensurate pulse.
The dimensionless waveform for this three-segment gate is
\begin{align}
    g(t)
    &=
    \begin{cases}
        1, & 0<t<\tau_3 \text{ or } 3\tau_3<t<4\tau_3, \\
        g_m, & \tau_3<t<3\tau_3,
    \end{cases}
    \qquad
    \tau_3=\frac{\pi}{\omega}.
    \label{eq:si-gateB-pulse}
\end{align}
For one mode of frequency $\nu$ and coherent shift $\beta$, the sequence is $0\to\beta\to g_m\beta\to\beta\to0$. 
Defining $u= e^{-i\nu\tau_3}$, the coherent state evolution for the three steps is described by the recursion from Eq.~\eqref{eq:si-recursion}:
\begin{align}
    \alpha_1^-
    &= -\beta u,
    \notag\\
    \alpha_2^+
    &= -\beta u + (1-g_m)\beta,
    \notag\\
    \alpha_2^-
    &= -\beta u^2\left(u+g_m-1\right),
    \notag\\
    \alpha_3^-
    &= -\beta u^4 + (1-g_m)\beta\left(u^3-u\right).
    \label{eq:si-gateB-recursion}
\end{align}
Hence, the coherent amplitude of the final state is
\begin{align}
    \alpha_f
    &= \beta\left[1-u^4+(1-g_m)(u^3-u)\right].
    \label{eq:si-gateB-alpha}
\end{align}
For the CM mode, $u=\exp(-i\pi)=-1$, so $\alpha_f=0$ for any $g_m$, implying that the CM mode closes automatically with the chosen gate time $T_G$, independent of the value $g_m$.
For the rel mode, we have $u=\exp(-i\pi \Omega/\omega)$, and imposing $\alpha_f=0$ fixes the waveform
\begin{align}
    g_m
    &= 1-2\cos(\pi \Omega/\omega).
    \label{eq:si-gateB-gm}
\end{align}
Thus the three-segment pulse closes both modes exactly for arbitrary $\eta^2<1$ within the harmonic approximation.

The corresponding phases can also be evaluated analytically. 
For the $\ket{01}$ branch, and equally the $\ket{10}$ branch, one finds
\begin{align}
    \Phi_{10}^{(B)}
    &= \frac{\pi}{2}\frac{2\eta^2}{1-\eta^2}\left(1+g_m^2\right)z^2
    - \frac{\sin(2\pi \Omega/\omega)}{(\Omega/\omega)^3}z^2,
    \label{eq:si-gateB-phi10}
\end{align}
while the entangling phase is
\begin{align}
    \Phi_{ZZ}^{(B)}
    &= \pi\frac{2\eta^2}{1-\eta^2}\left(1+g_m^2\right)z^2
    - 2\frac{\sin(2\pi \Omega/\omega)}{(\Omega/\omega)^3}z^2.
    \label{eq:si-gateB-phizz}
\end{align}
Choosing $z$ such that $\Phi_{ZZ}^{(B)}=\pi$ gives the bang--bang amplitude used in Fig.~\ref{fig:fig2}(c). 
For the value plotted there, $\eta^2=0.01$, one has $\Omega/\omega=\sqrt{0.99}$, $g_m\approx3.000$, and $z_{\mathrm{CZ}}\approx2.121$. 
Because the $\ket{11}$ branch has twice the relative coherent shift of the $\ket{01}$ and $\ket{10}$ branches, and the phase is quadratic in the shift, one has $\Phi_{11}^{(B)}=4\Phi_{10}^{(B)}$.
The single-particle phase for a CZ gate is therefore again $\Phi_{10}^{(B)}=\pi/2$.

\subsubsection{Five-segment sub-trap-period gate (Gate C)}

The two dynamical gates discussed previously have gate durations $T_G$ that are limited by the trap period $2\pi/\omega$.
However, one may achieve fast two-ion gates by having the ions not complete full oscillations~\cite{Garcia-Ripoll2003Speed}.
Instead, the trap is shifted further away and the dynamics follow only an arc of their phase-space trajectories. 
A combination of these arcs then allows us to achieve exact closure of the phase-space trajectories and the desired entangling phase.

We describe a realization of such a sub-trap-period gate for the ion-tweezer platform which we propose.
This Gate~C uses a symmetric five-segment waveform,
\begin{equation}
    g(t)=
    \begin{cases}
        1, & 0<t<\tau_5,\\
        b, & \tau_5<t<2\tau_5,\\
        c, & 2\tau_5<t<4\tau_5,\\
        b, & 4\tau_5<t<5\tau_5,\\
        1, & 5\tau_5<t<6\tau_5,
    \end{cases}
    \qquad
    \tau_5=\frac{T_G}{6}.
    \label{eq:si-five-pulse}
\end{equation}
For a mode of frequency $\nu$, closure of the phase-space trajectories is equivalent to the vanishing of the Fourier component of the drive at the mode frequency~\cite{Garcia-Ripoll2005Coherent},
\begin{equation}
    I_\nu[g]=\int_0^{T_G}g(t)e^{-i\nu t}\,\mathrm{d}t=0.
    \label{eq:si-five-I-def}
\end{equation}
Evaluating the integral for the pulse in Eq.~\eqref{eq:si-five-pulse} yields the expression
\begin{align}
    I_\nu[g]
    &=\frac{4\sin(\nu\tau_5/2)}{\nu}e^{-i\nu T_G/2}
    \left[\cos\!\left(\frac{5\nu\tau_5}{2}\right) +b\cos\!\left(\frac{3\nu\tau_5}{2}\right) +c\cos\!\left(\frac{\nu\tau_5}{2}\right)\right].
    \label{eq:si-five-I}
\end{align}
Thus the center-of-mass and relative closure equations can be expressed by two linear equations
\begin{align}
    u_3 b+u_1 c&=-u_5,\nonumber\\
    v_3 b+v_1 c&=-v_5,
    \label{eq:si-five-closure}
\end{align}
where we define
\begin{equation}
    u_n=\cos\!\left(\frac{n\omega T_G}{12}\right),
    \qquad
    v_n=\cos\!\left(\frac{n\Omega T_G}{12}\right),
    \qquad n\in\{1,3,5\}.
\end{equation}
Solving the linear equations in Eq.~\eqref{eq:si-five-closure} for the shift values $b$ and $c$ yields
\begin{equation}
    b=\frac{u_1v_5-u_5v_1}{u_3v_1-u_1v_3},
    \qquad
    c=\frac{u_5v_3-u_3v_5}{u_3v_1-u_1v_3}.
    \label{eq:si-five-bc}
\end{equation}
For a fixed gate duration $T_G$, the normalized waveform $g(t)$ is then completely specified by Eq.~\eqref{eq:si-five-bc}.

We remind ourselves that both the scalar part of the phase and the jump phases are quadratic in the overall displacement scale $z$.
It is therefore convenient to evaluate the mixed and doubly displaced branches once at unit amplitude ($z=1$), which yields the phases $\Phi_{10}^{(C)}(z=1;T_G)$ and $\Phi_{11}^{(C)}(z=1;T_G)$.
The entangling phase, defined in Eq.~\eqref{eq:si-phizz-def}, then takes the form
\begin{align}
    \Phi_{ZZ}^{(C)}(z;T_G)
    &= z^2\Bigl[\Phi_{11}^{(C)}(1;T_G)-2\Phi_{10}^{(C)}(1;T_G)\Bigr].
\end{align}
By setting $\Phi_{ZZ}^{(C)}(z;T_G)=\pi$, the value for the displacement $z_{\mathrm{CZ}}$ is obtained (taking $z_{\mathrm{CZ}}$ to be positive).

The single-particle phase is
\begin{align}
    \Phi_{10}^{(C)}(T_G)
    &= z^2_{\mathrm{CZ}}(T_G)\,\Phi_{10}^{(C)}(1;T_G),
    \label{eq:si-five-phi10-cz}
\end{align}
while Fig.~\ref{fig:fig3}(d) shows the magnitude of the phase.
The maximum physical shift plotted in Fig.~\ref{fig:fig3}(c) is
\begin{align}
    \xi_{\max}(T_G)
    &= 2l_0 z_{\mathrm{CZ}}(T_G)\max\{1,|b|,|c|\}.
    \label{eq:si-five-ximax}
\end{align}
Fig.~\ref{fig:fig3} considers the resulting $\xi_{\max}$ and $|\Phi_{10}|$ as $T_G$ is varied for $\eta^2=0.36$.

We include a marker corresponding to a gate time $T_G=\qty{450}{\nano\second}$ in Fig.~\ref{fig:fig3}(c,d). 
The panels Fig.~\ref{fig:fig3}(a,b) correspond to this gate time.
For that marker, Eq.~\eqref{eq:si-five-bc} gives $b\approx -2.642$ and $c\approx 1.672$, while the CZ condition gives $z_{\mathrm{CZ}}\approx 16.73$.
Then, the normalized shift values in Fig.~\ref{fig:fig3}(a) are
\begin{equation}
    \xi(t)/2l_0=z_{\mathrm{CZ}}(1,b,c,b,1)\approx (16.7,-44.2,28.0,-44.2,16.7).
\end{equation}
For $\omega/2\pi=\qty{1}{\mega\hertz}$ this corresponds to a maximum displacement of $\xi_{\max}\approx\qty{0.53}{\micro\metre}$, while the single-particle phase has magnitude $|\Phi_{10}|\approx 2.96\times 10^3$.
Thus, this point provides a sub-trap-period benchmark that remains within the half-depth criterion, although with a significantly reduced margin compared with Gates~A and~B.

\subsubsection{Dipole-blockade gate}

The blockade protocol has three stages: 
first ion~1 is shelved and its wavepacket is transferred into the displaced excited-state trap, where it is held; 
second, a resonant target pulse is applied to ion~2; 
third, the preparation of ion~1 is reversed. 
For the calculation shown in Fig.~\ref{fig:fig4}, the first and third stages are treated as ideal unitary preparation and reversal of that displaced control configuration, so the explicit numerical simulation considers only the middle target pulse. 
In the exact convention adopted here, the first stage ends with ion~1 held in the \emph{doubly} displaced trap; the parameters for that preparation step are summarized at the end of this subsection.

During the middle pulse, it is convenient to choose the motional basis centered on the branch in which ion~1 is already placed in the displaced trap and ion~2 has not been acted on. 
In that basis, shelving ion~2 adds exactly the same incremental center-of-mass and relative displacements as in Eq.~\eqref{eq:si-mixed-deltas}. 
For a constant displacement scale $z$ one therefore has
\begin{equation}
    \alpha_{\mathrm{CM}}
    = \frac{z}{\sqrt{2}}, \qquad
    \alpha_{\mathrm{rel}}
    = \frac{z}{\sqrt{2}(\Omega/\omega)^{3/2}}.
    \label{eq:si-blockade-alphas}
\end{equation}
The ordinary interaction-induced detuning of the target transition in the singly displaced geometry is
\begin{equation}
    \nu_z = C_{11}-2C_{10}+C_{00} = -\omega z^2\frac{\eta^2}{1-\eta^2}.
    \label{eq:si-blockade-nuz}
\end{equation}
For Fig.~\ref{fig:fig4}, however, ion~1 is already placed in the displaced trap before the target pulse. 
The detuned target transition then compares relative displacements $2\xi\to3\xi$, whereas the resonant transition in the branch without a displaced control ion compares $0\to\xi$. 
Since the scalar shift in Eq.~\eqref{eq:si-Crs} is quadratic in the relative displacement, the difference in the detuning of the target transition is therefore
\begin{equation}
    \nu_z^{(\mathrm{dyn})}
    = 2\nu_z
    = -2\omega z^2\frac{\eta^2}{1-\eta^2},
    \label{eq:si-blockade-nuz-catch}
\end{equation}
which reproduces Eq.~\eqref{eq:nuz-def} from the main text.
Hence, the blocked branch during the target pulse is detuned by $\nu_z^{(\mathrm{dyn})}$ from the unblocked one.

For the branch in which the control ion has been displaced, the target pulse couples an initial two-mode motional state $\ket{n_{\mathrm{CM}},n_{\mathrm{rel}}}$ to displaced Fock states.
The coupling is given through the two-mode Franck--Condon amplitudes
\begin{equation}
    A_{m_{\mathrm{CM}},m_{\mathrm{rel}}}
    = \bra{m_{\mathrm{CM}},m_{\mathrm{rel}}}
    D_{\mathrm{CM}}^\dagger(\alpha_{\mathrm{CM}})
    D_{\mathrm{rel}}^\dagger(\alpha_{\mathrm{rel}})
    \ket{n_{\mathrm{CM}},n_{\mathrm{rel}}}.
    \label{eq:si-blockade-fc}
\end{equation}
For Fig.~\ref{fig:fig4}(b) the initial state is the vacuum, the probabilities then factorize as
\begin{equation}
    |A_{m_{\mathrm{CM}},m_{\mathrm{rel}}}|^2
    = e^{-|\alpha_{\mathrm{CM}}|^2-|\alpha_{\mathrm{rel}}|^2}\frac{|\alpha_{\mathrm{CM}}|^{2m_{\mathrm{CM}}}}{m_{\mathrm{CM}}!}
    \frac{|\alpha_{\mathrm{rel}}|^{2m_{\mathrm{rel}}}}{m_{\mathrm{rel}}!}.
\end{equation}
We consider the commensurate perpendicular geometry with $\Omega/\omega=4/5$, equivalently $\eta^2=9/25$, together with a scaling parameter $z=2/3$.
For these parameters the dominant Franck--Condon channel for the unblocked branch starting from the vacuum is still the carrier.
The probability for the vacuum-to-vacuum transition is $|A_{0,0}|^2\approx0.52$, while the largest sideband probability is $|A_{0,1}|^2\approx0.23$.
For the pulse we take the pulse envelope to be a smooth function
\begin{align}
    \Omega_R(t)
    &= \Omega_{\max}\sin^2\!\left(\frac{\pi t}{T_{2\pi}}\right), \qquad 0<t<T_{2\pi},
    \label{eq:si-blockade-envelope}
\end{align}
to suppress residual excitation of neighboring sidebands.
The peak Rabi frequency is then chosen such that the unblocked branch performs a resonant smooth $2\pi$-rotation on the vacuum-to-vacuum transition.
Larger values of $z$ are possible, but they move the Franck--Condon weights to higher sidebands and make the dynamics more complex.

For the numerical simulation, the Schrödinger equation is solved in a truncated two-mode Fock basis and the resulting two-qubit operation is compared with an ideal controlled-phase gate after correcting local phases.
The displayed curve in Fig.~\ref{fig:fig4}(b) uses Fock cutoffs $(N_{\mathrm{CM}},N_{\mathrm{rel}})=(13,12)$, which are sufficient for the plotted parameters.\\

\paragraph*{Dynamical preparation pulse.}
A concrete implementation of steps~(1) and~(3) uses a piecewise-constant shift as a dynamical alternative to a resonant pulse.
As in the gates A--C discussed previously, we have a time-dependent shift for the trap center, but now we do not displace the ions jointly and only shift the trap of the control ion.
The shift is given as
\begin{equation}
    \xi_{\mathrm c}(t)=2l_0 z\, g_{\mathrm c}(t),
\end{equation}
where the parameter $g_{\mathrm c}=1$ is the singly displaced trap reached immediately after shelving and $g_{\mathrm c}=2$ corresponds to the final trap position in which the motion is placed.
For the control-only branch, the single-shift coherent displacements are
\begin{equation}
    \beta_{\mathrm{CM}}=\frac{z}{\sqrt{2}}, \qquad
    \beta_{\mathrm{rel}}=\frac{z}{\sqrt{2}(\Omega/\omega)^{3/2}}.
\end{equation}
During segment $j$, the induced CM and rel equilibrium shifts are therefore
\begin{equation}
    \delta_{\mathrm{CM},j}=g_{\mathrm c,j}\beta_{\mathrm{CM}}, \qquad
    \delta_{\mathrm{rel},j}=g_{\mathrm c,j}\beta_{\mathrm{rel}},
\end{equation}
where $g_{\mathrm c,j}$ is the value of $g_{\mathrm c}(t)$ on that segment.
After the final shift to the held trap we have $g_{\mathrm c}=2$, and the corresponding shifts are
\begin{equation}
    \delta_{\mathrm{CM},f} = 2\beta_{\mathrm{CM}}, \qquad
    \delta_{\mathrm{rel},f} = 2\beta_{\mathrm{rel}}.
\end{equation}
The recursion of Eq.~\eqref{eq:si-recursion} then again applies separately to each mode $\mu\in\{\mathrm{CM},\mathrm{rel}\}$, with frequency $\nu_\mu\in\{\omega,\Omega\}$:
\begin{align}
    \alpha_{\mu,1}^+ &= -\delta_{\mu,1}, \notag\\
    \alpha_{\mu,j}^- &= \alpha_{\mu,j}^+ e^{-i\nu_\mu t_j}, \notag\\
    \alpha_{\mu,j+1}^+ &= \alpha_{\mu,j}^- + \delta_{\mu,j}-\delta_{\mu,j+1}.
\end{align}
For a minimal preparation pulse, the ion is first held in the singly displaced trap, so $\delta_{\mu,1}=\beta_\mu$, and then instantaneously moved to the final held trap, so $\delta_{\mu,2}=2\beta_\mu$.
This gives
\begin{equation}
    \alpha_{\mu,f}
    = -\beta_\mu\!\left(1+e^{-i\nu_\mu T_{\mathrm{dyn}}}\right),
\end{equation}
which vanishes only if
\begin{equation}
    e^{-i\omega T_{\mathrm{dyn}}}
    = e^{-i\Omega T_{\mathrm{dyn}}}
    = -1,
\end{equation}
that is,
\begin{equation}
    \omega T_{\mathrm{dyn}}=(2m+1)\pi, \qquad
    \Omega T_{\mathrm{dyn}}=(2n+1)\pi, \qquad
    \frac{\Omega}{\omega}=\frac{2n+1}{2m+1}.
\end{equation}
Hence this one-step dynamical preparation pulse works only for odd/odd commensurate ratios and does not apply to the example $\Omega/\omega=4/5$ used for Fig.~\ref{fig:fig4}(b).
For arbitrary $\eta^2$, one may instead use an exact three-segment dynamical preparation pulse,
\begin{equation}
    g_{\mathrm c}(t)=
    \begin{cases}
        1, & 0<t<\tau_\alpha,\\
        g_1, & \tau_\alpha<t<\tau_\alpha+\tau_\beta,\\
        g_2, & \tau_\alpha+\tau_\beta<t<\tau_\alpha+\tau_\beta+\tau_\gamma,
    \end{cases}
\end{equation}
followed by the final shift to $g_{\mathrm c}=2$, with
\begin{equation}
    \tau_\alpha=\frac{\pi}{\omega}, \qquad
    \tau_\beta=\frac{2\pi n_1}{\omega}, \qquad
    \tau_\gamma=\frac{2\pi n_2}{\omega}.
\end{equation}
The center-of-mass mode then closes automatically, while the relative mode closure is linear in $g_1$ and $g_2$.
Defining
\begin{equation}
    u=e^{-i\pi \Omega/\omega}, \qquad
    v_j=e^{-i2\pi n_j \Omega/\omega},
\end{equation}
and
\begin{equation}
    A=v_2(1-v_1), \qquad
    B=1-v_2, \qquad
    C=2-(1-u)v_1v_2,
\end{equation}
one convenient exact solution is
\begin{equation}
    g_1=\frac{\operatorname{Im}(CB^*)}{\operatorname{Im}(AB^*)}, \qquad
    g_2=\frac{\operatorname{Im}(A^*C)}{\operatorname{Im}(A^*B)}.
\end{equation}
For the simplest choice $n_1=n_2=1$, the total duration of the dynamical preparation pulse is $T_{\mathrm{dyn}}=5\pi/\omega$.
At the commensurate point $\Omega/\omega=4/5$ relevant for Fig.~\ref{fig:fig4}(b), this gives
\begin{equation}
    g_1 \approx 2.44721, \qquad
    g_2 \approx 1.55279.
\end{equation}
The reversal in step~(3) is obtained by time-reversing the same sequence. 
The construction is included here to show that the preparation and reversal of the parked branch can be made exact within the harmonic model.
However, in the numerics in Fig.~\ref{fig:fig4}(b) we only simulate the middle target pulse.

\subsubsection{Parallel alternative geometry} \label{si:parallel-geometry}

For completeness we also note an alternative geometry in which the excited-state trap is displaced along the ion--ion axis rather than perpendicular to it. 
This is not the focus of the main text: for Gates~A and~B the perpendicular geometry is preferable because it yields a smaller single-particle phase at fixed entangling phase. 
We nevertheless mention the parallel case because, for Gate~C, it can slightly reduce that phase in part of the parameter range, and because the same algebra carries over with only a simple replacement of the mode ratio.

The mode ratio for the parallel case is
\begin{equation}
    s_{\parallel}
    = \frac{\Omega_{\parallel}}{\omega}
    = \sqrt{1+2\eta^2},
    \label{eq:app-sigma-ratios}
\end{equation}
instead of $\Omega_\perp/\omega=\sqrt{1-\eta^2}$ in the perpendicular case. 
We have dropped the $\perp$ subscript throughout the manuscript because the perpendicular case is the default case considered here.
Thus, in the parallel geometry the rel mode frequency $\Omega_{\parallel}$ is stiffened compared to the bare trap frequency $\omega$.

\subsection{Trap anharmonicity of the Gaussian tweezer} \label{si:gaussian-nonlinearity}

We make the harmonic approximation used in Sec.~\ref{si:hamiltonian} explicit and complement the brief discussion of anharmonicity in Sec.~\ref{sec:considerations} of the main text. 
The harmonic model that we consider describes each occupied branch by the local curvature of the optical trap around its minimum. 
For a Gaussian tweezer with minimum at $q=\xi$, the tweezer potential can be described by
\begin{equation}
    V_{\mathrm{tw}}(q) = -U_0\exp\!\left[-\frac{2(q-\xi)^2}{w_0^2}\right],
    \label{si:V_tweezer}
\end{equation}
with beam waist $w_0$.
Expanding around the minimum gives
\begin{equation}
    V_{\mathrm{tw}}(q)
    = -U_0 + \frac{1}{2}m\omega^2(q-\xi)^2 - \frac{m\omega^2}{2w_0^2}(q-\xi)^4 + \mathcal O\!\left(\frac{m\omega^2(q-\xi)^6}{w_0^4}\right),
    \qquad
    \omega^2=\frac{4U_0}{mw_0^2}.
\end{equation}
For the computational branch $a$ and ion $j$, this implies the condition $|q_j-\xi_a^{(j)}|\ll w_0$ to remain in the harmonic regime. 
The leading correction to the harmonic potential is a negative quartic term. 
For weak anharmonicity, the effect of this term therefore slightly reduces the local oscillation frequency.
In the two-ion problem, this translates into small shifts of the effective center-of-mass and relative mode frequencies that depend on how far the ions are displaced from the trap minimum, including both the branch-dependent displacement and the thermal motion.
Equivalently, if the wavepacket remains narrow compared with $w_0$ throughout the motion, the potential can be approximated by its local curvature along the trajectory. 
The resulting effective Hamiltonian remains quadratic in the motional operators $a_{\rm CM}$, $a_{\rm CM}^\dagger$, $a_{\rm rel}$, and $a_{\rm rel}^\dagger$, but acquires time-dependent coefficients, in particular time-dependent effective mode frequencies, which reduce to constants in the purely harmonic limit. 
The evolution therefore remains Gaussian and can be incorporated in the dynamics, providing a natural starting point for an optimization of the pulse sequence with numerical methods.

Within the analytical treatment we ignore the quartic and higher corrections to the trapping potential, which is reasonable for the trajectories considered with Gates~A and~B.
The constraint becomes more restrictive for the sub-trap-period Gate~C, because large trap-center shifts can bring the wavepacket into the anharmonic regime of the tweezer potential.
For the Gate~C benchmark shown in Fig.~\ref{fig:fig3}, the required displacement remains below the half-depth reference shift $\xi_{1/2}$ for barium ions, but with a noticeably reduced margin compared with Gates~A and~B, as we discuss in the next section.

\subsection{Recalibration of trap cross-talk}

We considered the displacement of the tweezers as perfectly state selective in the gate model.
In practice, the push tweezer might apply a residual force on the qubit manifold, and the qubit tweezer could contribute a potential for the $\ket{e}$ state. 
Here we discuss that this does not necessarily change the structure of the gate model.

To simplify the analysis we assume that the potentials remain harmonic and that the trap frequency is matched for the relevant internal states by adjusting the laser intensities.
Under these conditions the two optical potentials combine into a harmonic trap whose center depends on the internal state.
The effect of cross-talk is then simply a rescaling of the displacement that enters the expressions for the gate.

We let $Q_{t}^{(j)}(t)$ and $Q_{p}^{(j)}(t)$ be the centers of the trapping and push tweezers for ion $j$. 
We use the subscript $q$ to denote either qubit state, $\ket 0$ or $\ket 1$, and keep $\ket e$ for the shelved state. 
For $a\in\{q,e\}$, the local harmonic potential is
\begin{equation}
    V_a^{(j)}
    =
    \frac{1}{2}k_{a,t}
    \left(q_j-Q_{t}^{(j)}\right)^2
    +
    \frac{1}{2}k_{a,p}
    \left(q_j-Q_{p}^{(j)}\right)^2 .
\end{equation}
Completing the square gives another harmonic potential, with curvature $K_a=k_{a,t}+k_{a,p}$ and center
\begin{equation}
    \tilde{\xi}_a^{(j)} = Q_{t}^{(j)} + \gamma_a
    \left(Q_{p}^{(j)}-Q_{t}^{(j)}\right),
    \qquad \gamma_a = \frac{k_{a,p}}{k_{a,t}+k_{a,p}}.
\end{equation}
Thus the ideal trap minima $\xi_r^{(j)}$ are replaced by the effective minima $\tilde{\xi}_r^{(j)}$. 
We assume that the beam intensities are chosen such that
\begin{equation}
    K_{q}=K_e=m\omega^2,
\end{equation}
so that the mode frequencies are the same as in the ideal case.
The change compared to the ideal model is only in the trap centers.
When the ideal model uses $\xi_r^{(j)}$, we now have the effective minima $\tilde{\xi}_r^{(j)}$.
The remaining parameter that is needed to describe the cross-talk is then the differential response between the qubit states and the shelved state,
\begin{equation}
    \mathcal R=\gamma_e-\gamma_{q}.
\end{equation}
We first consider fixed trap centers for the qubit tweezer and opposite displacements of the push tweezers:
\begin{equation}
    Q_{t}^{(1)}=Q_{t}^{(2)}=0,
    \qquad
    Q_{p}^{(j)}=\sigma_j\Xi(t),
    \qquad
    \sigma_1=-1,
    \quad
    \sigma_2=+1.
\end{equation}
The corresponding effective minima are
\begin{equation}
    \tilde{\xi}_a^{(j)}(t)=\sigma_j\gamma_a\Xi(t),
    \qquad
    \gamma_0=\gamma_1=\gamma_{q}.
\end{equation}
The relevant state dependence is the difference between the effective trap center of $\ket{e}$ and that of the qubit state,
\begin{equation}
    \tilde{\xi}_e^{(j)}-\tilde{\xi}_q^{(j)}
    =
    \sigma_j(\gamma_e-\gamma_q)\Xi(t)
    =
    \sigma_j\mathcal R\,\Xi(t).
\end{equation}
We write $\gamma_a=\gamma_q+n_a\mathcal R$, with $n_0=n_1=0$ and $n_e=1$.
Thus the ideal state-dependent displacement is replaced by the effective displacement $\xi_{\mathrm{eff}}(t)=\mathcal R\Xi(t)$.
For an internal branch $\ket{r,s}$ with $r,s\in\{0,1,e\}$, the center-of-mass and relative equilibrium positions become
\begin{align}
    X_{r,s}^{\mathrm{eq}}
    &=
    \frac{\tilde{\xi}_r^{(1)}+\tilde{\xi}_s^{(2)}}{2}
    =
    \frac{\gamma_s-\gamma_r}{2}\,\Xi(t)
    =
    \frac{n_s-n_r}{2}\,\mathcal R\,\Xi(t),
    \\
    x_{r,s}^{\mathrm{eq}}
    &=
    \frac{\tilde{\xi}_s^{(2)}-\tilde{\xi}_r^{(1)}}{1-\eta^2}
    =
    \frac{\gamma_s+\gamma_r}{1-\eta^2}\,\Xi(t)
    =
    \frac{2\gamma_q+(n_r+n_s)\mathcal R}{1-\eta^2}\,\Xi(t).
\end{align}
These expressions show explicitly where the cross-talk enters. 
The response $\gamma_q$ produces a branch-independent rel mode shift, while $\gamma_e$ adds a differential contribution $\mathcal R$. 
The part $\gamma_q$ can change global and single-qubit phases, but it cancels from the expression that yields the entangling phase, as we explain now.
In the presence of trap cross-talk,
\begin{equation}
    C_{r,s}^{\mathrm{ct}} = -\omega\frac{\eta^2}{8(1-\eta^2)} \left[\frac{\left(2\gamma_q+(n_r+n_s)\mathcal R\right)\Xi(t)}{l_0}\right]^2.
\end{equation}
The nonlocal contribution is then
\begin{align}
    C_{e,e}^{\mathrm{ct}} -C_{e,0}^{\mathrm{ct}} -C_{0,e}^{\mathrm{ct}} + C_{0,0}^{\mathrm{ct}}
    &=
    -\omega\frac{\eta^2}{8(1-\eta^2)}
    \left(\frac{\Xi(t)}{l_0}\right)^2
    \big[(2\gamma_q+2\mathcal R)^2 -2(2\gamma_q+\mathcal R)^2 +(2\gamma_q)^2\big]
    \notag\\
    &=
    -\omega\frac{\eta^2}{4(1-\eta^2)}
    \left(\frac{\mathcal R\Xi(t)}{l_0}\right)^2 .
\end{align}
This is the ideal scalar offset where $\Xi(t)$ has been replaced by the effective displacement $\xi_{\mathrm{eff}}(t)=\mathcal R\Xi(t)$. 
The same replacement similarly applies to the jump phases in piecewise-constant sequences, because the entangling phase depends on the same branch-dependent trap-center differences.

Writing the applied push displacement as
\begin{equation}
    \Xi(t)=2l_0 z_{\mathrm{app}} g(t),
\end{equation}
we obtain
\begin{equation}
    \Phi_{ZZ}^{\mathrm{ct}}(z_{\mathrm{app}})
    =
    \Phi_{ZZ}^{\mathrm{ideal}}
    \left(\mathcal R z_{\mathrm{app}}\right).
\end{equation}
Since the ideal entangling phase is quadratic in the displacement amplitude, the controlled-phase condition is recovered, for $\mathcal R\ne0$, by using $z_{\mathrm{app}} = z_{\mathrm{ideal}}/|\mathcal R|$.
The pulse-shape parameters that close phase space trajectories are unchanged and only the overall amplitude and the local single-qubit phases need to be recalibrated.

If both tweezer centers can be shifted, the effective trap minima can be made equal to the ideal ones directly. 
Choosing target values for the effective minima
\begin{equation}
    \tilde{\xi}_{0}^{(j)}
    =
    \tilde{\xi}_{1}^{(j)}
    =
    0,
    \qquad
    \tilde{\xi}_{e}^{(j)}
    =
    \sigma_j\xi_{\mathrm{ideal}}(t),
\end{equation}
for $\mathcal R\ne0$, the required beam center positions are
\begin{equation}
    Q_{t}^{(j)}(t)
    =
    -\frac{\gamma_{q}}{\mathcal R}
    \sigma_j\xi_{\mathrm{ideal}}(t),
    \qquad
    Q_{p}^{(j)}(t)
    =
    \frac{1-\gamma_{q}}{\mathcal R}
    \sigma_j\xi_{\mathrm{ideal}}(t).
\end{equation}
One then reproduces the ideal harmonic Hamiltonian, up to local phases. 

If the curvature is not state independent, the normal-mode frequencies become branch dependent and the simple amplitude rescaling is no longer exact. 
In that case the mismatch may be included in a numerical optimization of the pulse.

\subsection{Experimental parameters for barium ions}

We collect benchmark numbers for the Ba$^+$ implementation discussed in the manuscript.
Unless noted otherwise, we assume here a trap frequency of $\omega/2\pi=\qty{1}{\mega\hertz}$, an ion mass of $m\approx 138\,\mathrm{u}\approx\qty{2.29e-25}{\kilogram}$, and a beam waist $w_0=\qty{1}{\micro\metre}$.
We first summarize how the geometry influences the coupling, and then discuss optical power requirements.

First, we state the inter-ion distances $d$ with their corresponding $\eta^2$ values in Table~\ref{tab:feas-distances}, and recall the formula for the interaction parameter
\begin{equation}
    \eta^2 = \frac{e^2}{2\pi\varepsilon_0 m d^3 \omega^2}.
\end{equation}
A value of $\eta^2=0.36$ is used in Gate~A in Fig.~\ref{fig:fig2}(b), Gate~C in Fig.~\ref{fig:fig3}, and the dipole-blockade gate in Fig.~\ref{fig:fig4}, while $\eta^2=0.01$ is used for Gate~B in Fig.~\ref{fig:fig2}(c).
\begin{table}[H]
    \centering
    \begin{tabular}{ccc}
        \toprule
        $d$ (\si{\micro\metre}) & $\eta^2$ & $\Omega/\omega$\\
        \midrule
        5.00 & 0.408 & 0.769\\
        5.21 & 0.360 & 0.800\\
        10.0 & 0.0510 & 0.974 \\
        17.2 & 0.0100 & 0.995\\
        20.0 & $6.38\times 10^{-3}$ & 0.997\\
        30.0 & $1.89\times 10^{-3}$ & 0.999\\
        \bottomrule
    \end{tabular}
    \caption{Distance--coupling values for the perpendicular geometry at $\omega/2\pi=\qty{1}{\mega\hertz}$.}
    \label{tab:feas-distances}
\end{table}

We next compare the displacements during the different protocols with the available shift distance.
For the protocols, $\xi_{\mathrm{CZ}}=2l_0 z_{\mathrm{CZ}}$ is the displacement scale for realizing an entangling gate, while $\xi_{\max}=\max_t|\xi(t)|$ is the largest displacement reached within the pulse.
The oscillator length is $l_0=\sqrt{\hbar/(2m\omega)}\approx\qty{6.05}{\nano\metre}$.
For a Gaussian tweezer with waist $w_0=\qty{1}{\micro\metre}$, we define a reference shift $\xi_{1/2}$ by $V(\xi_{1/2})=-U_0/2$, so that
\begin{equation}
    \xi_{1/2}=w_0\sqrt{\ln 2/2}\approx\qty{0.589}{\micro\metre}.
\end{equation}
This reference shift provides a simple benchmark for keeping the motion within the harmonic part of the trapping potential while avoiding loss from the trap.
In Table~\ref{tab:feas-gates} we compare this value with the required displacements for the two-ion gates.
For Gate~C, we include the plotted benchmark $T_G=\qty{450}{\nano\second}$ at $\eta^2=0.36$ shown in Fig.~\ref{fig:fig3}.
\begin{table}[H]
    \centering
    \begin{tabular}{lcccc}
        \toprule
        Protocol & $\eta^2$ & $z_{\mathrm{CZ}}$ & $\xi_{\mathrm{CZ}}$ (\si{\micro\metre}) & $\xi_{\max}$ (\si{\micro\metre}) \\
        \midrule
        Gate~A & 0.36 & 0.422 & 0.0051 & 0.0051 \\
        Gate~B & 0.01 & 2.121 & 0.0257 & 0.0770 \\
        Gate~C ($T_G=\qty{450}{\nano\second}$) & 0.36 & 16.7 & 0.202 & 0.535 \\
        \bottomrule
    \end{tabular}
    \caption{Gate displacements compared with the half-depth shift $\xi_{1/2}\approx\qty{0.589}{\micro\metre}$ for $w_0=\qty{1}{\micro\metre}$.}
    \label{tab:feas-gates}
\end{table}

Thus, Gates~A and~B require a displacement of the order of nanometers or tens of nanometers and remain comfortably within the bound $\xi_{1/2}$.
For the plotted Gate~C benchmark at $T_G=\qty{450}{\nano\second}$ and $\eta^2=0.36$, the maximum displacement is $\xi_{\max}\approx\qty{0.53}{\micro\metre}$ and therefore remains below $\xi_{1/2}\approx\qty{0.589}{\micro\metre}$, although with noticeably reduced margin.
Substantially shorter sub-trap-period gates would exceed $\xi_{1/2}$ unless the confinement is increased or the trap is engineered to be more nearly harmonic.

The displacement values in Table~\ref{tab:feas-gates} are set by $\eta^2$, the gate time, and the pulse shape.
They do not depend on the effective polarizability $\alpha$ of the ions.
The role of $\alpha$ is instead to determine the optical power $P$ required to realize the assumed confinement~\cite{Grimm2000Optical},
\begin{equation}
    P = \frac{\pi \varepsilon_0 mc}{4\alpha}\,\omega^2 w_0^4.
\end{equation}
We consider $\alpha=\qty{200}{\atomicunit}$ and $\alpha=\qty{500}{\atomicunit}$ as two representative scenarios, motivated by values for Ba$^+$ discussed in the Methods.
Table~\ref{tab:feas-alpha} summarizes the optical requirements $P_{\mathrm{req}}$ for the nominal benchmark $\omega/2\pi=\qty{1}{\mega\hertz}$ and $w_0=\qty{1}{\micro\metre}$.
\begin{table}[H]
    \centering
    \begin{tabular}{lcc}
        \toprule
        $\alpha$ (\si{\atomicunit}) & $P_{\mathrm{req}}$ (\si{\watt}) & $\omega/2\pi$ at $P=\qty{2}{\watt}$ (\si{\mega\hertz}) \\
        \midrule
        200 & 5.72 & 0.591\\
        500 & 2.29 & 0.935\\
        \bottomrule
    \end{tabular}
    \caption{Optical requirements for the two representative effective polarizabilities at the nominal benchmark $\omega/2\pi=\qty{1}{\mega\hertz}$ and $w_0=\qty{1}{\micro\metre}$.}
    \label{tab:feas-alpha}
\end{table}

One may also ask what stronger confinement would be required to obtain a faster sub-trap-period Gate~C while keeping the displacement at the half-depth scale $\xi_{1/2}$.
The required values are listed in Table~\ref{tab:feas-fast200} for a shorter reference time $T_G=\qty{200}{\nano\second}$.
For $\eta^2=0.36$ and $\alpha=\qty{500}{\atomicunit}$, this corresponds to about $\qty{8}{\watt}$ per tweezer at an inter-ion distance of $d\approx\qty{3.4}{\micro\metre}$.
\begin{table}[H]
    \centering
    \begin{tabular}{cccccc}
        \toprule
        $\eta^2$ & $\xi_{\max}$ at $\qty{1}{\mega\hertz}$ (\si{\micro\metre}) & required $\omega/2\pi$ (\si{\mega\hertz}) & \multicolumn{2}{c}{required $P$ (\si{\watt})} & required $d$ (\si{\micro\metre}) \\
        \cmidrule(lr){4-5}
         &  &  & $\alpha=\qty{200}{\atomicunit}$ & $\alpha=\qty{500}{\atomicunit}$ &  \\
        \midrule
        0.010 & 24.4 & 3.37 & 64.9 & 25.9 & 7.66 \\
        0.360 & 4.07 & 1.90 & 20.8 & 8.30 & 3.39 \\
        \bottomrule
    \end{tabular}
    \caption{Gate~C ($T_G=\qty{200}{\nano\second}$) at $w_0=\qty{1}{\micro\metre}$. The required $\omega/2\pi$ is chosen such that $\xi_{\max}=\xi_{1/2}$, and the corresponding power is shown for both representative effective polarizabilities.}
    \label{tab:feas-fast200}
\end{table}

We conclude by stating that we expect Gates~A and~B to be realizable with current laser technology.
By contrast, gates that are much faster than the trap period would require substantially stronger confinement and more optical power per tweezer, depending on the target gate time, the effective polarizability and the ion distance.
In the Methods, we also consider a trap frequency of $\omega/2\pi=\qty{100}{\kilo\hertz}$, which at fixed waist and polarizability requires one percent of the optical power per tweezer needed for $\omega/2\pi=\qty{1}{\mega\hertz}$.

\subsection{Gate diagnostics}

This section collects several formulas used in the numerical analysis of the gates.

\subsubsection{Thermal-state fidelity} \label{si:thermal-fidelity}

We analyze the gate fidelity of the two-ion gates for initial states that are not in the motional vacuum, but have some residual temperature.
The main physical point is that an exactly closed harmonic gate does \emph{not} decohere when the initial motion is thermal~\cite{Garcia-Ripoll2005Coherent}. 
The temperature dependence in the robustness panels of Fig.~\ref{fig:fig2} therefore comes only from deviations from exact closure and from a two-qubit phase error.

After removing local $\sigma^z$ phases, the branch operators for the perpendicular protocols can be written as
\begin{align}
    V_{00}
    &= \openone_{\mathrm{CM}}\otimes \openone_{\mathrm{rel}},
    \notag\\
    V_{01}
    &= D_{\mathrm{CM}}(+\Delta\alpha_{\mathrm{CM}})D_{\mathrm{rel}}(\Delta\alpha_{\mathrm{rel}}),
    \notag\\
    V_{10}
    &= D_{\mathrm{CM}}(-\Delta\alpha_{\mathrm{CM}})D_{\mathrm{rel}}(\Delta\alpha_{\mathrm{rel}}),
    \notag\\
    V_{11}
    &= -\mathrm e^{i\delta\Phi}D_{\mathrm{rel}}(2\Delta\alpha_{\mathrm{rel}}),
    \label{eq:app-heatmap-ops}
\end{align}
where $\Delta\alpha_{\mathrm{CM}}$ and $\Delta\alpha_{\mathrm{rel}}$ are the residual final coherent displacements of the $\ket{01}$ branch relative to perfect closure. 
The $\ket{10}$ branch differs only by the sign of CM displacement.
If the protocol closes exactly, then $\Delta\alpha_{\mathrm{CM}}=\Delta\alpha_{\mathrm{rel}}=0$ and only a phase remains. 
The entangling-phase error is
\begin{equation}
    \delta\Phi = \Phi_{11}-2\Phi_{10}-\pi.
    \label{eq:app-dphi-def}
\end{equation}
We assume an initial product of thermal states,
\begin{equation}
    \rho_{\mathrm{th}} = \rho_{\mathrm{th}}^{(\mathrm{CM})}\otimes\rho_{\mathrm{th}}^{(\mathrm{rel})},
\end{equation}
whose characteristic function is
\begin{equation}
    \operatorname{tr}\!\left[\rho_{\mathrm{th}}D_{\mathrm{CM}}(\zeta_{\mathrm{CM}})D_{\mathrm{rel}}(\zeta_{\mathrm{rel}})\right]
    = \exp\!\left[-\left(\bar n_{\mathrm{CM}}+\frac{1}{2}\right)|\zeta_{\mathrm{CM}}|^2
    -\left(\bar n_{\mathrm{rel}}+\frac{1}{2}\right)|\zeta_{\mathrm{rel}}|^2\right].
    \label{eq:app-thermal-char}
\end{equation}
We define the shorthands
\begin{equation}
     \lambda_{\mathrm{CM}} = \left(\bar n_{\mathrm{CM}}+\frac{1}{2}\right)|\Delta\alpha_{\mathrm{CM}}|^2, \qquad
     \lambda_{\mathrm{rel}} = \left(\bar n_{\mathrm{rel}}+\frac{1}{2}\right)|\Delta\alpha_{\mathrm{rel}}|^2,
     \qquad \Lambda = \lambda_{\mathrm{CM}}+\lambda_{\mathrm{rel}},
     \label{eq:app-thermal-lambdas}
\end{equation}
and the overlaps between computational basis branches $\mu, \nu$:
\begin{equation}
    \gamma_{\mu,\nu} = \operatorname{tr}[\rho_{\mathrm{th}} V_\nu^\dagger V_\mu].
    \label{eq:gamma-def}
\end{equation}
One finds the following nontrivial overlaps:
\begin{align}
    \gamma_{1,2}=\gamma_{1,3}
    &= \mathrm e^{-\Lambda}, \qquad&
    \gamma_{2,4}=\gamma_{3,4}
    &= -\mathrm e^{-i\delta\Phi}\,\mathrm e^{-\Lambda},
    \notag\\
    \gamma_{2,3}
    &= \mathrm e^{-4\lambda_{\mathrm{CM}}},
    \qquad&
    \gamma_{1,4}
    &= -\mathrm e^{-i\delta\Phi}\,\mathrm e^{-4\lambda_{\mathrm{rel}}}.
    \label{eq:app-thermal-gammas}
\end{align}
For a four-dimensional channel, the average gate fidelity is related to the entanglement fidelity by~\cite{Nielsen2002Simple}.
Then, for a target controlled-$Z$ gate, $U_{\rm CZ} = \operatorname{diag}(1,1,1,-1) = (u_1,u_2,u_3,u_4)$, the corresponding average fidelity is
\begin{equation}
    \bar F
    = \frac{4F_e+1}{5},\qquad
    F_e = \frac{1}{16}\sum_{\mu,\nu=1}^{4}\widetilde{\gamma}_{\mu\nu}, \qquad \widetilde{\gamma}_{\mu\nu}
    = u_\mu^* u_\nu\,\gamma_{\mu\nu}.
\end{equation}
We obtain
\begin{align}
    \bar F
    &= \frac{2}{5}
    + \frac{1}{10}\operatorname{Re}\!\left[2\gamma_{1,2}+\gamma_{2,3}-2\gamma_{2,4}-\gamma_{1,4}\right]
    \notag\\
    &= \frac{2}{5}
    + \frac{1}{10}\left[
    2\mathrm e^{-\Lambda}
    + \mathrm e^{-4\lambda_{\mathrm{CM}}}
    + \left(2\mathrm e^{-\Lambda}+\mathrm e^{-4\lambda_{\mathrm{rel}}}\right)\cos(\delta\Phi)
    \right].
    \label{eq:app-thermal-fid}
\end{align}
This expression is evaluated from the numerically propagated residual displacements and the phase mismatch.

In the plotted heatmaps, the two occupations correspond to a common physical temperature. 
Writing
\begin{align} \label{eq:si-temp-fid}
    \bar n_{\mathrm{CM}}
    &= \frac{1}{\mathrm e^{\beta\omega}-1}, \qquad \bar n_{\mathrm{rel}}
    = \frac{1}{\mathrm e^{\beta\Omega}-1},
    \qquad
    \beta
    = \frac{1}{k_{\mathrm B}T},
\end{align}
one obtains
\begin{align}
    \bar n_{\mathrm{rel}}
    &= \frac{1}{\left(1+1/\bar n_{\mathrm{CM}}\right)^{\Omega/\omega}-1}.
    \label{eq:app-common-T}
\end{align}
This relation is used to generate the secondary temperature axis in Fig.~\ref{fig:fig2}.

\subsubsection{Robustness panels for Gate~B}

For the three-segment pulses, the closure conditions enforce $\Delta\alpha_{\mathrm{CM}}=\Delta\alpha_{\mathrm{rel}}=0$, so the ideal Gate~B is temperature independent within the harmonic model. 
However, if the trap center has a small imperfection, the phase-space trajectories of the modes do not close perfectly anymore, which affects the gate fidelity.
An initial state with higher temperature is more sensitive to such imperfections, as can be seen from Eq.~\eqref{eq:app-thermal-lambdas}.

The perturbation model used in Fig.~\ref{fig:fig2}(c) is
\begin{align}
    (1,g_m,1)
    &\to
    \Bigl(1+\chi_0\varepsilon,\; g_m(1+\chi_1\varepsilon),\; 1+\chi_0\varepsilon\Bigr),
    \qquad
    \chi_0,\chi_1\in\{-1,+1\}.
    \label{eq:app-three-error-model}
\end{align}
We assume that the two outer segments share the same sign because they correspond to an identical trap center position.
For each $\varepsilon$, the $\ket{01}$ and $\ket{11}$ branches are recomputed with the same calibrated $z_{\mathrm{CZ}}$, and the residual displacements together with the phase error are inserted into Eq.~\eqref{eq:app-thermal-fid}.
The plotted value is the pointwise worst case fidelity over the four sign choices in Eq.~\eqref{eq:app-three-error-model}.
The horizontal axis in Fig.~\ref{fig:fig2}(c) is the relative position error $\varepsilon$.

\subsubsection{Cross-talk analysis}

For Fig.~\ref{fig:fig5}(a) we simulate two simultaneously driven perpendicular bang--bang gates, each identical to Fig.~\ref{fig:fig2}(c), in a four-ion harmonic model.
The ion positions are
\begin{align}
    \bm R_1 &= \left(-\frac{d}{2},0\right), &
    \bm R_2 &= \left(+\frac{d}{2},0\right), \notag\\
    \bm R_3 &= \left(-\frac{d}{2},L\right), &
    \bm R_4 &= \left(+\frac{d}{2},L\right),
    \label{eq:si-fourion-geometry}
\end{align}
so that ions $(1,2)$ and $(3,4)$ form the two active gates. 
The distance between two ions that form an active gate is $d$, while the distance between the gates is $L$.
We keep the one-dimensional gate model and consider motion along the common push direction $\hat{\bm e}_{\perp}$.
The angle between the push direction and the square array axis is $\phi$.

Let $q_i$ denote the dimensionless displacement of ion $i$ along that direction, measured in units of $d$, and let $V_0$ be a potential measured in units of $m\omega^2 d^2$. 
Expanding about the branch-independent equilibrium then gives
\begin{equation}
    V_0(\bm q)
    = \frac{1}{2}\sum_{i=1}^{4} q_i^2
    + \frac{1}{2}\sum_{i<j} c_{ij}(q_i-q_j)^2, \qquad
    c_{ij} = \frac{\eta^2}{2}\frac{3(\hat{\bm R}_{ij}\cdot\hat{\bm e}_{\perp})^2-1}{(R_{ij}/d)^3},
    \label{eq:si-fourion}
\end{equation}
with $\bm q=(q_1,q_2,q_3,q_4)^T$, $\bm R_{ij}=\bm R_i-\bm R_j$, $R_{ij}=\lVert\bm R_{ij}\rVert$, and $\hat{\bm R}_{ij}=\bm R_{ij}/R_{ij}$.
Thus the inter-gate couplings are dipolar and scale as $(L/d)^{-3}$.
We define a real symmetric stiffness matrix $K$, as
\begin{align}
    V_0(\bm q)=\tfrac{1}{2}\bm q^T K\bm q.
\end{align}
For a computational branch $\bm b=(b_1,b_2,b_3,b_4)$, define the corresponding sign vector $\bm u(\bm b)=(b_1,-b_2,b_3,-b_4)^T$.
The three segments of the bang--bang gate act with amplitudes $g_j\in\{1,g_m,1\}$, and each normal mode obeys the same piecewise-displaced-oscillator recursion as in Eq.~\eqref{eq:si-recursion}.
Hence, for the vacuum initial state used here, each branch $\bm b$ yields a total phase $\Phi_{\bm b}$ and a multimode coherent-state displacement $\bm\alpha_{\bm b}$,
\begin{align}
    U_{\mathrm{full}}\ket{\bm b}\ket{0}_{\mathrm{mot}}
    &= \mathrm e^{i\Phi_{\bm b}}\ket{\bm b}\ket{\bm\alpha_{\bm b}}.
    \label{eq:si-fourion-branch-map}
\end{align}
Tracing out the neighboring driven pair and the motion gives a diagonal reduced channel for the chosen ion pair,
\begin{equation}
    \mathcal E_{\mathrm{pair}}(\ket{a}\bra{a'})
    = \Gamma_{a,a'}\ket{a}\bra{a'}, \quad \text{with}\quad
    \Gamma_{a,a'}
    = \frac{1}{4}\sum_{t\in\{00,01,10,11\}}
    \mathrm e^{i[(\Phi_{a,t}-\varphi_a^{\mathrm{iso}})-(\Phi_{a',t}-\varphi_{a'}^{\mathrm{iso}})]}
    \langle \bm\alpha_{a',t}|\bm\alpha_{a,t}\rangle,
    \label{eq:si-fourion-gamma}
\end{equation}
where $a$ and $a'$ denote the branch of the pair of interest, $t$ the branch of the neighboring driven pair, and $\varphi_a^{\mathrm{iso}}$ is the isolated-gate reference phase of branch $a$. The plotted gate infidelity is $1-\bar F_{\mathrm{pair}}$, with
\begin{align}
    F_e^{(\mathrm{pair})}
    &= \frac{1}{16}\sum_{a,a'} \Gamma_{a,a'},
    &
    \bar F_{\mathrm{pair}}
    &= \frac{4F_e^{(\mathrm{pair})}+1}{5}.
    \label{eq:si-fourion-fidelity}
\end{align}

For Fig.~\ref{fig:fig5}(b) we apply the same construction to an odd $N_s\times N_s$ square array of simultaneously driven pairs, all with common orientation angle $\phi$. 
Let $N_p=N_s^2$ be the total number of driven pairs, and let $c$ denote the central pair whose fidelity is plotted. 
The computational states of that central pair are labeled by $a,a'\in\{00,01,10,11\}$, while the spectator-pair states are collected into
\begin{align}
    \mathbf t = \{t_q\}_{q\neq c},
    \qquad
    t_q\in\{00,01,10,11\}.
\end{align}

As in panel~(a), the branch dependence enters only through the global sign vector $\bm u$. 
Within the harmonic model, the equations of motion are linear in the drive, so the final multimode displacement is linear in $\bm u$. 
The branch-dependent phase is generated by completed-square offsets and by the phases from sudden trap-center changes, each quadratic in those same linear displacements. 
We may therefore write
\begin{equation}
    \bm\alpha(\bm u) = A\bm u, \qquad
    \Phi(\bm u) = \bm u^T M\bm u,
    \label{eq:si-parallel-linear-quadratic}
\end{equation}
ignoring the branch-independent constants.
We note that Eq.~\eqref{eq:si-parallel-linear-quadratic} is structurally the same as in Ref.~\cite{Garcia-Ripoll2005Coherent}.
For the vacuum initial state, the reduced channel of the central pair is again diagonal:
\begin{equation}
    \mathcal E_c(\ket{a}\bra{a'})
    = \Gamma^{(c)}_{a,a'}\ket{a}\bra{a'}, \quad \text{with}\quad
    \Gamma^{(c)}_{a,a'}
    = \frac{1}{4^{N_p-1}}\sum_{\mathbf t} W_{a,a'}(\mathbf t),
    \label{eq:si-central-gamma}
\end{equation}
where
\begin{equation}
    W_{a,a'}(\mathbf t)
    = \mathrm e^{i[(\Phi_{a,\mathbf t}-\varphi_a^{\mathrm{iso}})-(\Phi_{a',\mathbf t}-\varphi_{a'}^{\mathrm{iso}})]}
    \langle \bm\alpha_{a',\mathbf t}|\bm\alpha_{a,\mathbf t}\rangle.
    \label{eq:si-central-weight}
\end{equation}
Direct evaluation of Eq.~\eqref{eq:si-central-gamma} would require evaluating $4^{N_p-1}$ spectator configurations and is therefore impractical already for moderate array sizes.

It turns out that we may significantly reduce the number of terms that need to be evaluated. 
We split the sign vector into two contributions as
\begin{equation}
    \bm u(a,\mathbf t)
    = \bm V_a + \bm U(\mathbf t),
    \qquad
    \bm U(\mathbf t)=\sum_{q\neq c}\bm U_q(t_q),
    \label{eq:si-parallel-sign-split}
\end{equation}
where $\bm V_a$ has support only on the central pair and $\bm U_q(t_q)$ only on spectator pair $q$. 
Inserting the sign vector from Eq.~\eqref{eq:si-parallel-sign-split} into the phase from Eq.~\eqref{eq:si-parallel-linear-quadratic} yields
\begin{equation}
    \Phi_{a,\mathbf t}
    = \bm V_a^T M\bm V_a + 2\bm V_a^T M\bm U(\mathbf t) + \bm U(\mathbf t)^T M\bm U(\mathbf t).
\end{equation}
We note that the term $\bm U^T M\bm U$, which only has spectator contributions, cancels exactly in the phase difference in Eq.~\eqref{eq:si-central-weight}. 
Writing $G=A^\dagger A$, the exponent in the coherent-state overlap in Eq.~\eqref{eq:si-central-weight} is
\begin{equation}
     \ln\langle \bm\alpha_{a',\mathbf t}|\bm\alpha_{a,\mathbf t}\rangle=-\frac{1}{2}\bm u^T G\bm u - \frac{1}{2}\bm u'^T G\bm u' + \bm u'^T G\bm u,
\end{equation}
with $\bm u=\bm V_a+\bm U(\mathbf t)$ and $\bm u'=\bm V_{a'}+\bm U(\mathbf t)$. 
The terms quadratic in $\bm U(\mathbf t)$ therefore cancel here as well. 
The dependence on the spectator terms is therefore only linear in Eq.~\eqref{eq:si-central-weight}, so each weight can be written as
\begin{equation}
    W_{a,a'}(\mathbf t) = C_{a,a'}\exp\!\left[i\,\bm\lambda_{a,a'}\!\cdot\!\bm U(\mathbf t)\right] = C_{a,a'}\prod_{q\neq c}\exp\!\left[i\,\bm\lambda_{a,a'}\!\cdot\!\bm U_q(t_q)\right],
\end{equation}
with spectator-independent coefficients $C_{a,a'}$ and the complex $\bm\lambda_{a,a'}$ determined by the harmonic evolution of the full array. 
Averaging over spectator pairs then gives
\begin{equation}
    \Gamma^{(c)}_{a,a'}
    = C_{a,a'}\prod_{q\neq c}
    \Bigg[
    \frac{1}{4}\sum_{t_q\in\{00,01,10,11\}}
    \mathrm e^{i\bm\lambda_{a,a'}\cdot \bm U_q(t_q)}
    \Bigg].
    \label{eq:si-parallel-factorized}
\end{equation}
This factorization is exact within the harmonic model and reduces the computation from $4^{N_p-1}$ spectator configurations to one spectator-independent prefactor and $4(N_p-1)$ single-pair terms. 

The blue curves in Fig.~\ref{fig:fig5}(b) then show the infidelity $1-\bar F_c$, where 
\begin{align}
    F_e^{(c)}=\tfrac{1}{16}\sum_{a,a'}\Gamma^{(c)}_{a,a'}, \qquad
    \bar F_c=\frac{4F_e^{(c)}+1}{5}.
\end{align}
The green curves additionally optimize over local $\sigma^z$ rotations on the ions of the central gate.

\end{document}